\documentclass{aa}  
\usepackage{graphicx}
\usepackage[toc,page]{appendix}
\usepackage{caption}
\usepackage{float}

\usepackage{txfonts}

\usepackage{array}
\newcolumntype{P}[1]{>{\centering\arraybackslash}p{#1}}
\newcolumntype{M}[1]{>{\centering\arraybackslash}m{#1}}

\usepackage{natbib}
\bibpunct{(}{)}{;}{a}{}{,}

\usepackage{color}
\usepackage{ulem}

\begin{document} 

   \title{Clustering dependence on Ly$\alpha$ luminosity from MUSE surveys at $3<z<6$}
   \author{Yohana Herrero Alonso,
          \inst{1}
          T. Miyaji,
          \inst{2}
          L. Wisotzki,
          \inst{1}
          M. Krumpe,
          \inst{1}
          J. Matthee,
          \inst{4}
          J. Schaye,
          \inst{3}
          H. Aceves, 
          \inst{2}
          H. Kusakabe,
          \inst{5}
          \and
          T. Urrutia
          \inst{1}
          }

   \institute{Leibniz-Institut f\"ur Astrophysik Potsdam (AIP), An der Sternwarte 16, D-14482 Potsdam, Germany\\
              \email{yherreroalonso@aip.de}
         \and
            Universidad Nacional Aut\'onoma de M\'exico, Instituto de Astronom\'ia (IA-UNAM-E), AP 106, Ensenada 22860, BC, M\'exico
        \and
             Leiden Observatory, Leiden University, P.O. Box 9513, 2300 RA, Leiden, The Netherlands
        \and
             Department of Physics, ETH Zurich, Wolfgang-Pauli-Strasse 27, 8093 Zurich, Switzerland
        \and
             Observatoire de G\`eneve, Universit\'e de G\`eneve, 51 Chemin de P\'egase, 1290 Versoix, Switzerland
             }

   \date{Received xxx/Accepted xxx} 
 
  \abstract{We investigate the dependence of Ly$\alpha$ emitter (LAE) clustering on Ly$\alpha$ luminosity and connect the clustering properties of $\approx L^{\star}$ LAEs with those of much fainter ones, namely, $\approx0.04L^{\star}$. We use 1030 LAEs from the MUSE-Wide survey, 679 LAEs from MUSE-Deep, and 367 LAEs from the to-date deepest ever spectroscopic survey, the MUSE Extremely Deep Field. All objects have spectroscopic redshifts of $3<z<6$ and cover a large dynamic range of Ly$\alpha$ luminosities: $40.15<\log (L_{\rm{Ly}\alpha}/\rm{erg \:s}^{-1})<43.35$. We apply the Adelberger et al. K-estimator as the clustering statistic and fit the measurements with state-of-the-art halo occupation distribution (HOD) models. 
  We find that the large-scale bias factor increases weakly with an increasing line luminosity. For the low-luminosity ($ \log\langle L_{\rm{Ly}\alpha}/[\rm{erg\: s}^{-1}] \rangle=41.22$) and intermediate-luminosity ($ \log\langle L_{\rm{Ly}\alpha}/[\rm{erg\: s}^{-1}] \rangle=41.64$) LAEs,  we compute consistent bias factors $b_{\rm{low}}=2.43^{+0.15}_{-0.15}$ and $b_{\rm{interm.}}=2.42^{+0.10}_{-0.09}$, whereas for the high-luminosity ($ \log\langle L_{\rm{Ly}\alpha}/[\rm{erg\: s}^{-1}] \rangle=42.34$) LAEs we calculated $b_{\rm{high}}=2.65^{+0.13}_{-0.11}$. Consequently, high-luminosity LAEs occupy dark matter halos (DMHs) with typical masses of $\log (M_h/[h^{-1}M_\odot])=11.09^{+0.10}_{-0.09}$, while low-luminosity LAEs reside in halos of $\log (M_h/[h^{-1}M_\odot])=10.77^{+0.13}_{-0.15}$. The minimum masses to host one central  LAE, $M_{\rm{min}}$, and (on average) one satellite  LAE, $M_1$, also vary with Ly$\alpha$ luminosity, growing from $\log (M_{\rm{min}}/[h^{-1}M_\odot])=10.3^{+0.2}_{-0.3}$ and $\log (M_1/[h^{-1}M_\odot])=11.7^{+0.3}_{-0.2}$ to $\log (M_{\rm{min}}/[h^{-1}M_\odot])=10.7^{+0.2}_{-0.3}$ and $\log (M_1/[h^{-1}M_\odot])=12.4^{+0.4}_{-0.6}$ from low- to high-luminosity samples, respectively. The satellite fractions are $\lesssim10$\% ($\lesssim20$\%) at $1\sigma$ ($3\sigma$) confidence level, supporting a scenario in which DMHs typically host one single LAE. We next bisected the three main samples into disjoint subsets to thoroughly explore the dependence of the clustering properties on $L_{\rm{Ly}\alpha}$. We report a strong ($8\sigma$) clustering dependence on Ly$\alpha$ luminosity, not accounting for cosmic variance effects, where the highest luminosity LAE subsample ($ \log (L_{\rm{Ly}\alpha}/\rm{erg \:s}^{-1})\approx42.53$) clusters more strongly ($b_{\rm{highest}}=3.13^{+0.08}_{-0.15}$) and resides in more massive DMHs ($\log(M_{\rm{h}} / [h^{-1}\rm{M}_{\odot}])=11.43^{+0.04}_{-0.10}$) than the lowest luminosity one ($ \log (L_{\rm{Ly}\alpha}/\rm{erg \:s}^{-1})\approx40.97$), which presents a bias of $b_{\rm{lowest}}=1.79^{+0.08}_{-0.06}$ and occupies $\log(M_{\rm{h}} / [h^{-1}\rm{M}_{\odot}])=10.00^{+0.12}_{-0.09}$ halos. We discuss the implications of these results for evolving Ly$\alpha$ luminosity functions, halo mass dependent Ly$\alpha$ escape fractions, and incomplete reionization signatures.   }

    \keywords{large-scale structure -- high-redshift galaxies -- HOD models -- dark matter halo -- satellite galaxies}

   \titlerunning{Strong clustering dependence on Ly$\alpha$ luminosity at $3<z<6$}
   \authorrunning{Yohana Herrero Alonso et al.}
 
   \maketitle
%

\section{Introduction}
\label{sec:introduction}

Dark matter halos (DMHs) serve as sites of galaxy formation but their co-evolution is still a matter of investigation. Observations deliver snapshots of the luminosities of galaxies at given redshifts, while numerical analyses succeed at simulating the evolution and copiousness of DMHs. Linking these two constituents is not straightforward but, because the spatial distribution of baryonic matter is biased against that of dark matter (DM), the former indirectly traces the latter. The evolutionary stage of the two distributions depends on both the epoch of galaxy formation and the physical properties of galaxies (see \citealt{wechsler} for a review). Thus, studying the dependence of the baryonic-DM relation on galaxy properties is essential for better understanding the evolution of the two components.

Exploring the spatial distribution of high-redshift ($z>2$) galaxies and its dependence on physical properties provides an insight into the early formation and evolution of the galaxies we observe today. Clustering statistics yield observational constraints on the relationship between galaxies and DMHs, as well as on their evolution. Traditional studies of high-$z$ galaxies \citep{steidel96, hu98,ouchi03, gawiser07,ouchi10,khostovan19} model the large-scale ($R\gtrsim1-2\;h^{-1}$cMpc) clustering statistics with a two parameter power-law correlation function that takes the form $\xi=(r/r_0)^{-\gamma}$ \citep{davispeebles} to derive the large-scale linear galaxy bias and the associated typical DMH mass. To make full use of the clustering measurements, the smaller separations of the nonlinear  regime ($R\lesssim1-2\;h^{-1}$cMpc) are modeled by relating galaxies to DMHs within the nonlinear framework of halo occupation distribution (HOD) modeling. In this context, the mean number of galaxies in the DMH is modeled as a function of DMH mass, further assessing whether these galaxies occupy the centers of the DMHs or whether they are satellite galaxies. 

Although clustering studies of high-redshift galaxies are plentiful, HOD modeling has been rarely used to interpret the results. While several works have focused on Lyman-break galaxy (LBG) surveys, only one study fit a sample of Lyman-$\alpha$ emitters (LAEs) with HOD models \citep{ouchi17}. \cite{durkalec14,malkan,hatfield,harikane18} applied the full HOD framework to sets of LBGs to put constraints on the central and satellite galaxy populations, while \cite{ouchi17} partially exploited the power of HOD models in a sample of LAEs to infer the threshold DMH mass for central galaxies. 

The number of studies that have investigated the correlations between clustering strength and physical properties of high-redshift galaxies is slightly higher. In [\ion{O}{ii}] and [\ion{O}{iii}] emission-line-selected galaxy samples, \cite{khostovan18} found a strong halo mass dependence on the line luminosity and stellar mass. \cite{durkalec18} also observed a correlation with stellar mass, together with a further dependence on UV luminosity, in a sample of LBGs. However, these correlations become somewhat unclear near the epoch of reionization ($z\approx6$). Based on LAEs surveys, \cite{ouchi03,bielby,haruka} revealed tentative trends ($\approx1\sigma$) between  luminosity (both UV and Ly$\alpha$) and clustering strength, while only \cite{khostovan19} reported a clear (5$\sigma$) correlation between inferred DMH mass and Ly$\alpha$ luminosity. 

In a previous study \citep{yohana}, we used 68 MUSE-Wide fields to measure the LAE clustering with the K-estimator method presented in \cite{adelberger}. We computed the clustering at large scales ($R>0.6\;h^{-1}$Mpc) to derive the linear bias factor and the typical DMH mass of LAEs. By splitting our main sample into subsets based on physical properties of LAEs, we also found a tentative $2\sigma$ dependence on Ly$\alpha$ luminosity. Here, we extend this work with larger and more deeply spectroscopically confirmed samples and a refined set of analysis methods. We measured the clustering at smaller scales, applied full HOD modeling, and studied the dependence of the clustering properties on Ly$\alpha$ luminosity.

The paper is structured as follows. In Sect.~\ref{sec:data}, we describe the data used for this work and we characterize the LAE samples. In Sect.~\ref{sec:methods}, we explain our method for measuring and analyzing the clustering properties of our galaxy sets. We present the results of our measurements in Sect.~\ref{sec:results}. In Sect.~\ref{sec:discussion}, we  discuss our results and their implications, and we investigate the clustering dependence on Ly$\alpha$ luminosity. We give our conclusions in Sect.~\ref{sec:conclusions}. 

Throughout this paper, all distances are measured in comoving coordinates and given in units of $h^{-1}$Mpc (unless otherwise stated), where $h = H_0/100=0.70$ km s$^{-1}\; \rm{Mpc}^{-1}$. We  assume the same $h$ to convert line fluxes to luminosities. Thus, there are implicit $h_{70}^{-2}$ factors in the line luminosities. We use a $\Lambda$CDM cosmology and adopt $\Omega_M =$ 0.3, $\Omega_{\Lambda} =$ 0.7, and $\sigma_8 =$ 0.8 \citep{constants}. All uncertainties represent 1$\sigma$ (68.3\%) confidence intervals.


\section{Data}
\label{sec:data}

The MUSE spectroscopic surveys are based on a wedding cake design, namely: a first spatially wide region (bottom of the cake) is observed with a short exposure time (1~hour), while deeper observations (10~hours exposure) are carried out within the first surveyed area (middle tier of the cake). Contained in the last observed region, an even deeper survey (140~h) is then built up (at the top of the cake). These three surveys are known as: MUSE-Wide \citep{herenz17,urrutia19}, MUSE-Deep \citep{bacon17,inami,bacon22}, and MUSE Extremely Deep Field (MXDF; \citealt{bacon22}). Each of them can be seen as a different layer of a wedding cake, where higher layers become spatially smaller and correspond to deeper observations. In what follows, we give further details on survey and galaxy sample construction.

\subsection{MUSE-Wide}
\label{sec:muse-wide}

\begin{figure*}
\centering
\begin{tabular}{c c}
  \centering
  \includegraphics[width=.45\linewidth]{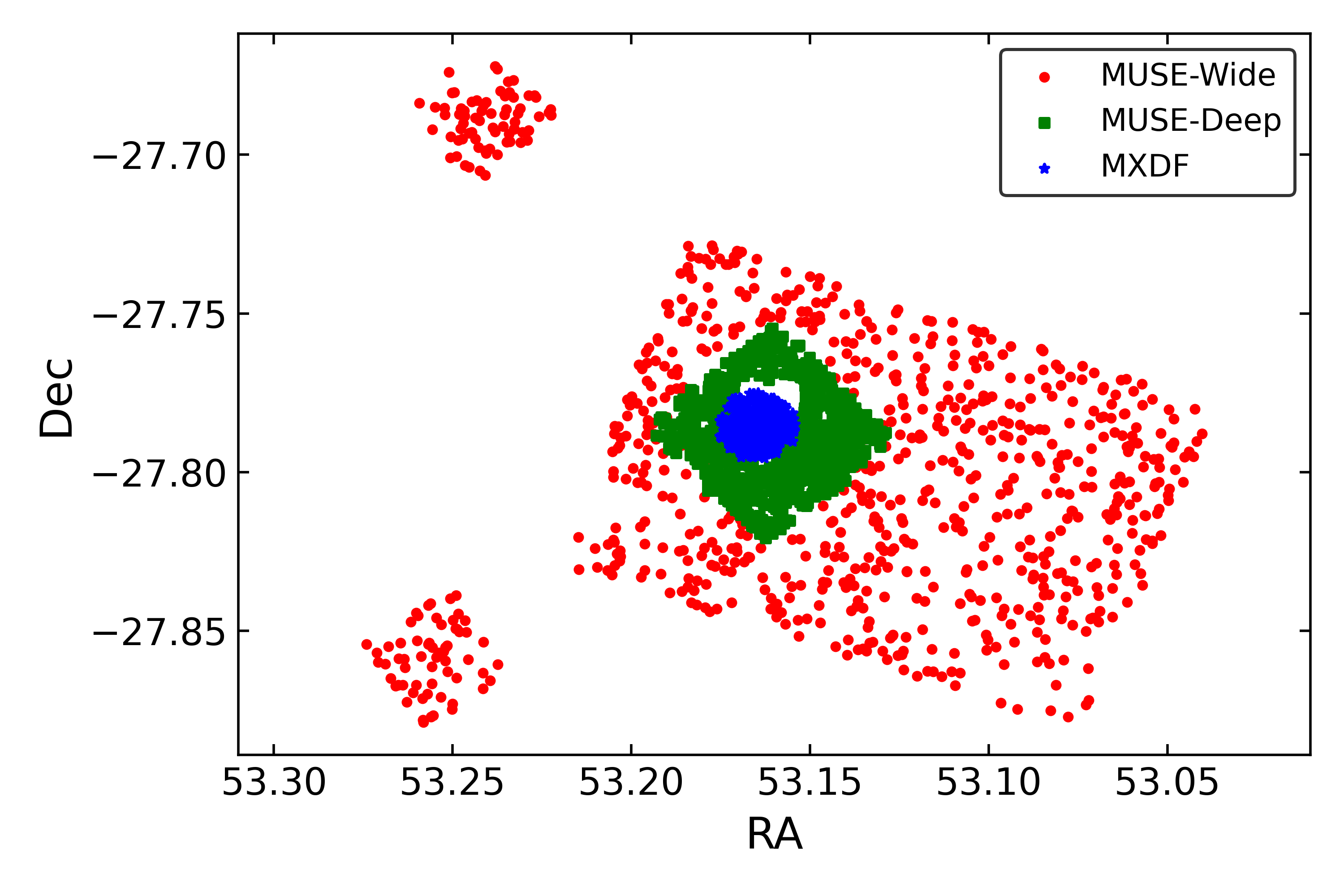}
\end{tabular}
\begin{tabular}{c c}
  \centering
  \includegraphics[width=.45\linewidth]{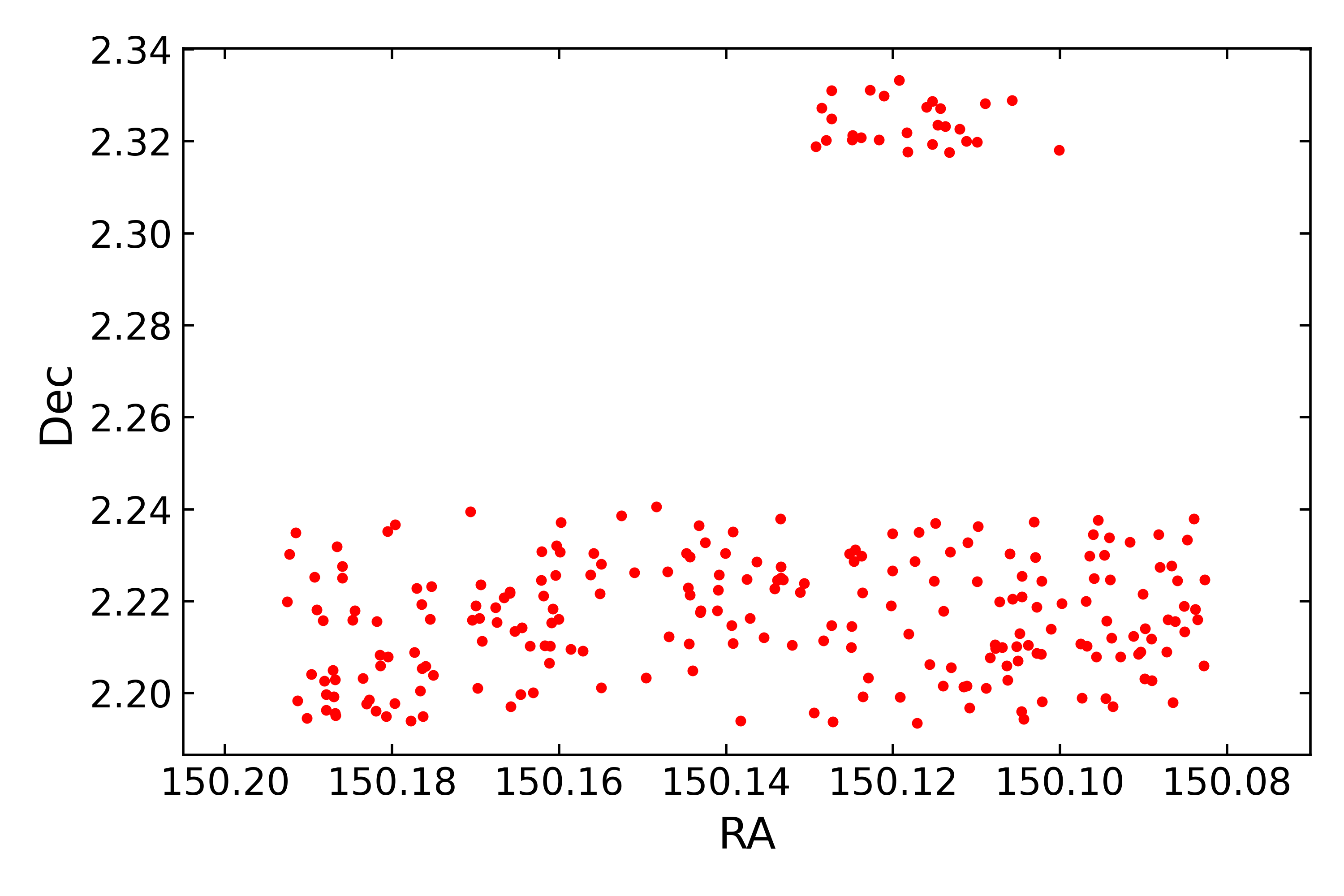}
\end{tabular}
\caption{Spatial distribution of the LAEs from the MUSE-Wide survey (red circles), MUSE-Deep (green squares) and MXDF (blue stars). The overlapping objects between the MXDF and MUSE-Deep samples have been removed from the MUSE-Deep LAE set, while those LAEs overlapping in MUSE-Deep and MUSE-Wide have been removed from the MUSE-Wide LAE sample. The MUSE-Wide survey covers part of the CANDELS/GOODS-S region and the HUDF parallel fields (left panel) as well as part of the CANDELS/COSMOS region (right panel). See Figure 1 in \cite{urrutia19} for the layout of the MUSE-Wide survey without individual objects, Figure 1 in \cite{bacon17} for that of MUSE-Deep, and Figure 2 in \cite{bacon22} for that of MUSE-Deep (MOSAIC) and MXDF together.}
\label{fig:RA-DEC}
\end{figure*}

The spectroscopic MUSE-Wide survey  \citep{herenz17,urrutia19}  comprises 100 MUSE fields distributed in the CANDELS/GOODS-S, CANDELS/COSMOS and the Hubble Ultra Deep Field (HUDF) parallel field regions.
Each MUSE field covers 1~arcmin$^2$. While 91 fields were observed with an exposure time of one hour, nine correspond to shallow (1.6~hours) reduced subsets of the MUSE-Deep data \citep[see next section; as well as][]{bacon17}, located within the HUDF in the CANDELS/GOODS-S region. However, we do not include the objects from this region since they overlap with the MUSE-Deep sample (see next section and gap in the left panel of Fig.~\ref{fig:RA-DEC}). The slight overlap between adjacent fields leads to a total spatial coverage of 83.52 arcmin$^2$. The red circles in Fig.~\ref{fig:RA-DEC}
display the spatial distribution of the LAEs from the MUSE-Wide survey. We refer to \cite{urrutia19} for further details on the survey build up, reduction and flux calibration of the MUSE data cubes. 

\begin{table*}[htbp]
\caption[]{Properties of the LAE samples.} \label{table:properties}
\centering
    \begin{tabular}{l@{\qquad}cccccccc}
        \hline \hline
           \noalign{\smallskip}
          &Area/[arcmin$^2$] & Number LAEs   & $\langle z \rangle$ &  $   n /[h^3\rm{Mpc}^{-3}] $ & $ \log(L_{\rm{Ly}\alpha}/[\rm{erg\: s}^{-1}])\: \rm{range} $ & $ \log\langle L_{\rm{Ly}\alpha}/[\rm{erg\: s}^{-1}] \rangle$ \\
             \noalign{\smallskip} 
            \hline \hline
            \noalign{\smallskip}
              MUSE-Wide &83.52 &1030  &  4.0 & $1\cdot10^{-3}$ & 40.92 -- 43.35
              &\;\;\;\;\;\;42.34 ($\approx L^{\star}$) \\
              MUSE-Deep &\;9.92 &\;679  & 4.1   & $8\cdot10^{-3}$  & 40.84 -- 43.12 & \;\;41.64 ($\approx 0.2L^{\star}$) \\
              MXDF &\;1.47 &\;367  & 4.2   &  $3\cdot10^{-2}$ & 40.15 -- 43.09 & 41.22 ($\approx 0.08L^{\star}$) \\
            \noalign{\smallskip}
        \hline 
        \multicolumn{7}{l}{%
          \begin{minipage}{13cm}%
          \vspace{0.3\baselineskip}
            \small \textbf{Notes}: Properties marked with $\langle\rangle$ represent median values for the galaxies in the samples.
          \end{minipage} 
          }\\
    \end{tabular}
\end{table*}
\begin{table*}[htbp]
\caption[]{Properties of the LAE subsamples.} \label{table:subsamples}
\centering
    \begin{tabular}{l@{\qquad}ccccc}
        \hline \hline
           \noalign{\smallskip}
          && Number LAEs   & $\langle z \rangle$ &   $ \log\langle L_{\rm{Ly}\alpha}/[\rm{erg\: s}^{-1}] \rangle$ \\
             \noalign{\smallskip} 
            \hline \hline
            \noalign{\smallskip}
              MUSE-Wide low L  & ($\log L_{\rm{Ly}\alpha} <42.34$) &  515  &  3.7 & \;\;42.06 ($\approx 0.5L^{\star}$) \\
              MUSE-Wide high L& ($\log L_{\rm{Ly}\alpha} >42.34$) &  515 & 4.1 & \;\;42.53 ($\approx 1.5L^{\star}$)\\
              MUSE-Deep low L& ($\log L_{\rm{Ly}\alpha} <41.64$) &   340 & 3.7  & \;\;41.46 ($\approx 0.1L^{\star}$)\\
              MUSE-Deep high L& ($\log L_{\rm{Ly}\alpha} >41.64$) & 339  &  4.5 & \;\;41.89 ($\approx 0.3L^{\star}$)\\
              MXDF low L &($\log L_{\rm{Ly}\alpha} < 41.22$) & 183  &  4.0 & 40.97 ($\approx 0.04L^{\star}$)\\
              MXDF high L&($\log L_{\rm{Ly}\alpha} > 41.22$) & 184  & 4.5  & \;\;41.54 ($\approx 0.2L^{\star}$)\\
              
            \noalign{\smallskip}
        \hline 
        \multicolumn{5}{l}{%
          \begin{minipage}{12cm}%
          \vspace{0.3\baselineskip}
            \small \textbf{Notes}: Properties marked with $\langle\rangle$ represent median values for the galaxies in the subsamples.
          \end{minipage} 
          }\\
    \end{tabular}
\end{table*}

In this paper, we extend (x2 spatially, 50\% more LAEs) the sample used in \cite{yohana}  and include all the 1~h exposure fields from the MUSE-Wide survey. Despite the somewhat worse seeing (generally) in the COSMOS region (right panel of Fig.\ref{fig:RA-DEC}), we demonstrate in Appendix~\ref{appendix:k-fields} that adding these fields does not significantly impact our clustering results but helps in minimizing the effects of cosmic sample variance. 

\begin{figure}[h]
\centering
\includegraphics[width=\columnwidth]{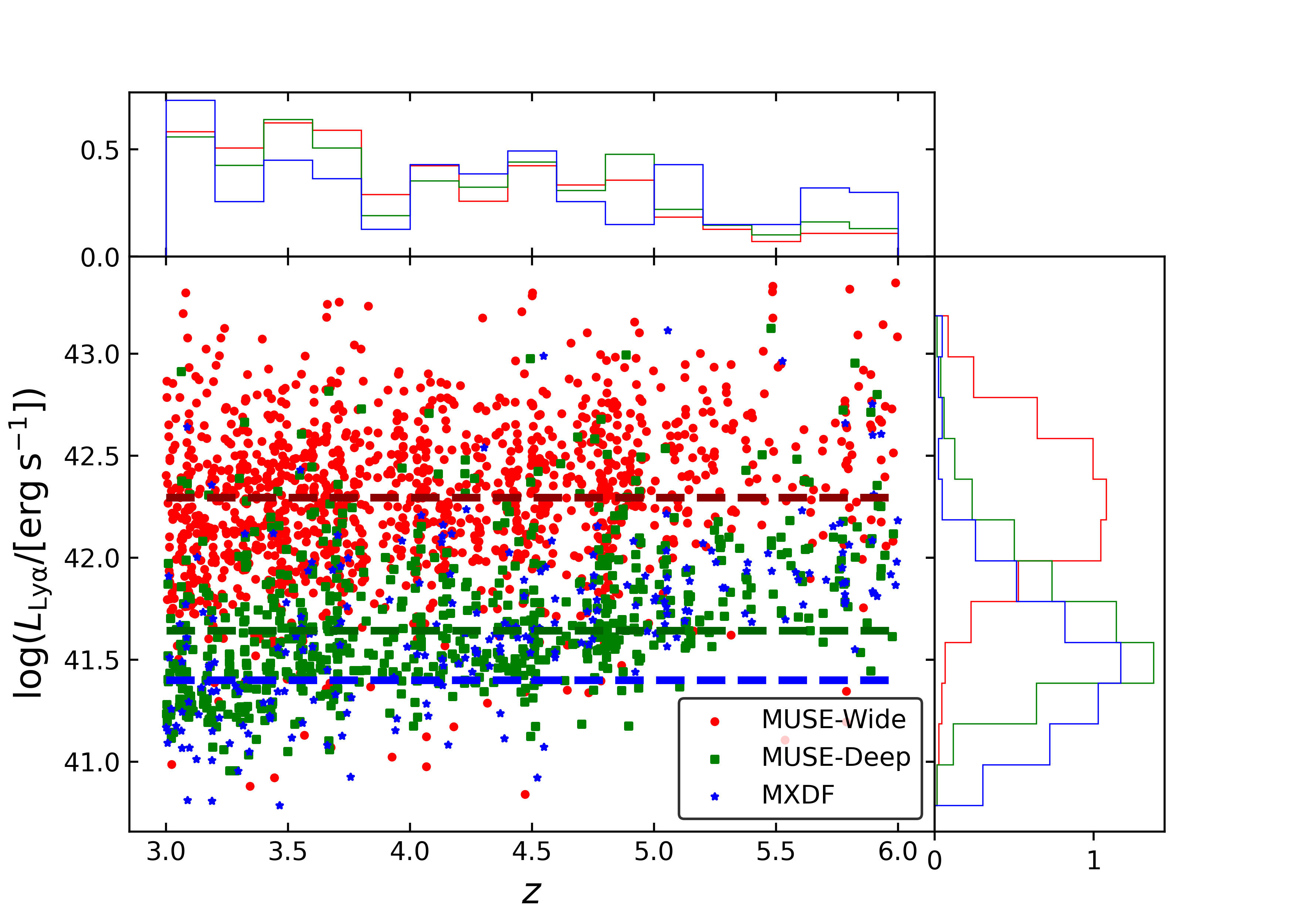}
\caption{Ly$\alpha$ luminosity-redshift for the LAEs in MUSE-Wide (red circles), MUSE-Deep (green squares) and MXDF (blue stars). The dashed colored lines correspond to the median $\log{L}_{\rm{Ly}\alpha}$ values of the corresponding samples. The redshift and $L_{\rm{Ly}\alpha}$ distributions are shown in the top and right panel, respectively.} 
\label{fig:luminosities}
\end{figure}

 We also expanded the redshift range of the sample. While MUSE spectra cover 4750--9350~\AA, implying a Ly$\alpha$ redshift interval of $2.9 \la z \la 6.7$, we limited the redshift range to $3 < z < 6$ (differing from the more conservative range of \citealt{yohana}; $3.3<z<6$) as the details of the selection function near the extremes are still being investigated. Section~2 of \cite{yohana} describes the aspects relevant to our analysis on the construction of a sample of LAEs, as well as the strategy to measure line fluxes and redshifts. 
 The redshift distribution of the sample is shown in red in the top panel of Fig.~\ref{fig:luminosities}. Systematic uncertainties introduced in the redshift-derived 3D positions of the LAEs have negligible consequences for our clustering approach (see Sect. 2.2 in \citealt{yohana}).

 

Within 83.52 arcmin$^2$ and in the selected redshift interval, we detected a total of 1030 LAEs. This implies a LAE density of more than 13 objects per arcmin$^2$ or $n \approx1\cdot10^{-3}\;h^3\rm{Mpc}^{-3}$ (for $3<z<6$). At the median redshift of the sample $\langle z\rangle=4.0$, the transverse extent of the footprint is $\approx43$ $h^{-1}$Mpc. 
 The range of Ly$\alpha$ luminosities is 40.92 $<$ log(\textit{L}$_{\rm{Ly}\alpha}/[\rm{erg\:s}^{-1}])$ $<$ 43.35 (see red circles in Fig.~\ref{fig:luminosities}), with a median value of $ \log\langle L_{\rm{Ly}\alpha}/[\rm{erg\:s}^{-1}]\rangle  = 42.34$ (or $\approx L^{\star}$ in terms of characteristic luminosity $L^{\star}$; \citealt{herenz19}), which makes this sample the highest luminosity data set of our three considered surveys. The Ly$\alpha$ luminosity distribution is shown in red in the right panel of Fig.~\ref{fig:luminosities}. The main properties of the MUSE-Wide LAEs are summarized in Table~\ref{table:properties}.

\subsection{MUSE-Deep}
\label{sec:MUSE-Deep}

MUSE-Deep \citep[10 hour MOSAIC;][]{bacon17,inami,bacon22} encompasses nine fields located in the CANDELS/GOODS-S region of the HUDF, each spanning 1 arcmin$^2$ and observed with a 10~h exposure time. The total spatial coverage is 9.92 arcmin$^2$. We represent the spatial distribution of the survey in green in Fig.~\ref{fig:RA-DEC}. We did, however, remove the MUSE-Deep objects that are selected in the deepest survey, described in the next section. We refer to \cite{bacon17,bacon22} for a detailed description on survey construction and data reduction.

The sources in MUSE-Deep were blindly detected and extracted using ORIGIN \citep{mary}, based on a matched filtering approach and developed to detect faint emission lines in MUSE datacubes. While the redshift measurements and line classifications were carried out with pyMarZ, a python version of the redshift
fitting software MarZ \citep{hinton}, the line flux determination was conducted with pyPlatefit, which is a python module optimized to fit emission lines of high-redshift spectra.  The redshift distribution of the sample is shown in green in the top panel of Fig.~\ref{fig:luminosities}, also within $3<z<6$.

The LAE density of the MUSE-Deep sample is $8\cdot10^{-3}\;h^3\rm{Mpc}^{-3}$ (68 LAE per arcmin$^2$ in the whole redshift range). The survey spans $\approx 8.7$ $h^{-1}$Mpc transversely. The range of Ly$\alpha$ luminosities is $40.84 < \log({L}_{\rm{Ly}\alpha}/[\rm{erg\:s}^{-1}])$ $<$ 43.12, represented with green squares in Fig.~\ref{fig:luminosities}, together with its distribution (right panel). MUSE-Deep is our intermediate luminous dataset, with a median luminosity of $ \log\langle L_{\rm{Ly}\alpha}/[\rm{erg\:s}^{-1}]\rangle  = 41.64$. 
The sample properties are recorded in Table~\ref{table:properties}. 




\subsection{MUSE Extremely Deep}
\label{sec:MXDF}

The MUSE Extremely Deep Field \citep{bacon22} is situated in the CANDELS/GOODS-S region and overlaps with MUSE-Deep and MUSE-Wide. It is composed of a single quasi circular field with inner and outer radii of 31'' and 41'', respectively. While a 140 hour exposure was employed to observe the totality of the field, the inner field is 135 hours deep, decreasing to 10 hours depth at the outer radius. This makes MXDF the deepest spectroscopic survey to date. For further details see  \cite{bacon22} and the blue data points in Fig.\ref{fig:RA-DEC}, where the MXDF field is overplotted on the previous surveys.

The survey assembly and data reduction is described in \cite{bacon22} and is similar to the one applied to MUSE-Deep \citep{bacon17}. The source extraction in MXDF and the redshift and flux measurements are conducted following the same procedure as was done for MUSE-Deep.  The redshift distribution of the sample is shown in blue in the top panel of Fig.~\ref{fig:luminosities}.

Contained within $\approx$1.47 arcmin$^2$ and over the same redshift range as for the previous catalogues, we detected 367 LAEs, corresponding to a LAE density of $  n \approx3\cdot10^{-2}\;h^3\rm{Mpc}^{-3}$ (432 LAEs per arcmin$^2$ at $3<z<6$). With a median redshift of $\langle z\rangle=4.2$, the footprint covers $\approx 2.8$ $h^{-1}$Mpc (transversely). The  Ly$\alpha$ luminosities span 40.15 $<$ log(\textit{L}$_{\rm{Ly}\alpha}/[\rm{erg\:s}^{-1}])$ $<$ 43.09 (see blue stars in Fig.~\ref{fig:luminosities} and its distribution in the right panel). The median Ly$\alpha$ luminosity is $ \log\langle L_{\rm{Ly}\alpha}/[\rm{erg\:s}^{-1}]\rangle  = 41.22$ (or $\approx 0.08L^{\star}$), more than one order of magnitude fainter than for MUSE-Wide. This makes MXDF the faintest ever observed sample of non-lensed LAEs.  The main properties are listed in Table~\ref{table:properties}.




\subsection{LAE subsamples}
\label{sec:subsamples}

We bisected the main samples into disjoint subsets based on their median Ly$\alpha$ luminosity to investigate the clustering dependence on this quantity. We did not merge the main LAE datasets because their distinct Ly$\alpha$ luminosities, together with their slightly different location on the sky, might introduce systematics in the clustering measurements. The subsample properties are summarized in Table~\ref{table:subsamples} and described in the following.

We split the MUSE-Wide sample at the median Ly$\alpha$ luminosity $ \log\langle L_{\rm{Ly}\alpha}/[\rm{erg\: s}^{-1}] \rangle=42.34$. The two subsamples consist of 515 LAEs each. The low-luminosity subset has a median redshift and Ly$\alpha$ luminosity  of $\langle z_{\rm{low}} \rangle=3.7$ and $ \log\langle L_{\rm{Ly}\alpha\rm{low}}/[\rm{erg\: s}^{-1}] \rangle=42.06$, while the high-luminosity subsample has $\langle z_{\rm{high}} \rangle=4.1$ and $ \log\langle L_{\rm{Ly}\alpha\rm{high}}/[\rm{erg\: s}^{-1}] \rangle=42.53$. The median redshift of the number of galaxy pairs for the low-luminosity subset is $z_{\rm pair}\approx3.4$, and that for the high-luminosity one is $z_{\rm pair}\approx4.1$.

We next bisected the MUSE-Deep set at the median Ly$\alpha$ luminosity $ \log\langle L_{\rm{Ly}\alpha}/[\rm{erg\: s}^{-1}] \rangle=41.64$. The low-luminosity subsample has 340 LAEs and presents a median redshift and Ly$\alpha$ luminosity of $\langle z_{\rm{low}} \rangle=3.7$ and $ \log\langle L_{\rm{Ly}\alpha\rm{low}}/[\rm{erg\: s}^{-1}] \rangle=41.46$. The high-luminosity subset is formed by 339 LAEs with $\langle z_{\rm{high}} \rangle=4.5$ and $ \log\langle L_{\rm{Ly}\alpha\rm{high}}/[\rm{erg\: s}^{-1}] \rangle=41.89$. While for the low-luminosity subsample $z_{\rm pair}\approx3.5$, for the high-luminosity one $z_{\rm pair}\approx4.4$.

We also divide the sample with the largest dynamic range of Ly$\alpha$ luminosities (MXDF) at the median Ly$\alpha$ luminosity $ \log\langle L_{\rm{Ly}\alpha}/[\rm{erg\: s}^{-1}] \rangle=41.22$. While the lower luminosity subset contains 183 LAEs with $\langle z_{\rm{low}} \rangle=4.0$ and $ \log\langle L_{\rm{Ly}\alpha\rm{low}}/[\rm{erg\: s}^{-1}] \rangle=40.97$, the higher luminosity subsample consists of 184 LAEs with $\langle z_{\rm{high}} \rangle=4.5$ and $ \log\langle L_{\rm{Ly}\alpha\rm{high}}/[\rm{erg\: s}^{-1}] \rangle=41.54$. For the low-luminosity subset, we have $z_{\rm pair}\approx3.9$, and for the high-luminosity one, we have $z_{\rm pair}\approx4.8$.

\begin{figure*}[h]
\centering
\begin{tabular}{c c c}
  \centering
  \includegraphics[width=.3\linewidth]{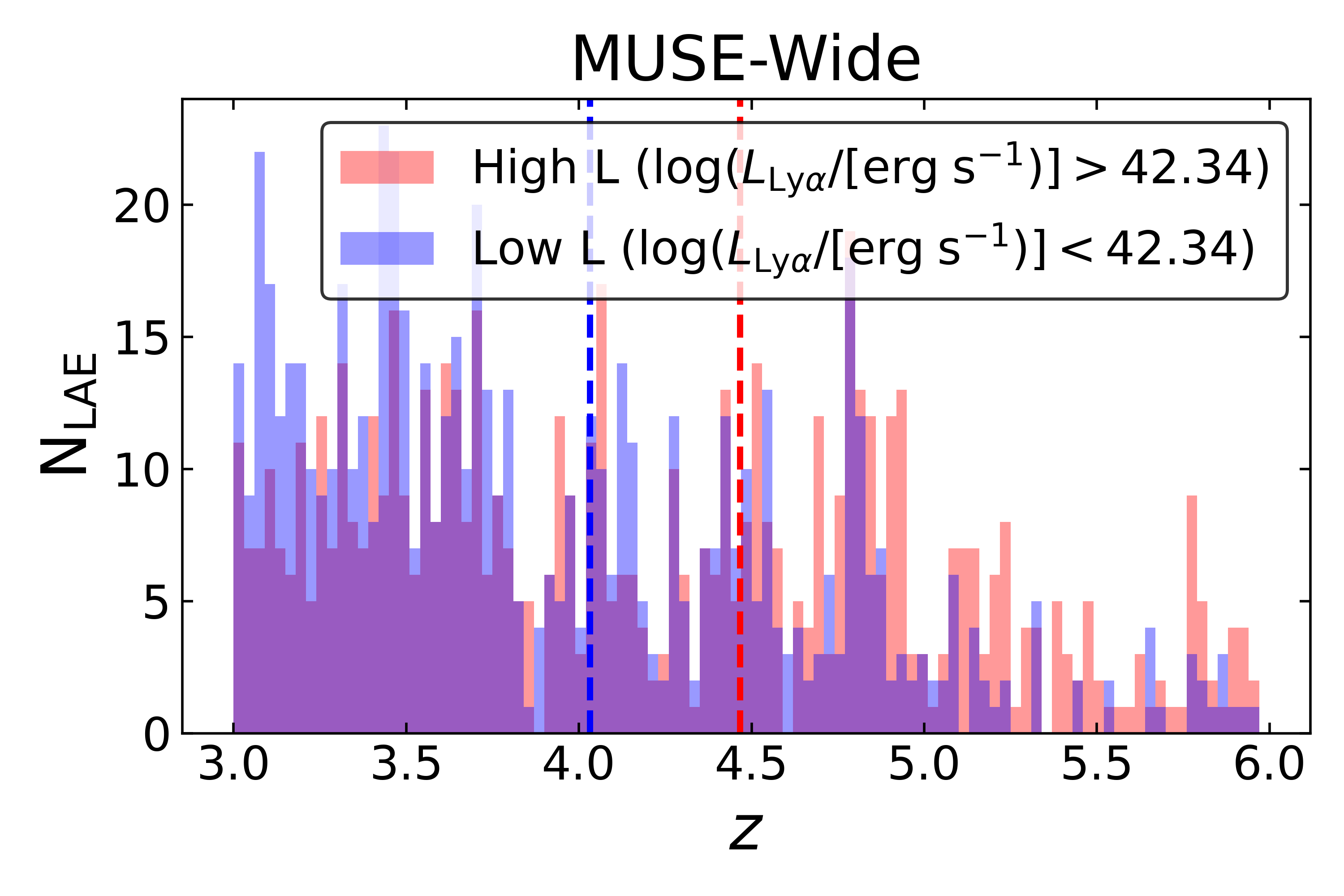}
\end{tabular}%
\begin{tabular}{c c c}
  \centering
  \includegraphics[width=.3\linewidth]{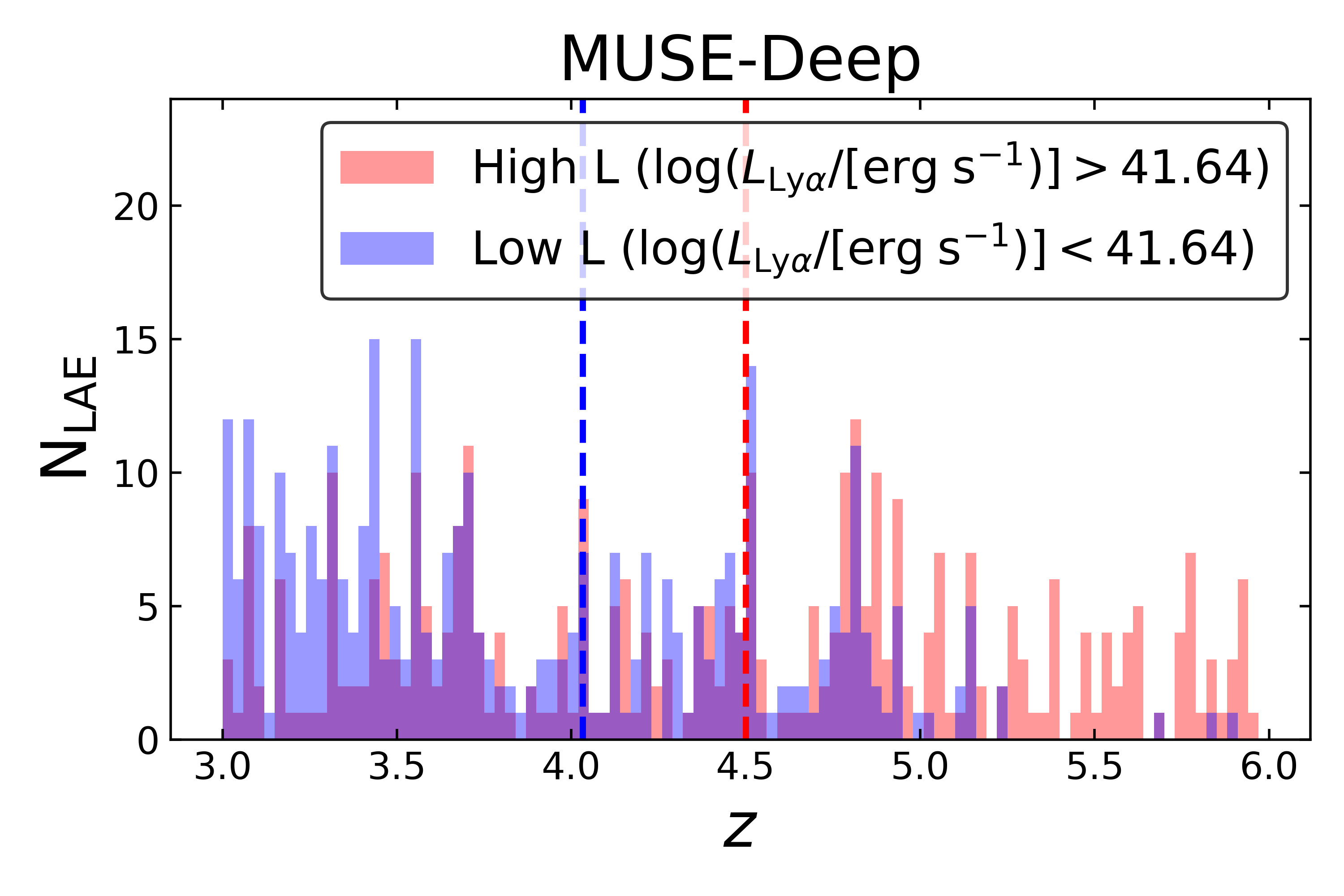}
\end{tabular}
\begin{tabular}{c c c}
  \centering
  \includegraphics[width=.3\linewidth]{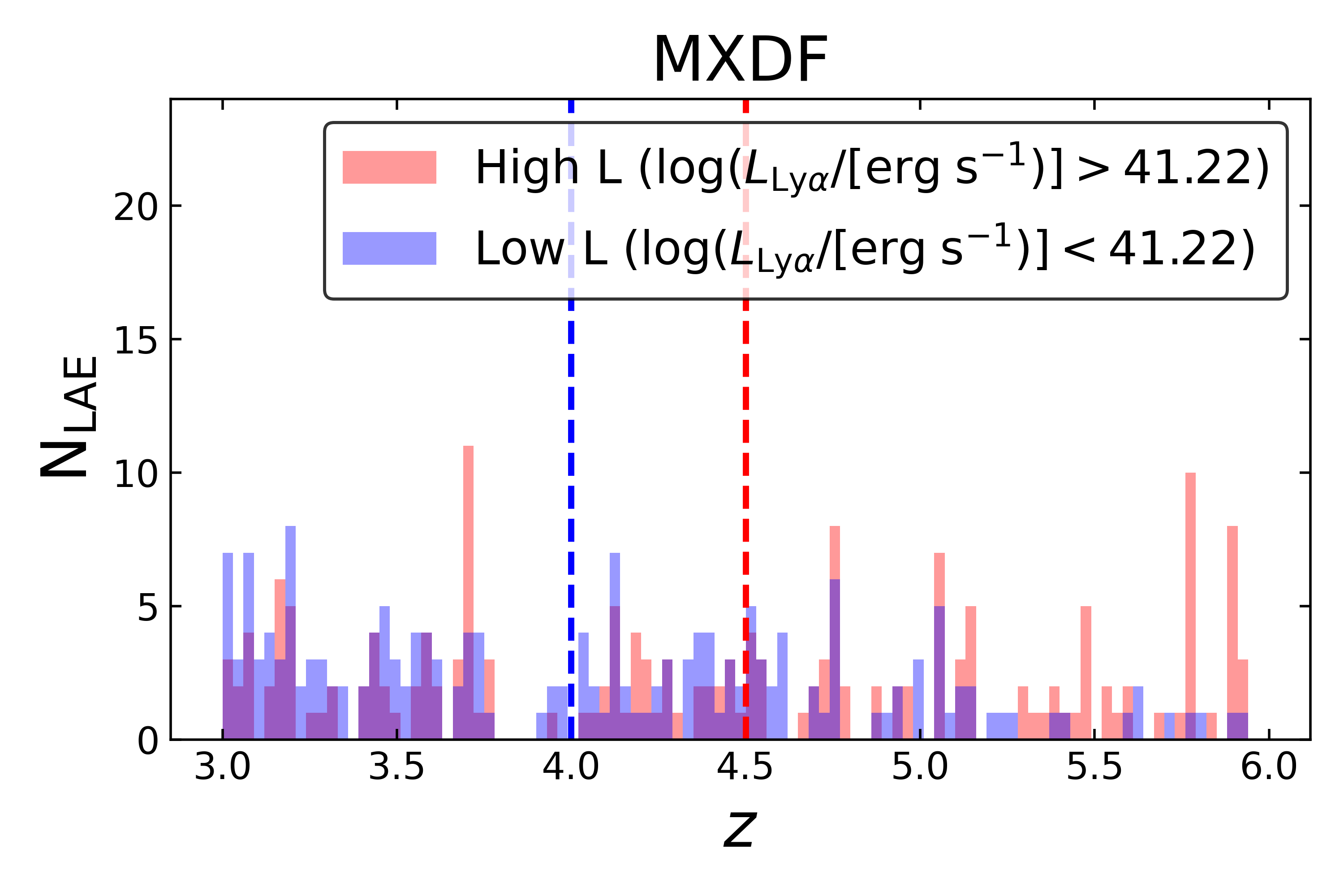}
\end{tabular}
\caption{Redshift distribution of the subsamples bisected at the median Ly$\alpha$ luminosity of MUSE-Wide, MUSE-Deep and MXDF (panels from left to right). Blue (red) colors show the low- (high-) luminosity subsets. The vertical dashed lines represent the median redshift of the corresponding subsample.}
\label{fig:zdistribution}
\end{figure*}

The redshift distribution of each subsample is shown in Fig.~\ref{fig:zdistribution}. The corresponding median redshifts are represented with a vertical dashed line. Despite the similar median redshifts between the subsample pairs, the redshift distributions are significantly different, with a higher amount of spike-trough contrasts in the high-luminosity subsets.

\section{Methods}
\label{sec:methods}

\subsection{K-estimator}
\label{sec:k-estimator}

Galaxy clustering is commonly measured by two-point correlation function (2pcf) statistics. Samples investigated by this method typically span several square degrees on the sky. With MUSE, we encounter the opposite scenario. By design, MUSE surveys cover small spatial extensions on the sky and provide a broad redshift range. Although the MUSE-Wide survey is the largest footprint of all MUSE samples, its nature is still that of a pencil-beam survey. Its transverse scales are of the order of $40 \;h^{-1}$Mpc, while in redshift space it reaches almost 1500 $h^{-1}$Mpc. If we consider the deeper samples, the difference is even more prominent: $ 8.7$ vs $ 1500\;h^{-1}$Mpc for MUSE-Deep and $2.8$ versus $ 1500\;h^{-1}$Mpc for MXDF. It is thus paramount to exploit the radial scales and utilize alternative methods to the traditional 2pcf.

In \cite{yohana} we applied the so-called K-estimator, introduced by  \cite{adelberger}, to a subset of our current sample. Here, we build on our previous work by extending the dataset and measuring the small-scale clustering required to perform full HOD modeling. The details of the K-estimator are given in Sect.~3.1 of \cite{yohana}. In the following, we provide a brief description of the method.

\begin{figure}[h]
\centering
\includegraphics[width=\columnwidth]{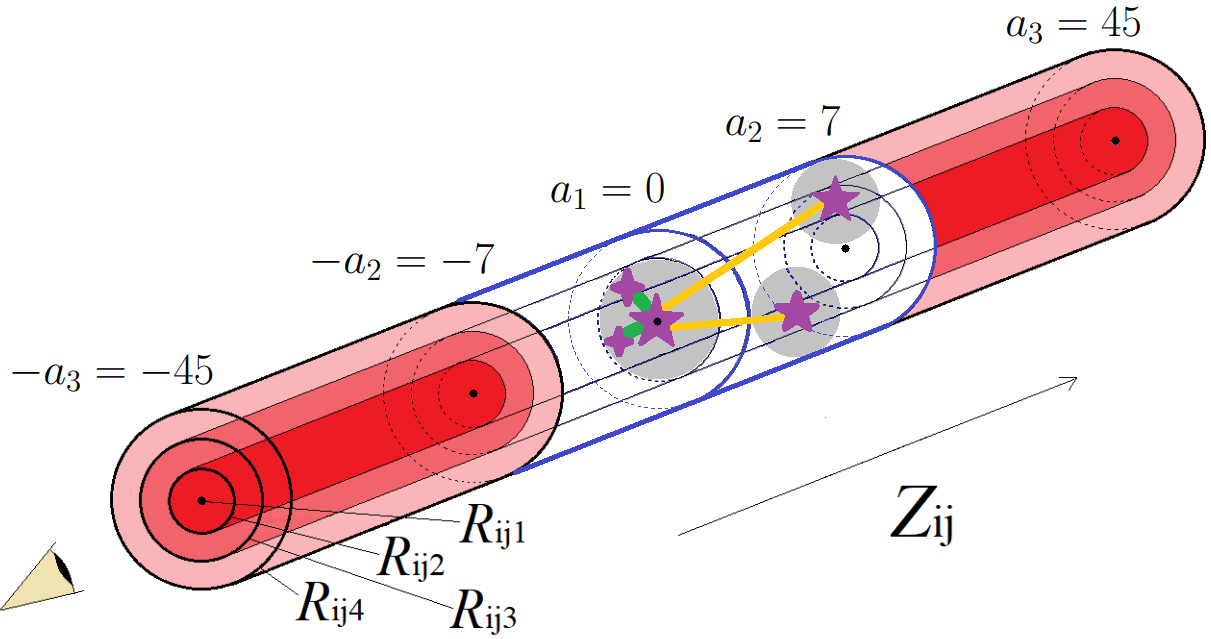}
\caption{Sketch of the K-estimator, representing the relative geometry that probe the one- and two-halo term scales. The empty blue and filled red cylinders, delimited by $\lvert a_2\rvert=7\;h^{-1}$Mpc and $\lvert a_3\rvert=45\;h^{-1}$Mpc respectively, illustrate the line-of-sight distance $Z_{ij}$ intervals within which we count galaxy pairs at fixed transverse separations $R_{ij}$, represented by nested cylinders. Pairs of LAEs connected with green lines within the same DMH (filled gray circle) contribute to the one-halo term (small $R_{ij}$ scales), while pairs belonging to two different DMHs (yellow lines) probe the two-halo term (larger $R_{ij}$ separations).} 
\label{fig:K-estimator}
\end{figure}

The K-estimator measures the radial clustering along line-of-sight distances, $Z_{ij}$, by counting galaxy pairs (formed by galaxy $i$ and galaxy $j$) in redshift space at fixed transverse separations, $R_{ij}$.  Although the K-estimator does not need a random sample to carry out the clustering measurements, its nature is very similar to that of the projected two-point correlation function. We bin by $R_{ij}$, shown with distinct radii in the cylinders of Fig.~\ref{fig:K-estimator}, and count the number of pairs within individual transverse bins, for two different ranges of $Z_{ij}$, represented in red and blue in Fig.~\ref{fig:K-estimator}. The K-estimator as a function of $R_{ij}$ is then defined as the ratio of galaxy pairs within the first $Z_{ij}$ interval (blue cylinder) and the total $Z_{ij}$ range (red and blue cylinder), quantifying the excess of galaxy pairs in the first $Z_{ij}$ bin with respect to the total one. We optimize the choice of the $Z_{ij}$ ranges, and thus the K-estimator, by seeking out the estimator that delivers the best sensitivity for the clustering signal (i.e., the highest signal-to-noise  ratio, S/N; see Sect. 3.1.2 in \citealt{yohana}). Although slightly different than in \cite{yohana}, we find nearly identical K-estimators for each of the current samples ($K_{7,45}^{0,7}$ for MUSE-Wide, $K_{7,45}^{0,7}$ for MUSE-Deep, and $K_{7,40}^{0,7}$ for MXDF), whose clustering signals only differ in their S/N. We chose the same K-estimator for the three data sets, $K_{7,45}^{0,7}$.

The K-estimator is directly related to the average underlying correlation function (see Eq. 2 in \citealt{yohana}). In fact, its definition is proportional to a combination of projected two-point correlation functions corresponding to the blue and red cylinders of Fig.~\ref{fig:K-estimator}. While the traditional 2pcf method integrates the correlation function $\xi(R_{ij}, Z_{ij})$ over line-of-sight separations up to a maximum line-of-sight distance $\pi_{\rm{max}}$, the K-estimator integrates up to $a_2$ and $a_3$. 
The correlation function $\xi(R_{ij}, Z_{ij})$ can be approximated with a power-law following \cite{limber} equations as we did in \cite{yohana}, or modeled with a halo occupation distribution (HOD) model (see Sect.~\ref{sec:hod}). 
For reference, randomly distributed galaxies in space ($\xi(R_{ij}, Z_{ij})=0$) provide $ K_{7, 45}^{0, 7}(R_{ij}) $ values equal to 7/45 (see Eq.~2 in \citealt{yohana}). Samples with data points significantly above 7/45 dispense clustering signals.

\subsection{Error estimation}

\subsubsection{Error estimation for the MUSE-Wide survey}
\label{sec:errorsK}

Applying clustering statistics delivers correlated data points. One single galaxy might be part of more than one galaxy pair and can therefore contribute to several $R_{ij}$ bins, especially if they are adjacent.  In order to quantify the actual correlation between data points, we applied the jackknife resampling technique, followed by the computation of the covariance matrix \citep[see e.g.,][]{mirko10,miyaji11}. For the MUSE-Wide sample, we employed ten logarithmic bins in the range $0.16 < R_{ij} < 27.5$~$h^{-1}$Mpc, discarding lower $R_{ij}$ scales since they host very few galaxy pairs. 

We then found a compromise between the number of independent regions (jackknife zones) and the size of the jackknife zones and divide the sky coverage into $N_{\rm{jack}}=10$ regions, each of which extends $\approx4\;h^{-1}$Mpc in both RA and Dec directions (see Appendix~\ref{appendix:covariance} for a visual representation of the sky division). The limited spatial extent of the survey does not allow for a higher number of jackknife zones. We then constructed $N_{\rm{jack}}$ jackknife subsamples, excluding one jackknife zone at a time, and computed the K-estimator for each of the subsets. The K-estimator measurements are then used to derive the covariance matrix $M_{ij}$, which quantifies the correlation between bins $i$ and $j$. The matrix is expressed as
\begin{equation}
\label{eq:covariance}
    \begin{split} M_{ij}  = \frac{N_{\rm{jack}}-1}{N_{\rm{jack}}} \left[ \displaystyle\sum_{k=1}^{N_{\rm{jack}}}\Bigg(\:  K_k(R_i) \:\:-\left< K(R_i)\right>\Bigg)\; \right. \\
     \times\left. \Bigg(\:K_k(R_j)\:\:-\langle K(R_j)\rangle \Bigg) \right],
    \end{split}
\end{equation}
where $K_k(R_i)$, $K_k(R_j)$ are the K-estimators from the k-th jackknife samples and $\left<K(R_i) \right >$, $\langle K(R_j) \rangle$ are the averages over all jackknife samples in the $i$, $j$ bins, respectively. The error bar for the K-estimator at the \textit{i}th bin comes from the square root of the diagonal element ($\sqrt{ M_{ii}}$) of the covariance matrix, our so-called "jackknife uncertainty." This approach could not be followed in \cite{yohana} because of the smaller sky coverage. Instead, we used a galaxy bootstrapping approach. In Appendix~\ref{appendix:err_comparison}, we compare the two techniques and show that bootstrapping uncertainties are $\approx 50$\% larger than the jackknife error bars, in agreement with \cite{norberg09}, who found that boostrapping overestimates the uncertainties.

We next search for the best-fit parameters by minimizing the correlated $\chi^2$ values according to
\begin{equation}
\label{eq:chi2}
    \begin{split} \chi^2   = \displaystyle\sum_{i=1}^{N_{\rm{bins}}}\displaystyle\sum_{j=1}^{N_{\rm{bins}}} \Bigg(\: & K(R_i) \:\:- K(R_i)^{\rm{HOD}}\Bigg)  \\
    & \times\;M_{ij}^{-1}\Bigg(\:K(R_j)\:\:- K(R_j)^{\rm{HOD}}\Bigg),
    \end{split}
\end{equation}
where $N_{\rm{bins}}=10$ is the number of $R_{ij}$ bins, $K(R_i)$, $K(R_j)$ are the measured K-estimators and $K(R_i)^{\rm{HOD}}$, $K(R_j)^{\rm{HOD}}$ are the K-estimators predicted by the HOD model for each $i$, $j$ bin, respectively.

Regardless of the larger sample  considered in this work, we are still limited by the spatial size of the survey, which only permits a small number of jackknife zones. The insufficient statistics naturally lead to a higher noise contribution in the covariance matrix, which cause the $\chi^2$ minimization to mathematically fail (i.e., cases of $\chi^2<0$) when the full covariance matrix is included. Hence, we only incorporated the main diagonal of the matrix and its two contiguous diagonals. In Appendix~\ref{appendix:covariance}, we discuss the high level of noise in the matrix elements corresponding to bins that are significantly apart from each other. We also verify the robustness of our approach and show that our clustering results are not altered (within 1$\sigma$) by this choice. 

 
\subsubsection{Error estimation for the deeper surveys}
\label{sec:errorsK_deeper}

The small sky coverage of the deeper surveys does not allow us to follow the same error estimation approach as for the MUSE-Wide survey. In Appendix~\ref{appendix:err_comparison}, we not only compare the bootstrapping technique applied in \cite{yohana} to the jackknife approach performed in MUSE-Wide, but we also consider the Poisson uncertainties. We demonstrate that Poisson and jackknife errors are comparable in our sample. In fact, we show that while bootstrapping uncertainties are $\approx 50$\% larger than jackknife errors, Poisson uncertainties are only $\approx 7$\% higher. Thus, and similarly to \cite{adelberger,catrina,khostovan18}, we stick to Poisson uncertainties for the MUSE-Deep and MXDF samples. For these datasets, we measure the K-estimator in eight and six logarithmic bins in the ranges $0.09 < R_{ij}/[h^{-1}\rm{Mpc}] < 4.75$ and $0.09 < R_{ij}/[h^{-1}\rm{Mpc}] < 1.45$, respectively, constrained by the spatial extent of the surveys.

We then perform a standard $\chi^2$ minimization to find the best fitting parameters to the K-estimator measurements. Namely,
\begin{equation}
\label{eq:chi2deep}
    \chi^2   = \displaystyle\sum_{i=1}^{N_{\rm{bins}}} \Bigg(\: \frac{K(R_i) \:\:- K(R_i)^{\rm{HOD}}}{\sigma_i}\Bigg)^2,
\end{equation}
where $ K(R_i) $, $K(R_i)^{\rm{HOD}}$, and $\sigma_i$ denote the measured K-estimator, the HOD modeled K-estimator and the Poisson uncertainty in the $i$th bin, respectively.

We note that the standard $\chi^2$ minimization does not account for the correlation between bins. Although in Appendix~\ref{appendix:covariance} we show that only contiguous bins are moderately correlated, we should take the resulting fit uncertainties with caution. 

\subsection{Halo occupation distribution modeling}
\label{sec:hod}

The clustering statistics can be approximated with a power-law or modeled with state-of-the-art HOD modeling. Traditional clustering studies make use of power laws to derive the correlation length and slope, from which they infer large-scale bias factors and typical DMH masses. This simple approach deviates from the actual shape of the clustering statistic curve, even in the linear regime, and its inferred DMH masses suffer from systematic errors (e.g., \citealt{jenkins} and references therein). 
To overcome these concerns, physically motivated HOD models do not treat the linear and non-linear regime alike but differentiate between the clustering contribution from galaxy pairs that reside in the same DMH and pairs that occupy different DMHs. 

In \cite{yohana} we only
modeled the two-halo term of the K-estimator with HOD modeling, which only delivered
the large-scale bias factor and the typical DMH mass of the sample. We now extend into the non-linear regime (i.e., $R_{ij}<0.6\;h^{-1}$Mpc) of the one-halo term. We can then model the clustering measured by the K-estimator with a \textit{full} HOD model, combining the separate contributions from the one- (1h, i.e., galaxy pairs residing in the same DMH) and the two-halo (2h, i.e., galaxy pairs residing in different DMHs) terms:
\begin{equation}
\label{eq:halo-contributions}
    \xi = \xi_{1h}+\xi_{2h},
\end{equation}
where  $\xi$ is the correlation function.

The HOD model we used is the same as in \cite{yohana}, an improved version of that described by \cite{miyaji11,mirko12,mirko15,mirko18}.  We assumed that LAEs are associated with DMHs, linked by the bias-halo mass relation from \cite{tinker}. From \cite{tinker}, we also included the effects of halo-halo collisions and scale-dependent bias.
The mass function of DMHs, which is denoted by $\phi(M_{\rm{h}})\rm{d}M_{\rm{h}}$, is based on \cite{sheth}, and the DMH profile is taken from \citet{nfw97}. We use the concentration parameter from \citet{zheng07}, and the weakly redshift-dependent collapse overdensity from \cite{nfw97,vandenbosch13}. We further incorporated redshift space distortions (RSDs) in the two-halo term using linear theory (Kaiser infall; \citealt{kaiser} and \citealt{vandenbosch13}). We did not model RSDs in the one-halo term because  the peculiar velocity has negligible effects to our K-estimator as demonstrated in the following. The velocity dispersion ($\sigma_{\rm v}$) of satellites in a $M_{\rm h}$ halo can be estimated by $\sigma_{\rm v}^2\approx GM_{\rm h}/(2R_{\rm vir})$, where $R_{\rm vir}$ is the virial radius \citep{tinker07}. Its effect on the line-of-sight physical distance estimate is then $\sigma_{\rm v}/H(z)$. For $10^{11-12}\;h^{-1}M_\odot$ DMH masses, which are typical for our sample, with virial radii of $\approx0.02-0.05$ (physical) $h^{-1}$Mpc, the line-of-sight distance estimation is deviated by $\approx 0.15-0.30\;h^{-1}$Mpc, corresponding to a peculiar velocity dispersion of $\sigma_{\rm v}\approx 80-170$ km s$^{-1}$. This is significantly small compared to our $a_2=7\;h^{-1}$Mpc. We thus assume that the one-halo term contributes only to the $Z_{ij}=0-7\;h^{-1}$Mpc bin. We evaluated the HOD model at the median redshift of $N(z)^2$, where $N(z)$ is the redshift distribution of the sampled galaxy pairs. For our three main datasets, $z_{\rm pair}\approx3.8$.

The mean halo occupation function is a simplified version of the five parameter model by \citet{zheng07}. We fixed the halo mass at which the satellite occupation becomes zero to $M_0=0$ and the smoothing scale of the central halo occupation lower mass cutoff to  $\sigma_{\log M}=0$, due to sample size limitations.
We define the mean occupation distribution of the central galaxy $\langle N_{\text{c}}(M_{\rm{h}})\rangle$ as
\begin{equation}
\label{eq:Nc}
  \langle N_{\text{c}}(M_{\rm{h}})\rangle = 
  \begin{cases}
  \; 1 & (M_{\rm{h}}\geq M_{\text{min}}) \\
  \; 0 & (M_{\rm{h}}< M_{\text{min}})
  \end{cases}
\end{equation}
and that of satellite galaxies $\langle N_{\text{s}}(M_{\rm{h}}) \rangle$ as
\begin{equation}
\label{eq:Ns}
  \langle N_{\text{s}}(M_{\rm{h}}) \rangle = \langle N_{\text{c}}(M_{\rm{h}}) \rangle \cdot \left(\frac{M_{\rm{h}}}{M_1}\right)^\alpha,
\end{equation}
where $M_{\rm{min}}$ is the minimum halo mass required to host a central galaxy, $M_1$ is the halo mass threshold to host (on average) one satellite galaxy, and $\alpha$ is the high-mass power-law slope of the satellite galaxy mean occupation function. The total halo occupation is given by the sum of central and satellite galaxy halo occupations, $N(M_{\rm h})=N_{\rm c}(M_{\rm h})+N_{\rm s}(M_{\rm h})$. 

The dependencies of the HOD parameters on the shape of the K-estimator are detailed in Appendix~\ref{appendix:HODparameters}. In short, for the HOD parameters there selected, the clustering amplitude of the two-halo term is ascertained by the hosting DMHs and is thus very sensitive to their mass, $M_{\rm{min}}$, and to the fraction of galaxies in massive halos with respect to lower-mass halos, linked to $\alpha$. The clustering in the one-halo term regime, however, is affected by the three parameters in a complex manner; roughly $M_{\rm{min}}$ and $\alpha$ vary the amplitude, and $\alpha$ as well as (moderately) $M_1$ modify the slope.


To find the best-fit HOD model, we construct a 3D parameter grid for $M_{\rm{min}}$, $M_1$, and $\alpha$. We vary $\log (M_{\rm{min}}/[h^{-1}M_\odot])$ in the range $9.5-11.2$, $\log (M_1/M_{\rm{min}})$ from 0.5 to 2.5, and $\alpha$ within $0.2-4.3$, all in steps of 0.1. For each parameter combination, we computed $\xi$ (Eq.~\ref{eq:halo-contributions}), converted it to the K-estimator using Eq.~2 in \cite{yohana}, and computed a $\chi^2$ value (Eqs.~\ref{eq:chi2} or \ref{eq:chi2deep}). We then used the resulting 3D $\chi^2$ grid to estimate the confidence intervals for the HOD parameters. For each point on a 2D plane, we search for the minimum $\chi^2$ for the contouring along the remaining parameter. The contours we plot are at $\Delta\chi^2=3.53$ and 8.02, which correspond to Gaussian 68\% (1$\sigma$) and 95\% (2$\sigma$) confidence levels, respectively, applying the $\chi^2$ distribution for three degrees of freedom. The projections of the 68\% probability contours on the three interesting parameters are then used to compute the uncertainty of each HOD parameter.

For each point in the three parameter grid, we also computed
the large-scale galaxy bias factor, $b$, and the fraction of satellite galaxies per halo, $f_{\rm{sat}}$, as follows:


\begin{equation}
\label{eq:b}
  b =  \frac{\int\langle N (M_{\rm{h}})\rangle\, b_{\rm{h}}(M_{\rm{h}})\,\phi(M_{\rm{h}})\,{\rm d}M_{\rm{h}}}{\int \langle N (M_{\rm{h}})\rangle\, \phi(M_{\rm{h}}) \,{\rm d}M_{\rm{h}}},
\end{equation}

\begin{equation}
\label{eq:fsat}
  f_{\rm{sat}} = \frac{\int \langle N_{\text{s}}\,(M_{\rm{h}})\rangle\, \phi(M_{\rm{h}})\, {\rm d}M_{\rm{h}}}{\int \langle N(M_{\rm{h}})\rangle\, \phi(M_{\rm{h}}) \,{\rm d}M_{\rm{h}}},
\end{equation}
where $b_{\rm{h}}(M_{\rm{h}})$ denotes the
large-scale halo bias. The typical DMH mass is determined by the large-scale galaxy bias factor.
We ultimately compute the bias and $f_{\rm{sat}}$ distributions from the HOD models that fall within the 68\% confidence (for the three-parameter space) contours. These distributions are then used to assess the uncertainties in the bias and $f_{\rm{sat}}$.


\section{Results from HOD modeling}
\label{sec:results}
\subsection{Fit results from the MUSE-Wide survey}
\label{sec:fits_wide}

\begin{figure*}[h]
\centering
\begin{tabular}{c c }
  \centering
  \includegraphics[width=.45\linewidth]{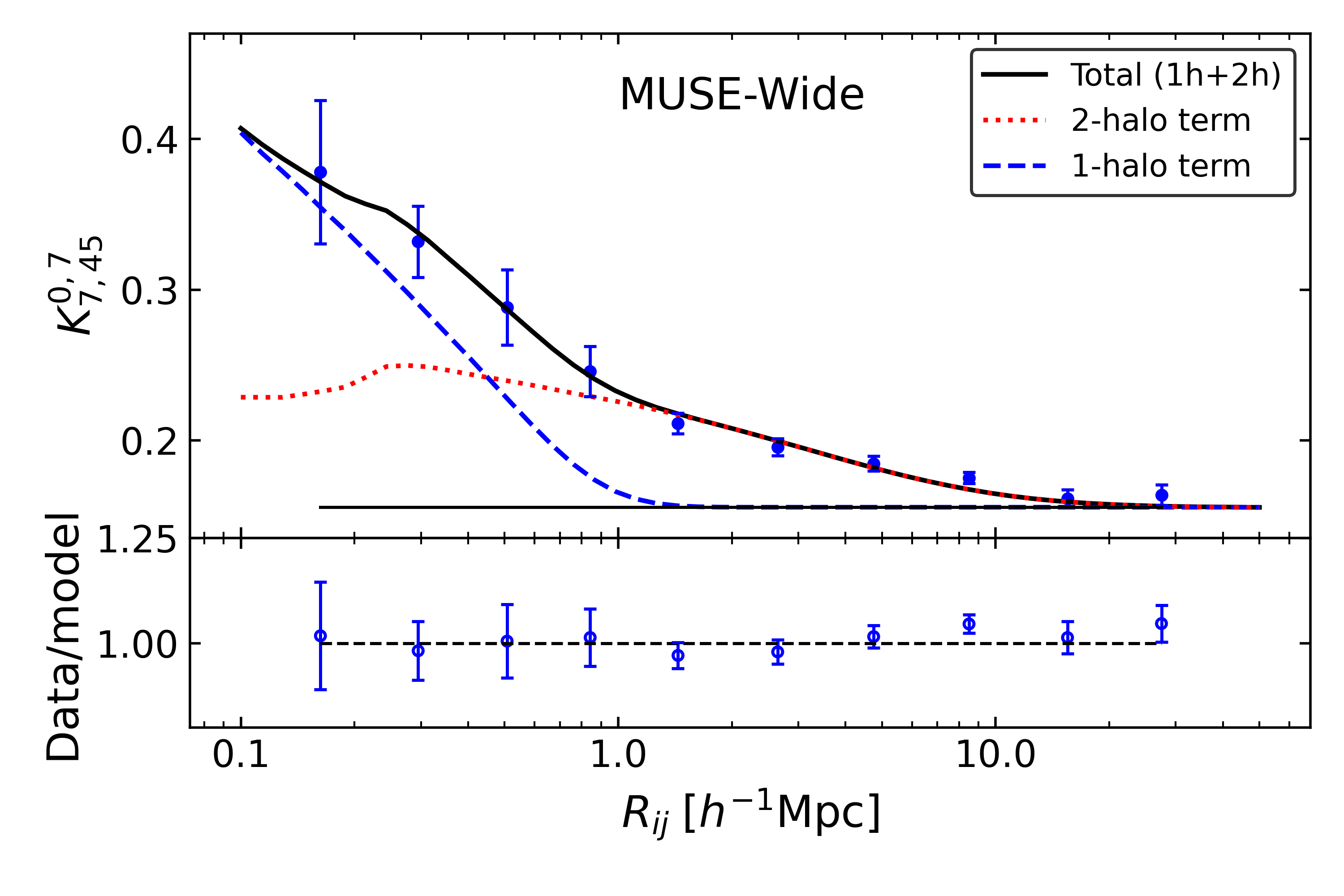}
\end{tabular}%
\begin{tabular}{c c }
  \centering
  \includegraphics[width=.45\linewidth]{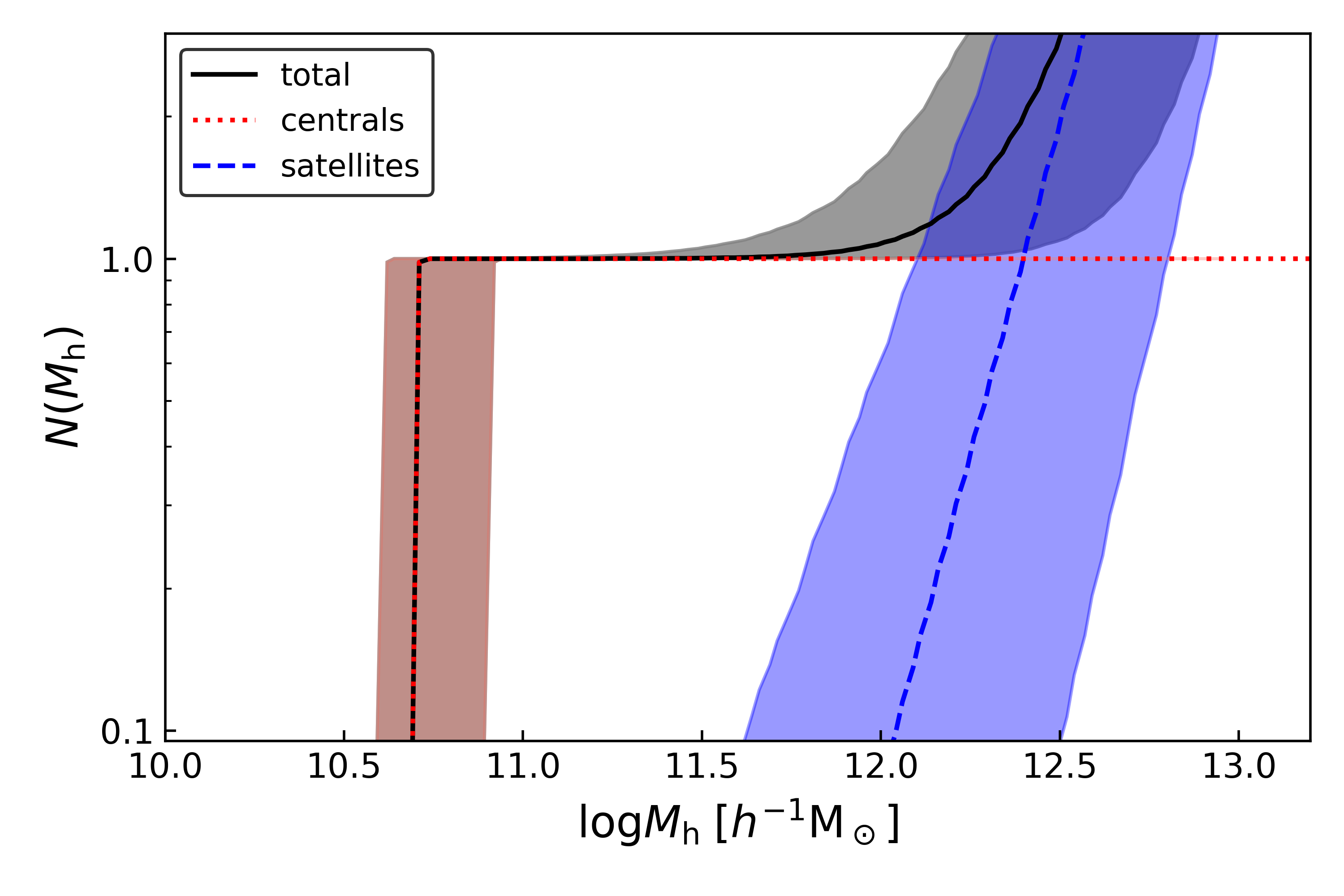}
\end{tabular}
\begin{tabular}{c c }
  \centering
  \includegraphics[width=.45\linewidth]{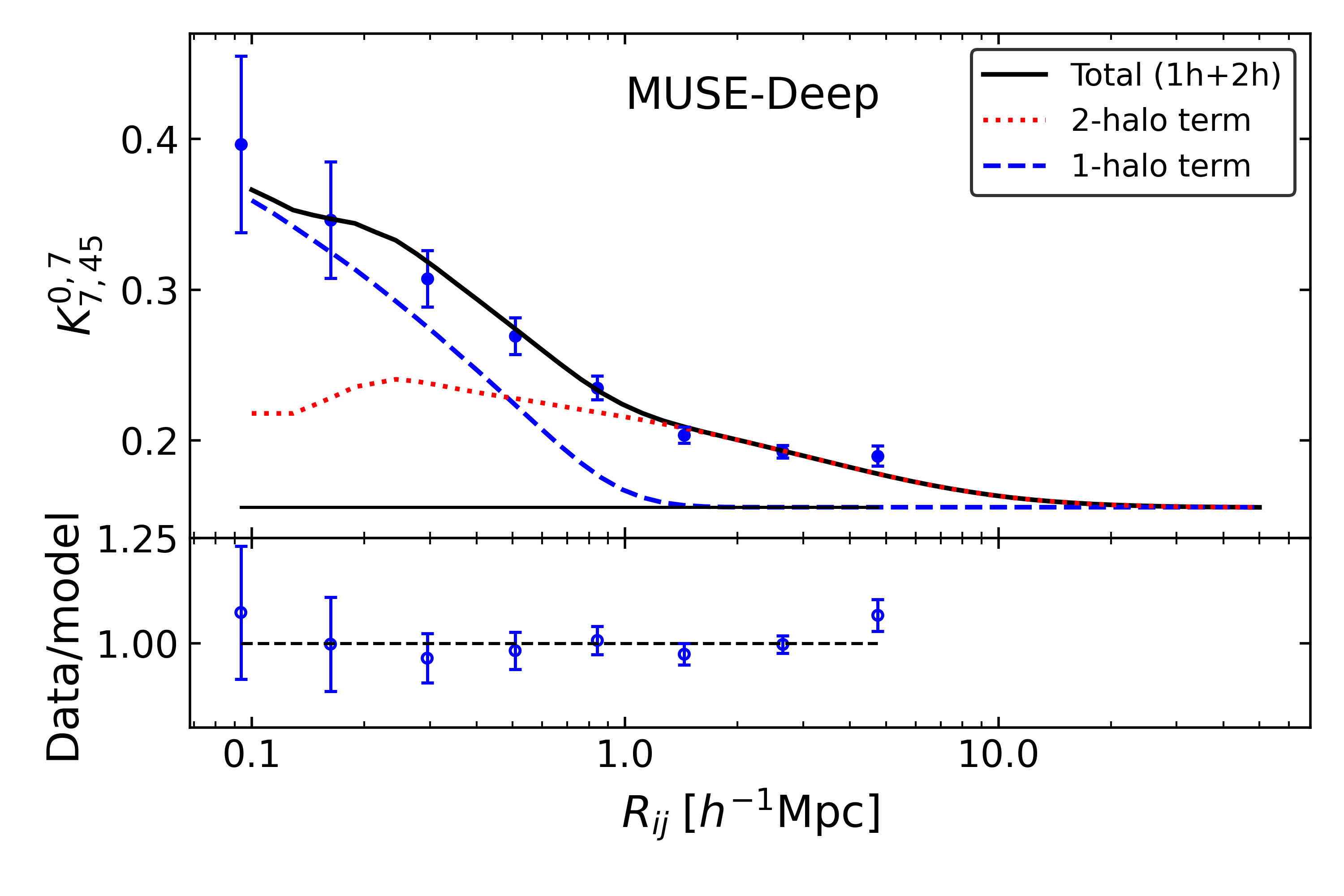}
\end{tabular}%
\begin{tabular}{c c }
  \centering
  \includegraphics[width=.45\linewidth]{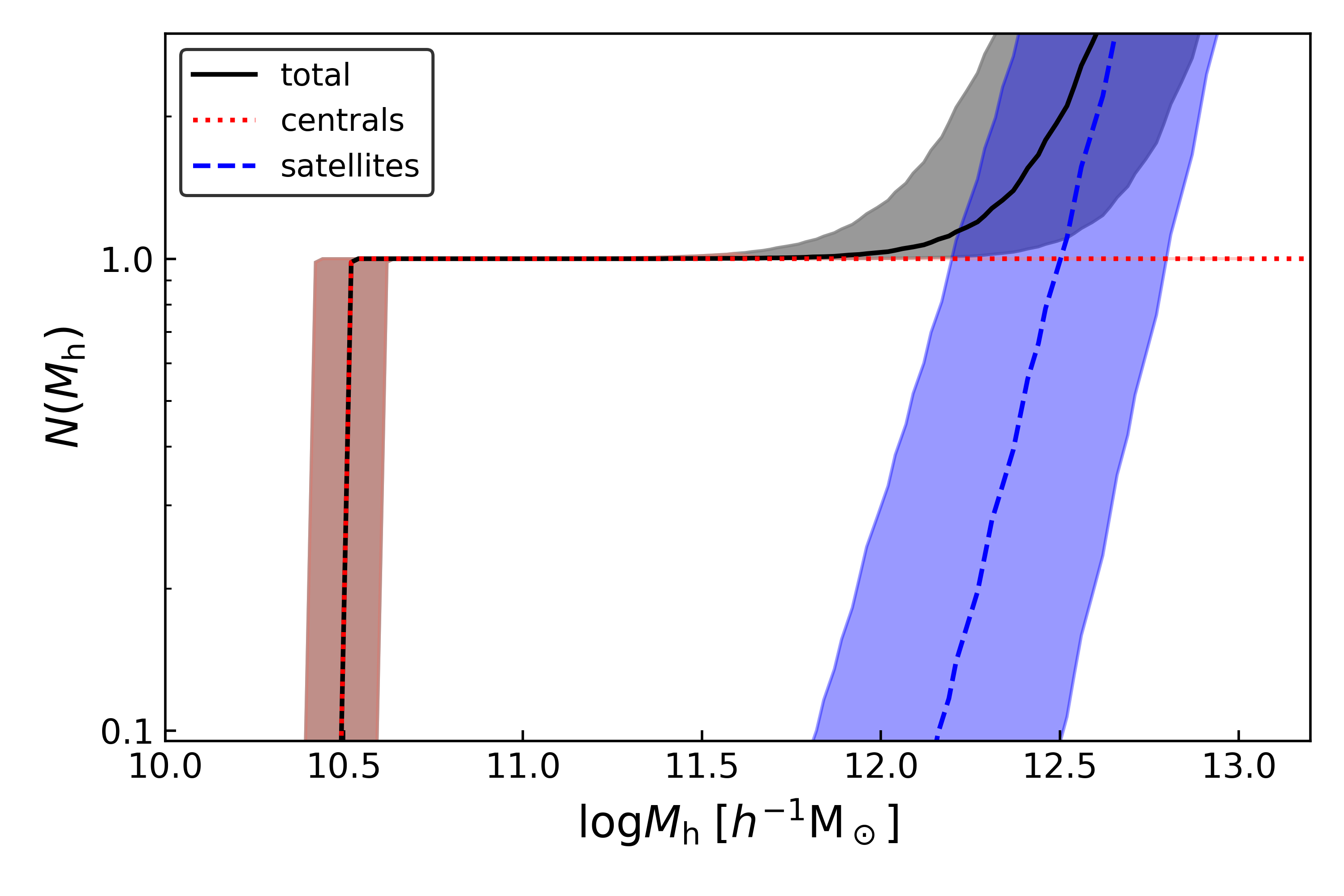}
\end{tabular}
\begin{tabular}{c c }
  \centering
  \includegraphics[width=.45\linewidth]{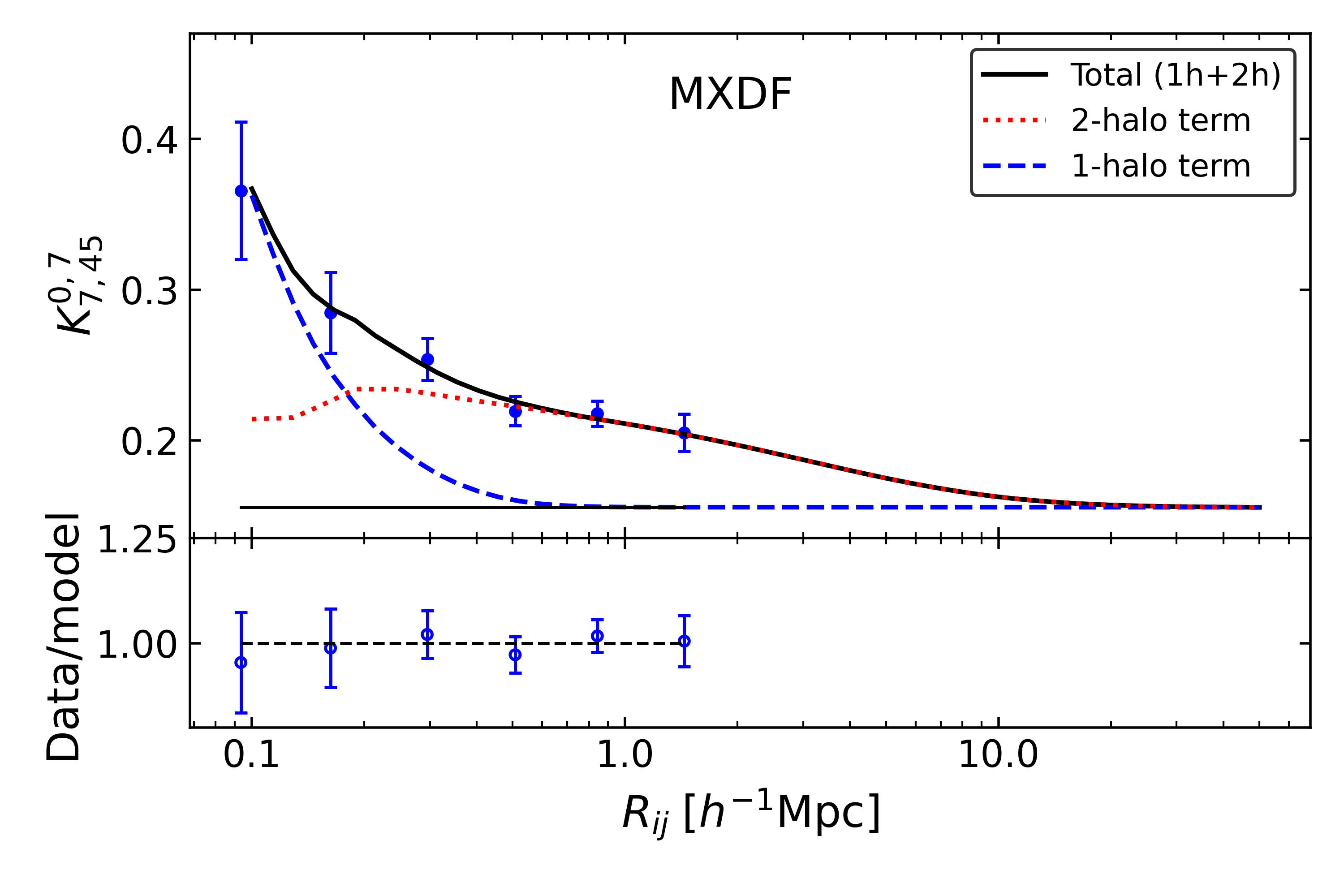}
\end{tabular}%
\begin{tabular}{c c }
  \centering
  \includegraphics[width=.45\linewidth]{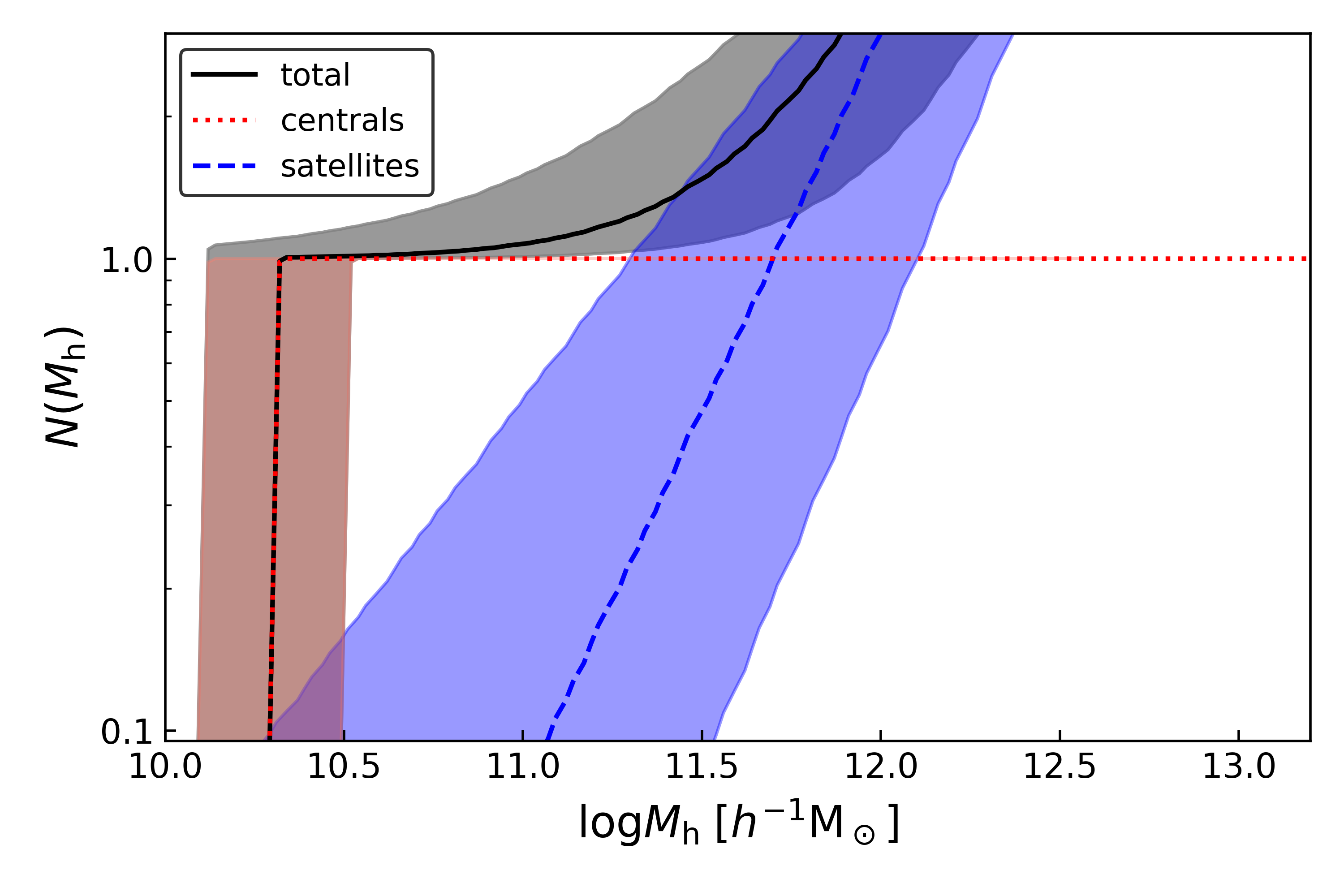}
\end{tabular}
\caption{Best-fit HOD models to the LAE clustering measurements (blue data points) from MUSE samples. Top left: Blue dashed, red dotted, and black continuous curves show the one-halo,  two-halo, and  total clustering terms from the MUSE-Wide sample, respectively. The black straight line shows the expected $K$ value of an unclustered sample. The residuals are shown below. The uncertainties are computed with the jackknife technique described in Sect.~\ref{sec:errorsK}. Top right: Best-fit HODs for central (red dotted), satellite (blue dashed), and total LAEs (black continuous) from the MUSE-Wide survey. Shaded regions correspond to $1\sigma$ confidence space. Middle: Same but for MUSE-Deep and using Poisson error bars. Bottom: Same but for MXDF and Poisson uncertainties.}
\label{fig:fit}
\end{figure*}

Using the K-estimator $K_{7,45}^{0,7}$, we compute the clustering of our LAE sample in ten logarithmic bins in the range $0.16 < R_{ij}/[h^{-1}\rm{Mpc}] < 27.5$, with error bars calculated following the jackknife resampling technique described in Sect.~\ref{sec:errorsK}. In the left top panel of Figure \ref{fig:fit}, we show the measured clustering signal, with all MUSE-Wide data points significantly above the 7/45 baseline, which represents the expected clustering of an unclustered population.

Following the procedure laid out in Sect.~\ref{sec:hod}, we obtain constraints on the HOD parameters. 
From the grid search and the $\chi^2$ minimization, we find the best HOD fit to the K-estimator, colored in black in the same figure and dissected into the one- and two-halo term contributions. It can be seen from the residuals (bottom) that the model is in remarkable agreement with the measurements. 

A somewhat intriguing feature, at least at first sight, is the kink in the two-halo term profile at $0.2<R_{ij}/[h^{-1}\rm{Mpc}]<0.4$. This reflects the effect of the halo-halo collision introduced in the HOD model formalism by \citet{tinker}, where the galaxy pairs within the same DMH cannot contribute to the two-halo term.

Our fitting allows us to find the best-fit HOD from Eqs.~\ref{eq:Nc} and \ref{eq:Ns}. In the right top panel of Fig.~\ref{fig:fit}, we represent the best HODs for the central, satellite, and total LAEs from the MUSE-Wide survey. While the halo mass needed to host one (central) LAE is $\log(M_{\rm{h}} / [h^{-1}\rm{M}_{\odot}])> 10.6$, satellite galaxies are only present if the DMHs are at least one order of magnitude more massive ($\log(M_{\rm{h}} / [h^{-1}\rm{M}_{\odot}])> 11.6$).
\begin{table*}[h]
\caption[]{Best-fit HOD parameters for the main samples of LAEs. } \label{table:hod_table}
\centering
\begin{tabular}{l@{\qquad}ccccccc}
        \hline \hline
           \noalign{\smallskip}
         & $\langle z \rangle$ & $\log(M_{\rm{min}} / [h^{-1}\rm{M}_{\odot}])$ &  $\log(M_1/M_{\rm{min}})$   & $\alpha$ &  $f_{\rm{sat}}$ & $b$ & $\log(M_{\rm{h}} / [h^{-1}\rm{M}_{\odot}])$\\
             \noalign{\smallskip} 
            \hline \hline
            \noalign{\smallskip}    MUSE-Wide &  4.0 &\;\;$10.7^{+0.2}_{-0.3}$ &   $1.7^{+0.4}_{-0.6}$ &  $2.8^{+0.9}_{-0.7}$ & $0.012^{+0.018}_{-0.009}$ &
            $2.65^{+0.13}_{-0.11}$ &
            $11.09^{+0.10}_{-0.09}$   \\
            MUSE-Deep & 4.1 & \;\;$10.5^{+0.2}_{-0.1}$ &   $1.9^{+0.3}_{-0.2}$ &  $3.0^{+0.4}_{-0.5}$ & $0.004^{+0.009}_{-0.002}$ &
            $2.42^{+0.10}_{-0.09}$ &
            $10.89^{+0.09}_{-0.09}$   \\
            MXDF & 4.2& \;\;$10.3^{+0.2}_{-0.3}$ &   $1.4^{+0.3}_{-0.2}$ &  $1.5^{+0.5}_{-0.5}$ & \;\;\;$0.08^{+0.02}_{-0.05}$ &
            $2.43^{+0.15}_{-0.15}$ &
            $10.77^{+0.13}_{-0.15}$   \\
            \noalign{\smallskip}
        \hline 
        \multicolumn{8}{l}{%
          \begin{minipage}{16cm}%
          \vspace{0.3\baselineskip}
            \small \textbf{Notes}: $\langle z \rangle$ is the median redshift of the sample. $M_{\rm{min}}$, $M_1$ are the threshold DMH masses to host a central and a satellite LAE, respectively. $\alpha$ is the high-mass power-law slope of the number of satellite galaxies, $f_{\rm{sat}}$ is the satellite fraction, $b$ is the large-scale bias factor and $M_h$ is the typical DMH mass of the galaxy sample.
          \end{minipage} 
          }\\
    \end{tabular}
\end{table*}

\begin{figure*}[h]
\centering
\includegraphics[width=0.8\textwidth,height=0.6\textheight]{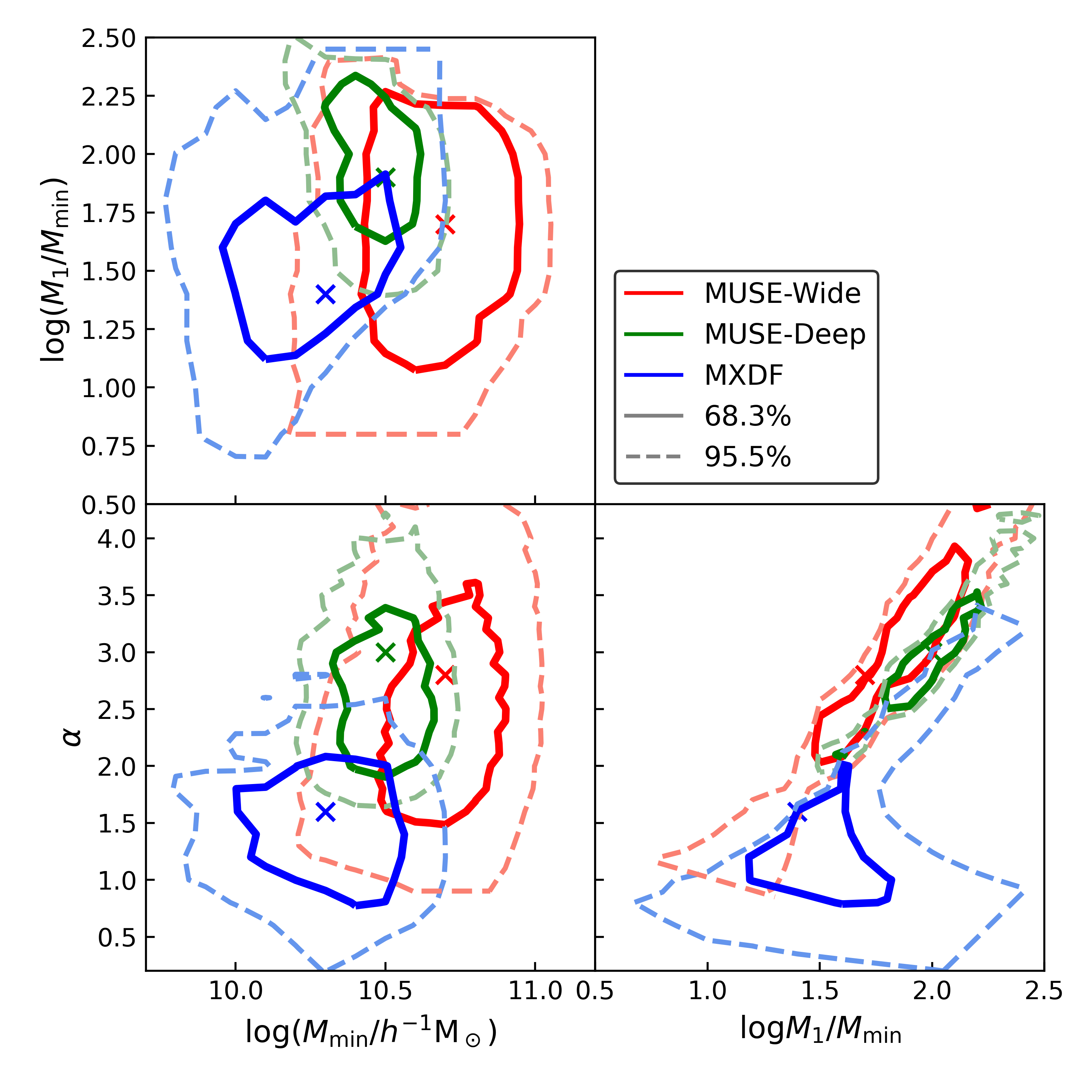}
\caption{Confidence contours in the three HOD parameter space. Red corresponds to MUSE-Wide, green to MUSE-Deep, and blue to MXDF. The thick (dashed) contours represent the 68.3\% (95.5\%) confidence, at $\Delta \chi^2=3.53$ (8.02) level.  The crosses stand for best-fit ($\chi^2_{\rm{min}}$), searched along the remaining parameter for each 2D parameter plane. }
\label{fig:contours}
\end{figure*}

As described in Sect.~\ref{sec:hod}, we also compute the confidence regions for the HOD parameters. We show the probability contours (red) in Fig.~\ref{fig:contours}.  The wobbliness of the curves, especially those involving $\alpha$, is caused by making use of a discrete grid. For our sample, the contours are constrained to have $\alpha>1$, $\log(M_1/M_{\rm{min}})>1$, and $\log(M_{\rm{min}}/[h^{-1}M_\odot])>10.4$. 

We list the best-fit HOD parameters in Table~\ref{table:hod_table}. While the minimum DMH mass required to host a central galaxy is $\log (M_{\rm{min}}/[h^{-1}M_\odot])=10.7^{+0.2}_{-0.3}$, that needed to host one central and (on average) one satellite is $\log (M_1/[h^{-1}M_\odot])=12.4^{+0.4}_{-0.6}$ (i.e., $\log (M_1/M_{\rm{min}})=1.7^{+0.4}_{-0.6}$). The power-law slope of the number of satellites is found to be $\alpha=2.8^{+0.9}_{-0.7}$. The inferred typical DMH mass is $\log( M_{\rm{h}}/[h^{-1}M_\odot])=11.09^{+0.10}_{-0.09}$, corresponding to a large-scale bias factor of $b=2.65^{+0.13}_{-0.11}$. The high values of $\log M_1$ and $\alpha$, considering the typical DMH mass of LAEs, suggest a low number of satellite galaxies detected in our sample. 

Seeking robust information about the number of satellite galaxies, we compute the satellite fraction $f_{\rm{sat}}$ (Eq.~\ref{eq:fsat}) for each parameter combination. 
We find $f_{\rm{sat}}\lesssim0.10$ at the 3$\sigma$ confidence level, being $f_{\rm{sat}}=0.012^{+0.018}_{-0.009}$. That is, $\approx 3$\% (1$\sigma$ upper limit) of the LAEs in the MUSE-Wide survey are satellites. In other words, at most $\approx 2$ out of $\approx 65$ DMHs in our sample host one satellite LAE.


\subsection{Fit results from MUSE-Deep}
\label{sec:fits_deep}

We measure the clustering of the MUSE-Deep LAE sample with the same K-estimator in eight logarithmic bins within $0.09 < R_{ij}/[h^{-1}\rm{Mpc}] < 4.75$. We compute Poisson uncertainties as laid out in Sect.~\ref{sec:errorsK_deeper} and display the result in the middle left panel of Fig.~\ref{fig:fit}. Overplotted on the clustering signal, we show the best HOD fit, split into the one- and two-halo term contributions. The good quality of the fit is quantified with the residuals in the bottom panel of the figure. 

Following the procedure described in Sect.~\ref{sec:errorsK_deeper}, we compute the confidence intervals for the HOD parameters and list them in Table~\ref{table:hod_table}. We plot the probability contours (green) in Fig.~\ref{fig:contours}, which overlap significantly  with those from the MUSE-Wide sample.  Central LAEs can occupy DMHs if these are at least as massive as $\log (M_{\rm{min}}/[h^{-1}\rm{Mpc}])=10.5^{+0.2}_{-0.1}$, whereas, in order to host satellite LAEs, the halos must have masses $\log (M_1/[h^{-1}\rm{Mpc}])=12.4^{+0.3}_{-0.2}$ ($\log (M_1/M_{\rm{min}})=1.9^{+0.3}_{-0.2}$).
These values correspond to a large-scale bias and typical DMH mass $b=2.42^{+0.10}_{-0.09}$ and $\log (M_{\rm{h}}/[h^{-1}\rm{M}_{\odot}])=10.89^{+0.09}_{-0.09}$, which are similar to those found in the MUSE-Wide survey. The derived satellite fraction is  $f_{\rm{sat}}=0.004^{+0.009}_{-0.002}$, consistent with that from the MUSE-Wide LAE sample.

We then compute the best-fit HOD for central, satellite and total LAEs (middle right panel of Fig.~\ref{fig:fit}). In line with the best-fit HOD parameters and somewhat lower than the values found for the MUSE-Wide survey, the smallest DMH that can host a central LAE has a mass of $\log (M_{\rm{h}}/[h^{-1}\rm{M}_{\odot}])>10.4$, more than one order of magnitude lower than that required to host one additional LAE (satellite).

\subsection{Fit results from the MUSE Extremely Deep Field}
\label{sec:fits_mxdf}

We make use of six logarithmic bins in the range $0.09 < R_{ij}/[h^{-1}\rm{Mpc}] < 1.45$ and Poisson errors (see Sect.~\ref{sec:errorsK_deeper}) to quantify the clustering of the sample of LAEs from MXDF. We show the K-estimator measurements in the bottom left panel of Fig.~\ref{fig:fit}, along with the corresponding best HOD fit. 

 The probability contours are plotted in blue in Fig.~\ref{fig:contours}, significantly apart from those of MUSE-Wide and MUSE-Deep. While the minimum DMH mass to host a central LAE is $\log (M_{\rm{min}}/[h^{-1}\rm{Mpc}])=10.3^{+0.2}_{-0.3}$, that to host one central and one satellite LAE is $\log (M_1/[h^{-1}\rm{Mpc}])=11.7^{+0.3}_{-0.2}$ ($\log (M_1/M_{\rm{min}})=1.4^{+0.3}_{-0.2}$). These values are somewhat lower than those found for the MUSE-Wide survey and correspond to a bias factor and typical halo mass of $b=2.43^{+0.15}_{-0.15}$ and $\log (M_{\rm{h}}/[h^{-1}\rm{M}_{\odot}])=10.77^{+0.13}_{-0.15}$, respectively. The inferred satellite fraction is  $f_{\rm{sat}}=0.08^{+0.02}_{-0.05}$ ($f_{\rm{sat}}\lesssim0.2 $ at the 3$\sigma$ confidence level), tentatively higher than that found in the MUSE-Wide survey. 

 From the best-fit HOD parameters, we calculate the HODs for central, satellite and total LAEs and show them in the bottom right panel of Fig.~\ref{fig:fit}. Significantly lower than in the MUSE-Wide survey, central LAEs reside in DMHs if these are more massive than $\log (M_{\rm{h}}/[h^{-1}\rm{M}_{\odot}])>10.2$. For the satellite case, and similarly to the previous LAE samples, they only exist if the halos are around one order of magnitude more massive.

It is worth pointing out that the three HOD parameters have some degree of degeneracy, printed out in the diagonally elongated probability contours 
 in $\log M_1/M_{\rm min}$ -- $\alpha$ space in the bottom right panel of Fig.~\ref{fig:contours}. This can be understood as follows: a higher $\alpha$ in the models causes an increase of satellites at high mass halos, but this can be compensated by producing less satellites by increasing $\log M_1/M_{\rm min}$.
 While this correlation is clearly visible for the MUSE-Wide and MUSE-Deep samples, the MXDF dataset only seems to be affected in the 95\% confidence contour. We did not observe clear correlations between other parameters with any of our samples. Appendix \ref{appendix:k-fields} shows how our K-estimator varies with the parameters. The causes of parameter degeneracies are also noticeable in Fig.~\ref{fig:k-parameters}. We note however that while the correlation between the HOD parameters leads to the perturbed shape of the probability contours, the lowest (MXDF) and highest luminosity (MUSE-Wide) sample contours are detached from each other. Thus, for the purposes of this study, simultaneously fitting the three HOD parameters and showing their correlations is preferable over, for instance, fixing $\alpha$ to a dubious value.

\section{Discussion}
\label{sec:discussion}

\subsection{Clustering dependence on Ly$\alpha$ luminosity}
\label{sec:dependence}

\begin{figure*}
\centering
\begin{tabular}{c c }
  \centering
  \includegraphics[width=.45\linewidth]{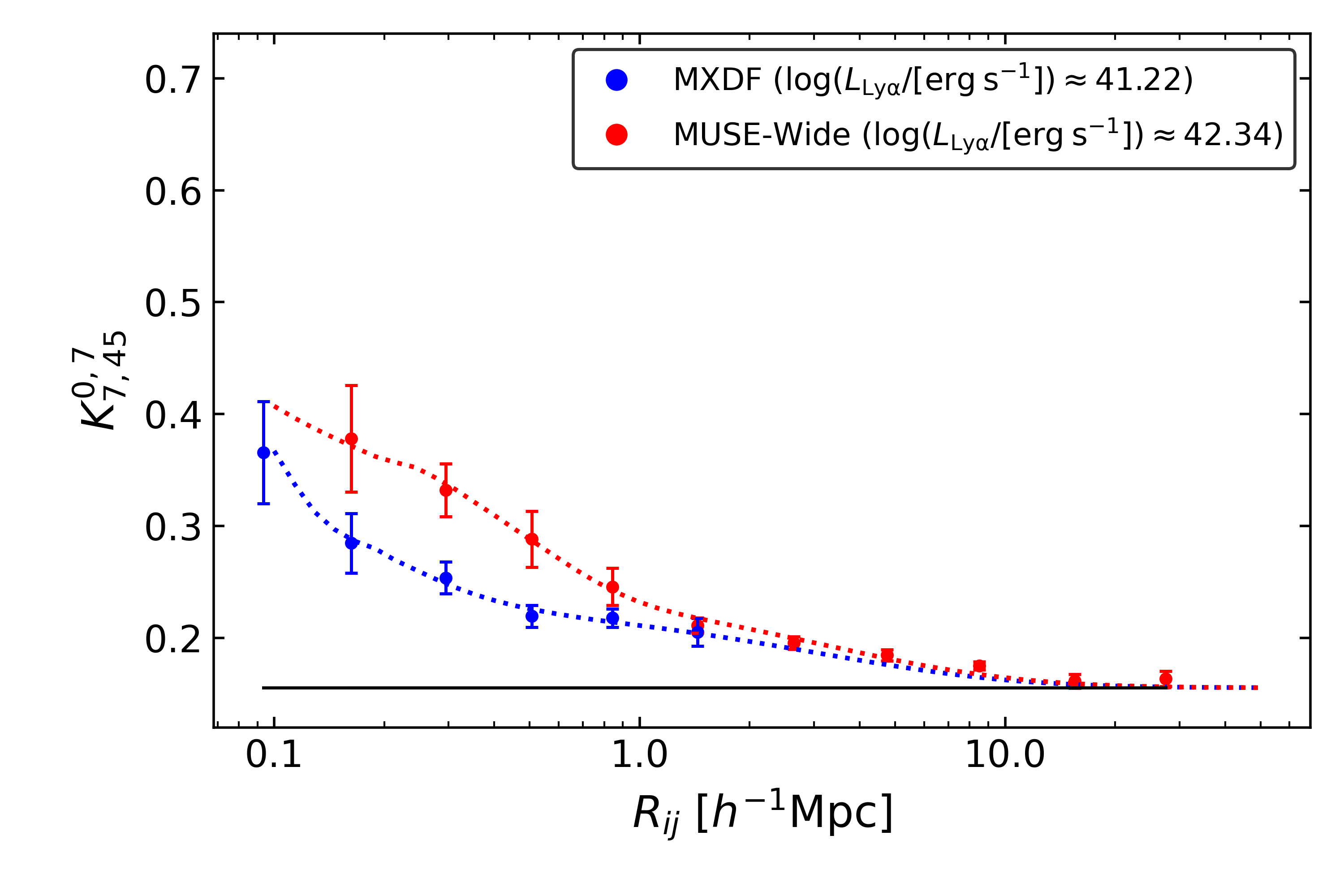}
\end{tabular}
\begin{tabular}{c c }
  \centering
  \includegraphics[width=.45\linewidth]{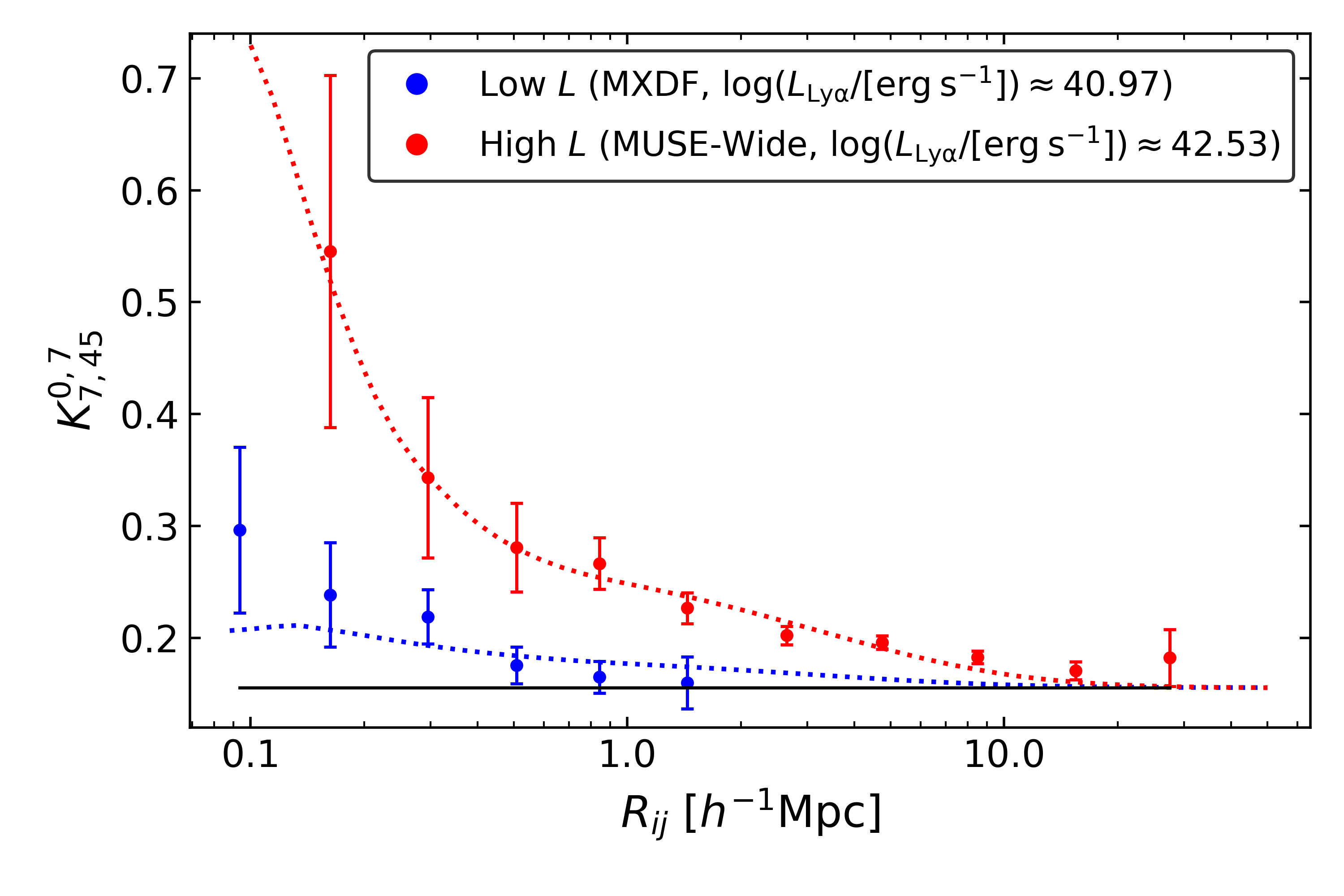}
\end{tabular}
\caption{Clustering dependence on Ly$\alpha$ luminosity. Left: K-estimator measurements in the MUSE-Wide survey (red; $\langle L_{\rm{Ly}\alpha}\rangle\approx10^{42.34}$ erg s$^{-1}$) and MXDF (blue; $\langle\log L_{\rm{Ly}\alpha}\rangle \approx10^{41.22}$ erg s$^{-1}$). The dotted curves represent the best HOD fits. The black straight line shows the expected K-estimator of an unclustered sample.   Right: Same for the high $L_{\rm{Ly}\alpha}$ subset (red) from the MUSE-Wide survey and the low $L_{\rm{Ly}\alpha}$  subsample (blue) from MXDF.} 
\label{fig:luminosity}
\end{figure*}

The complex radiative transfer processes that the Ly$\alpha$ photons are subject to make the search for correlations between Ly$\alpha$ luminosity and other physical properties a difficult task. Despite this complication, \cite{yajima} predicted a correlation between simulated $L_{\rm{Ly}\alpha}$ and halo mass based on halo merger trees and Ly$\alpha$
 radiative transfer calculations. 
\cite{khostovan19} is, however, the only study so far that has reported a clear (5$\sigma$) relation between these quantities using observational data. Motivated by these results, we exploited the large dynamic range of Ly$\alpha$ luminosities that we cover to investigate the relation between Ly$\alpha$ luminosity and DMH mass. 
As a first step, we compare the K-estimator measurements in the MUSE-Wide survey (highest luminosity LAE sample: $\langle L_{\rm{Ly}\alpha}\rangle \approx 10^{42.34}$ erg s$^{-1}$, but still fainter than those in \cite{khostovan19}) and in MXDF (faintest LAE sample; $\langle L_{\rm{Ly}\alpha}\rangle \approx 10^{41.22}$ erg s$^{-1}$) and show the outcome of this comparison in the left panel of Fig.~\ref{fig:luminosity}.

\begin{figure}[h]
\centering
\includegraphics[width=\columnwidth]{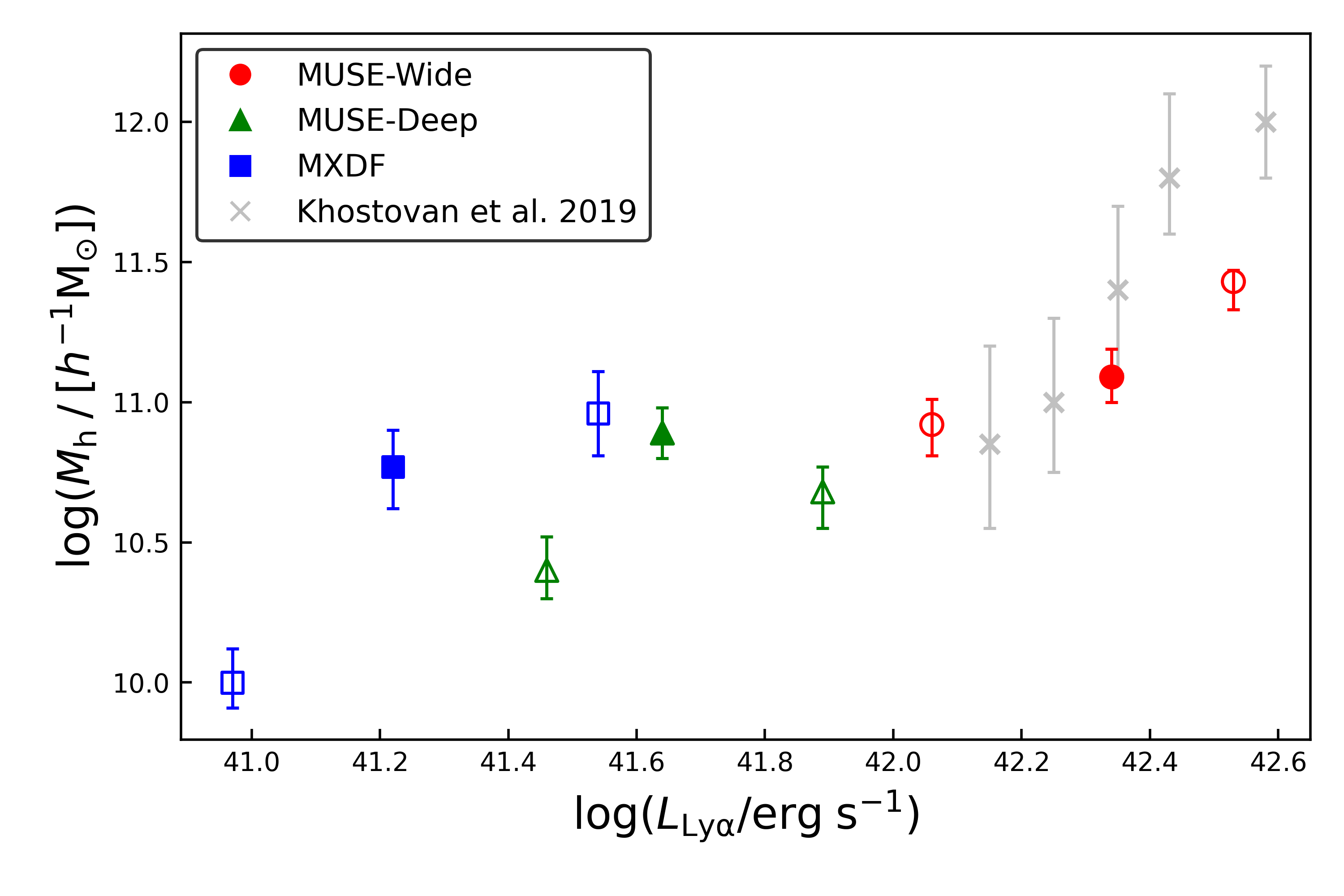}
\caption{Typical dark matter halo mass against observed median Ly$\alpha$ luminosity. Filled and unfilled symbols correspond to the values derived from the samples and subsamples described in Sect.~\ref{sec:data}, respectively. Red circles, green triangles and blue squares belong to MUSE-Wide, MUSE-Deep and MXDF, respectively. Gray crosses represent the results from \cite{khostovan19} in the Ly$\alpha$ luminosity interval relevant for this study.} 
\label{fig:bias-lum}
\end{figure}

The relatively luminous LAEs from the MUSE-Wide survey cluster slightly more strongly ($b_{\rm{Wide}}=2.65^{+0.13}_{-0.11}$) than the low-luminosity LAEs from MXDF ($b_{\rm{MXDF}}=2.43^{+0.15}_{-0.15}$). 
The clustering measurements and bias factor ($b=2.42^{+0.10}_{-0.09}$) in MUSE-Deep ($\log(L_{\rm{Ly}\alpha}/[\rm{erg\: s}^{-1}])=41.64$) fall between those from MUSE-Wide and MXDF. We convert the bias factors from the three main samples of this study into  typical DMH masses and plot them as a function of their median Ly$\alpha$ luminosity with colored symbols in Fig.~\ref{fig:bias-lum}. 

\begin{table}[htbp]
\caption[]{Best HOD fit large-scale bias factor and typical DMH mass for the LAE subsamples.} \label{table:subsamples_hod}
\centering
    \begin{tabular}{p{0.32\linewidth}p{0.08\linewidth}p{0.13\linewidth}p{0.28\linewidth}l}
        \hline \hline
           \noalign{\smallskip}
          Subsample &$\langle z \rangle$& \;\;\;\;$b$   & $\log(M_{\rm{h}} / [h^{-1}\rm{M}_{\odot}])$\\
             \noalign{\smallskip} 
            \hline \hline
            \noalign{\smallskip}
              
              MUSE-Wide high L &4.1& $3.13^{+0.08}_{-0.15}$  & \;\;\;\;\;$11.43^{+0.04}_{-0.10}$ \\
           	  MUSE-Wide low L &3.7&  $2.45^{+0.10}_{-0.12}$    & \;\;\;\;\;$10.92^{+0.09}_{-0.11}$  \\
	  MUSE-Deep high L &4.5&  $2.41^{+0.12}_{-0.10}$  & \;\;\;\;\;$10.40^{+0.12}_{-0.10}$  \\
              MUSE-Deep low L & 3.7&  $2.20^{+0.09}_{-0.11}$ & \;\;\;\;\;$10.68^{+0.09}_{-0.13}$ \\
              MXDF high L & 4.5& $3.10^{+0.24}_{-0.22}$  & \;\;\;\;\;$10.96^{+0.15}_{-0.15}$ \\
              MXDF low L & 4.0& $1.79^{+0.08}_{-0.06}$  &  \;\;\;\;\;$10.00^{+0.12}_{-0.09}$\\
            \noalign{\smallskip}
        \hline 
    \end{tabular}
    \tablefoot{$\langle z \rangle$ is the median redshift of the subsample. The uncertainties do not include cosmic sample variance.}
\end{table}

Although the three main datasets sample the same region of the sky, their transverse coverage is limited and somewhat differs. Therefore, our results are affected by cosmic sample variance. Ideally, this uncertainty is estimated from the variance of clustering measurements from simulated mocks in different lines of sight. Inferring cosmic variance from a large set of mocks that are able to reproduce the observed clustering of our LAEs is however beyond the scope of this paper. 

We further investigate the possible dependence on $L_{\rm{Ly}\alpha}$ by splitting the main LAE samples into disjoint subsets (see Table~\ref{table:subsamples}). We compute the K-estimator in each $L_{\rm{Ly}\alpha}$ subsample, find the best HOD fit and list the large-scale bias factors and the typical DMH masses in Table~\ref{table:subsamples_hod}. We also plot the typical DMH masses in Fig.~\ref{fig:bias-lum} (empty symbols) as a function of the median $L_{\rm{Ly}\alpha}$ of the subsamples. We find that typical halo mass increases from $10^{10.00}$ to $10^{11.43}M_\odot$ between $10^{40.97}$ and $10^{42.53}$ erg s$^{-1}$ in line luminosity. 

For each subsample pair, the high-luminosity subset always clusters more strongly than the low-luminosity one and, in this case, cosmic sample variance effects can be completely neglected because subset pairs span the exact same area on the sky. The most pronounced difference is found when splitting the MXDF sample, the dataset with the largest dynamic range of Ly$\alpha$ luminosity. The best HOD fits deliver $b_{\rm{low}}=1.79^{+0.08}_{-0.06}$ and $b_{\rm{high}}=3.10^{+0.24}_{-0.22}$ ($3.9\sigma$ significant). 

Despite its higher luminosity, we infer a less massive DMH for the MUSE-Deep high-luminosity subsample than for the main dataset. This is due to the higher $z_{\rm{pair}}$ of the subset (see Sect.~\ref{sec:subsamples} and \ref{sec:hod}). Because we evaluate the HOD model at $z_{\rm{pair}}$, a higher redshift corresponds to HOD models in which the halo mass function presents a lower number density of massive halos and, thus, deliver less massive typical DMHs. The same reasoning applies when comparing the high-luminosity MXDF and low-luminosity MUSE-Deep subsamples and the high-luminosity MUSE-Deep and low-luminosity MUSE-Wide subsets. While each subsample pair presents similar median luminosities,  the former also has similar $z_{\rm{pair}}$, unlike the latter one (see Sect.~\ref{sec:subsamples}). This translates into similar DMH masses for the first pair but significantly distinct masses for the second.

We last consider the most extreme cases, the low-luminosity subset from MXDF and the high-luminosity one from the MUSE-Wide survey. We show the measured clustering in the two subsamples in the right panel of Fig.~\ref{fig:luminosity}. 
The high-luminosity LAEs cluster $8\sigma$ more strongly than the low-luminosity LAEs, without accounting for cosmic variance. We find that LAEs with $ \log(L_{\rm{Ly}\alpha}/[\rm{erg\: s}^{-1}])\approx42.53$ reside in DMHs of $\log(M_{\rm{h}} / [h^{-1}\rm{M}_{\odot}])=11.43^{+0.04}_{-0.10}$ and that lower luminosity LAEs ($\log(L_{\rm{Ly}\alpha}/[\rm{erg\: s}^{-1}])\approx40.97$) are hosted by DMHs of masses ranging $\log(M_{\rm{h}} / [h^{-1}\rm{M}_{\odot}])=10.00^{+0.12}_{-0.09}$. These results fit well within the assumed framework in which star-forming galaxies that reside in more massive halos present higher star formation rates and thus show more luminous nebular emission lines \citep{haruka}. This dependence can then be weakened by low Ly$\alpha$ escape fractions in high mass halos.

Following Sect.~5.4.1 of \cite{yohana}, we matched the redshift distributions of the three main samples and of each subsample pair to verify that the difference in clustering amplitude is not driven by the different redshift distribution of the datasets. For each main sample, we compare individual bins between their corresponding $z$-distributions and select the one that contains a higher number of objects. We then randomly remove LAEs until we match the number counts of the non-selected samples in that bin. Once all bins have been inspected, we obtain "matched" $z$-distributions (i.e., equivalent), but with still different Ly$\alpha$ luminosity distributions. We ran the K-estimator in the three "matched" datasets and find consistent results with the original ones. We follow the same approach for the subsamples such that the low- and high-luminosity subsets have exactly the same $z$-distribution. We find that the clustering difference between the "matched" and original subsamples varies within $1\sigma$. Besides, as we did for $L_{\rm{Ly}\alpha}$, we also searched for a possible clustering dependence on redshift and found no trend. Thus, we discarded the possibility of a possible clustering dependence on Ly$\alpha$ luminosity driven by $z$.

Our results are not driven by AGN or low-redshift emission line contamination either. The Ly$\alpha$-emitting AGN fraction for $L_{\rm{Ly}\alpha}<10^{43}$ erg s$^{-1}$ is close to zero (\citealt{spinoso} and references therein) and the four known X-ray detected AGNs \citep{luo17}, which only affect MUSE-Wide and MUSE-Deep, were not included in our datasets. Besides, \cite{urrutia19} performed a stacking experiment of X-ray images centered on MUSE-Wide LAEs, yielding no signal. The presence of low-redshift interlopers in our spectroscopic samples is also unlikely. [\ion{O}{ii}] emitters are the typical contaminants of high-redshift LAE samples but the high resolution of the MUSE instrument allows to distinguish the [\ion{O}{ii}] emission line doublet with high confidence.

These results are in line with the tentative trends seen in \cite{ouchi03,haruka,yohana} and the clear dependence found in \cite{khostovan19}. While \cite{ouchi03} noted a slight difference in the correlation amplitude of two $L_{\rm{Ly}\alpha}$ subsamples (30 and 57 LAEs in each subset at $z = 4.86$ with $\log(L_{\rm{Ly}\alpha}/[\rm{erg\: s}^{-1}])>42.2$ and $\log(L_{\rm{Ly}\alpha}/[\rm{erg\: s}^{-1}])<42.2$, respectively), \cite{haruka} observed a tendency ($<2\sigma$) of larger bias factors corresponding to higher luminosity LAEs. They used four deep survey fields at $z=2$ with limiting Ly$\alpha$ luminosities within the range of $41.3<\log(L_{\rm{Ly}\alpha}/[\rm{erg\: s}^{-1}])<42$ computed from NB387 magnitudes.

More significant  is the dependence found in \cite{khostovan19} and \cite{yohana}.
While the latter measured a $2\sigma$ difference in bias factors or DMH masses between two subsets of 349 and 346 LAEs at $z\approx4$ with $\log(L_{\rm{Ly}\alpha}/[\rm{erg\: s}^{-1}])\approx42.14$ and $\log(L_{\rm{Ly}\alpha}/[\rm{erg\: s}^{-1}])\approx42.57$, the former used various surveys with discrete redshift slices between $2.5<z<6$ and $42.0<\log(L_{\rm{Ly}\alpha}/[\rm{erg\: s}^{-1}])<43.6$ to find that halo mass clearly ($5\sigma$) increases with increasing line luminosity. For a direct comparison, we plot in Fig.~\ref{fig:bias-lum} (gray crosses) the DMH masses computed by \cite{khostovan19} from samples with similar redshifts ($z\approx3$) and Ly$\alpha$ luminosities ($\log(L_{\rm{Ly}\alpha}/[\rm{erg\: s}^{-1}])\approx42$) to our current LAE samples. Our results are in good agreement and extend to much fainter Ly$\alpha$ luminosities. 

Our results, along with those from the literature, demonstrate that having a broad dynamic range of $L_{\rm{Ly}\alpha}$ (nearly extending two orders of magnitude) and a large number of LAEs in the samples is crucial to detect the clustering dependence on $L_{\rm{Ly}\alpha}$.


\subsection{Comparison to \cite{yohana}}
\label{sec:comparison_yohana}

\begin{figure}[h]
\centering
\includegraphics[width=\columnwidth]{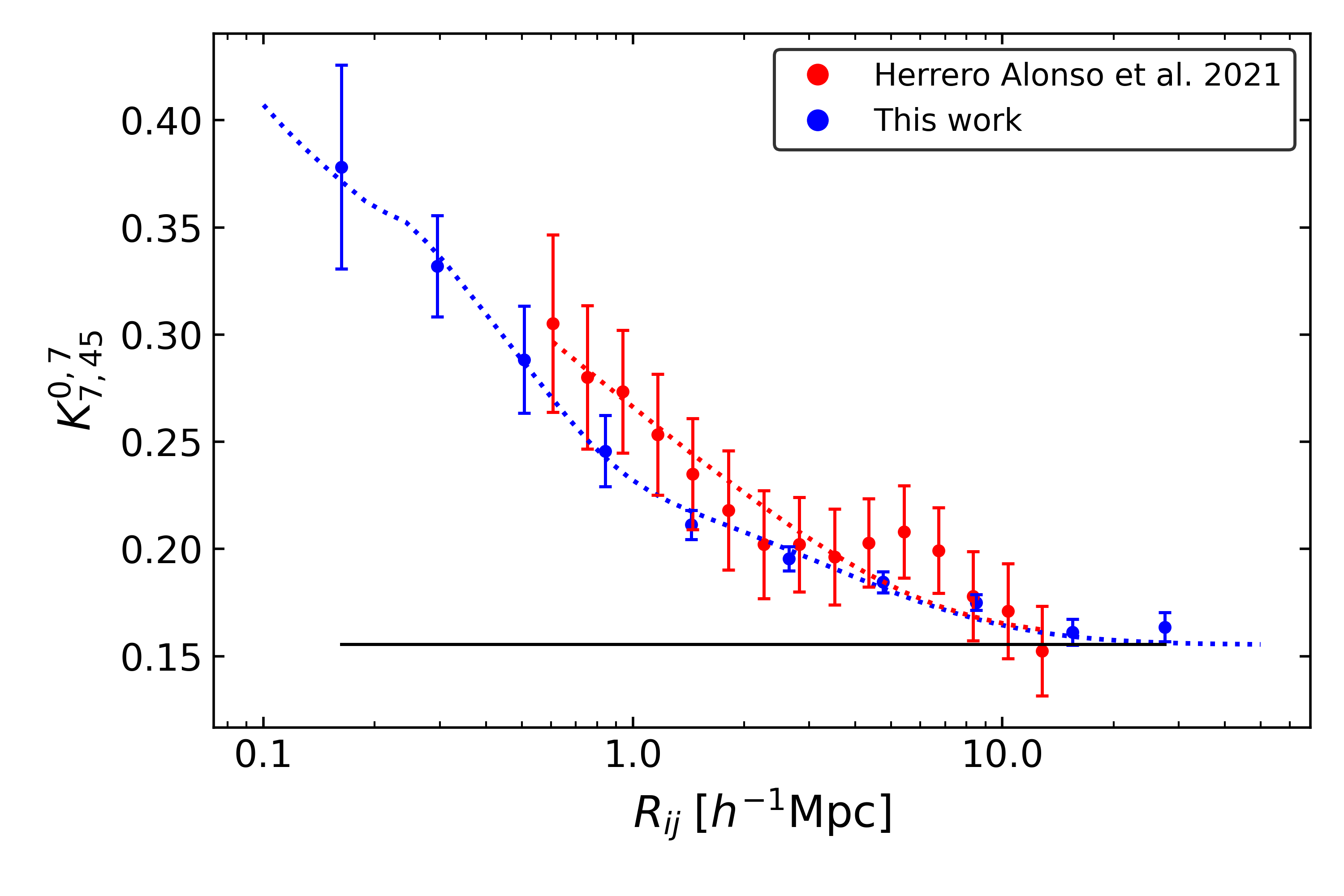}
\caption{Clustering of the full MUSE-Wide sample (blue; this work) compared to the subset considered in HA21 (red). The former measurements show jackknife uncertainties (see Sect.~\ref{sec:errorsK}) and the latter bootstrapping errors (see Sect.~3.1.3 in HA21). The blue dotted curve represents our best-fit from full  HOD modeling. The red dotted curve displays the two-halo term only best HOD fit found in Sect.~4.3 of HA21. The black straight line shows the expected $K$ value of an unclustered sample.} 
\label{fig:paper1}
\end{figure}

In this section we compare our results with the findings of our previous study \citep[][hereafter HA21]{yohana}, where we measured the clustering of a subset (68 fields of the MUSE-Wide survey) of our current sample (91 fields of the MUSE-Wide survey) and fitted the corresponding signal with a two-halo term only HOD modeling. In order to envisage the methodological and statistical improvement of our new investigation, we applied our $K_{7,45}^{0,7}$ estimator to the sample considered in HA21 (695 LAEs at $3.3<z<6$). We compare the outcome to our current clustering measurement in Fig.~\ref{fig:paper1}.

The two datasets show good agreement within the uncertainties, with smaller errors for the current sample. Besides the higher number of LAEs and larger spatial coverage, the error estimation was carried out following different procedures. While the spatial coverage of the full MUSE-Wide survey allows us to compute the covariance matrix from the jackknife resampling technique, the smaller transverse extent covered by the 68 fields did not allow the split of the surveyed area into a significant number of jackknife zones. Thus, in HA21, we chose bootstrapping error bars as our next most conservative and realistic approach.

The slightly puzzling hump seen in Sect.~4 of HA21 at $4 \lesssim R_{ij}/[h^{-1}\rm{Mpc}] \lesssim 7$ is no longer visible in our new dataset. This confirms the judgement in HA21 that the feature was consistent with a statistical fluctuation resulting from the correlation between datapoints.

In HA21, we limited the range of transverse separations to $R_{ij}>0.6\:h^{-1}\rm{Mpc}$, excluding the smallest scales of the one-halo term. Thus, we fitted the signal with a two-halo term only HOD model (red dotted curve in Fig.~\ref{fig:paper1}) in contrast to the full HOD modeling performed in this work (blue dotted curve). While the former only constrained the large-scale bias factor and the typical DMH mass of LAEs, the latter further determines the number of central and satellite galaxies, as well as the required DMH mass to host each type of galaxy. Despite these dissimilarities, the two fits are in good agreement: the bias factor ($b=2.80^{+0.38}_{-0.38}$) and the typical DMH mass of LAEs ($\log(M_{\rm{DMH}}$ / $[\it{h}^{-1}\rm{M}_{\odot}])=11.34^{+0.23}_{-0.27}$) from HA21 are consistent with those derived in this work ($b=2.65^{+0.13}_{-0.11}$ and $\log(M_{\rm{DMH}}$ / $[\it{h}^{-1}\rm{M}_{\odot}])=11.09^{+0.10}_{-0.09}$). The higher accuracy of our current measurements originates from the larger sample, the availability of more realistic error bars, and constraints from the one-halo term.

\subsection{Comparison to the literature}
\label{sec:literature}

A common way to infer the host DMH masses of LAEs is to quantify the galaxy clustering of the detected population through clustering statistics, which is then traditionally approximated with power-laws or fit with physically motivated HOD models.

Following the traditional approach, \cite{gawiser07}, \cite{ouchi10} and \cite{bielby} focused on the clustering of a few hundred LAEs at $z= 3.1-6.6$ to obtain typical DMH masses in the range $10^{10}-10^{11}\;M_\odot$. Similar masses were found by \cite{khostovan18} in a much larger sample ($\approx 5000$ LAEs) in discrete redshift slices within $2.5<z<6$, adopting the same procedure. A major improvement in terms of methodology  was presented in \cite{lee06,durkalec14,ouchi17,durkalec18}, who considered samples of high-$z$ galaxies (2000-3000 mainly LAEs and Lyman-break galaxies, LBGs) and quantified the clustering with HOD modeling. While \cite{ouchi17} found that their LAEs at $z=5.7$ (6.6) are hosted by DMHs with typical masses of $\log (M_h/M_\odot)=11.1^{+0.2}_{-0.4}$ ($10.8^{+0.3}_{-0.5}$), \cite{lee06} and \cite{durkalec14,durkalec18} computed $\log (M_h/h^{-1}M_\odot)\approx11.7$ for their sample of galaxies at $z=4-5$ and $z=3$, respectively. 
Considering that we have performed a full HOD modeling at the median redshift of our number of galaxy pairs ($z_{\rm pair}=3.8$) and that the DMH masses are predicted to evolve with cosmic time, our derived typical DMH masses $\log (M_h/h^{-1}M_\odot)\approx10.77-11.09$ are in good agreement with the literature.

Besides the computation of typical DMH masses, modeling the one-halo term of the clustering statistics with HOD models delivers the minimum DMH mass required to host a central galaxy, $M_{\rm{min}}$, that is needed for a satellite galaxy, $M_1$, and the power-law slope of number of satellites, $\alpha$. These three parameters constrain the satellite fraction, $f_{\rm{sat}}$. \cite{ouchi17} partially exploited the power of HOD models in a sample of $\approx2000$ LAEs to obtain $\log (M_{\rm{min}}/M_\odot)=9.5^{+0.5}_{-1.2}$ ($9.1^{+0.7}_{-1.9}$) at $z=5.7$ (6.6). Our derived minimum masses to host a central galaxy at $z_{\rm pair}=3.8$ are considerably larger ($\log (M_{\rm{min}}/M_\odot)\approx10.3-10.7$), which can be explained by the different Ly$\alpha$ luminosities covered in the two studies, and by the fact that several HOD parameters were fixed in \cite{ouchi17}, namely, $\sigma_{\log M}=0.2$, $\log M_0=0.76M_1+2.3$, $\log M_1=1.18\log M_{\rm{min}}-1.28$, and $\alpha=1$, which are not compatible with ours. This was the only previous study that performed HOD modeling in a sample of LAEs. 

\cite{lee06} and \cite{durkalec14} made use of the full potential of HOD models to reproduce the clustering of their LBG population at $z=4-5$ and $2.9<z<5$, respectively. Although it is still under debate whether LBGs and LAEs are the same galaxy population (\citealt{garel15} and references therein), \cite{lee06} computed a minimum DMH mass to host a central LBG of $\log (M_{\rm{min}}/M_\odot)\approx10.8$, to host a satellite LBG of $\log (M_1/M_\odot)\approx12.0$, and a power-law slope $\alpha$ for the number of satellites of $\alpha\approx0.7$, with considerable uncertainties.  Similarly, \cite{durkalec14} found  $\log (M_{\rm{min}}/M_\odot)=11.18^{+0.56}_{-0.70}$, $\log (M_1/M_\odot)=12.55^{+0.85}_{-0.88}$, and $\alpha=0.73^{+0.23}_{-0.30}$.
While their halo masses are in agreement with our findings, their slope is somewhat shallower. This is partially expected given the dissimilarities in the galaxy populations (i.e., disparate observational selection techniques detect distinct galaxy populations). 


\subsection{Satellite fraction}
\label{sec:fsat}

In the above discussions on HOD modeling, we limit ourselves to the HOD model form expressed by Eqs. \ref{eq:Nc} and \ref{eq:Ns}, which is rather restrictive. The underlying assumption of the model is that the center of the halo with mass $M_{\rm h}>M_{\rm min}$ is always occupied by one galaxy in the sample (or at least at a $M_{\rm h}$-independent constant probability). This form may be appropriate for instance, for luminosity or stellar mass thresholding samples, but there is no reason that this has to be the case for samples selected by other criteria.

We note that the inferred value of $f_{\rm sat}$ is sensitive to the form of the parameterized model of the central and satellite HODs. In this work and in the literature, a power-law form of the satellite HOD is customarily assumed. In this case, a lower $\alpha$ would increase the model $\langle N_{\rm s}(M_{\rm h})\rangle$ at the lower $M_{\rm h}$ end, near $M_{\rm min}$, and yield fewer satellites in higher mass halos. Since the halo mass function drops with increasing mass, $f_{\rm sat}$ is mainly determined by the HOD behavior around $M_{\rm h}\sim M_{\rm min}\sim 10^{10.5}\;h^{-1} {\rm M_\odot}$, where the halo mass function is large and the virial radius is $r_{\rm vir}\approx 0.08 \;h^{-1} {\rm Mpc}$ at $z\sim 3.8$ \citep{zheng07}. These scales are too small to be well constrained by our observations. 
Our observed one-halo term mainly constrains the satellite fraction at larger mass halos ($M_{\rm h}\sim M_{\rm min}\sim 10^{13}\;h^{-1} {\rm M_\odot}$, where $r_{\rm vir}\approx 0.5 \;h^{-1} {\rm Mpc}$ at the same redshift).  Thus, the $f_{\rm sat}$ values from the HOD modeling should be viewed with caution and may well reflect the artefacts of the assumed form of the model. On the other hand, the sheer presence of a significant one-halo term indicates the existence of some satellites at higher halo masses. The extent of the one-halo term up to $R_{ij}\approx 0.5\; h^{-1} {\rm Mpc}$ shows that there are indeed satellites up to $M_{\rm h}\sim 10^{13}\;h^{-1} {\rm M_\odot}$. 

In spite of the above caveats, the small satellite fraction of the LAEs is likely to be robust. The small $f_{\rm sat}$ values for the assumed HOD model indicate that not only central-satellite pairs are rare, but also satellite-satellite pairs are as well, suggesting that only a small fraction of halos contain multiple LAEs. The small $M_{\rm min}$ values themselves are also an indication that a large majority of the halos (at the low mass end) that contain a LAE are indeed dominated by one galaxy and in this case, the LAE is probably the central galaxy.


\subsection{Implications}
\label{sec:implications}

The clustering results of this study do not only have implications on the baryonic-DM relation, but also on evolving Ly$\alpha$ luminosity functions, signatures of incomplete reionization, and halo mass-dependent Ly$\alpha$ escape fractions. We address these aspects in the following. 

The relation between halo mass (or clustering strength) and Ly$\alpha$ luminosity (Table~\ref{table:subsamples_hod} and Fig.~\ref{fig:bias-lum}) demonstrates that high-luminosity LAEs tend to reside in higher density environments than lower luminosity ones. As a result, overdense regions contain a larger fraction of high-luminosity sources (and a lower fraction of less luminous ones) than environments of lower density. These inferences affect the Ly$\alpha$ LF measurements at $3<z<6$. While we expect a shallower faint-end slope of the Ly$\alpha$ LF in overdense regions, the slope should steepen in average or low density environments. As a consequence, surveys for relatively high-luminosity ($L_{\rm{Ly}\alpha}\approx10^{42}$ erg s$^{-1}$) LAEs are implicitly biased against the lowest density regions and thus gives a biased shape for the LF, which should not be extrapolated towards lower Ly$\alpha$ luminosities. 

Assuming that our $L_{\rm{Ly}\alpha}-M_h$ relation still holds at higher redshifts, the Ly$\alpha$ LF at $z\geq6$ would be even more affected, not only because of the above discussion but also because higher redshift bins are mainly populated by high-luminosity sources, contrary to lower redshift bins (typical case for telescopes with higher sensitivity at bluer wavelengths). Thus, it is important to be careful when interpreting Ly$\alpha$ LFs, especially near the epoch of reionization (EoR), where a shallow to steep variation in the slope of the LF from higher ($z\approx7$) to lower redshifts ($z\approx5.7$) is commonly interpreted as a sign of incomplete reionization \citep{konno14,matthee15,santos16}. 

Simulations at those higher redshifts also tend to find that high-luminosity LAEs are more likely to be observed  than low-luminosity ones because they are able to ionize their surroundings and form \ion{H}{ii} regions around them \citep[i.e., ionized bubbles; ][]{matthee15,hutter15,yoshioka}. These allow Ly$\alpha$ photons to redshift out of the resonance wavelength and escape the region. Lower luminosity LAEs are then observed if they reside within the ionized bubbles of higher luminosity LAEs or if they are able to transmit enough flux through the IGM \citep{matthee15}. If our $L_{\rm{Ly}\alpha}-M_h$ relation is still valid at these redshifts, our results would support this simulation paradigm since high-luminosity LAEs (situated in overdense regions) could form large ionized bubbles more efficiently than low-luminosity sources which tend to be located in lower density environments \citep{tilvi20}. 

Theoretical studies \cite[e.g.,][]{furlanetto,mcquinn} have modeled the size distribution of these \ion{H}{ii} regions and predicted an increase in the apparent clustering signal of LAEs towards the epoch of reionization (i.e., towards a more neutral IGM). Large ionized bubbles become rarer as the ionizing fraction declines. This patchy distribution of \ion{H}{ii} regions, which mostly surrounds large galaxy overdensities, boosts the apparent clustering of LAEs. This is commonly interpreted as another sign of incomplete reionization \citep[e.g.,][]{matthee15,hutter15}. Comparisons between observed intrinsic LAE clustering and model predictions have  therefore been used to infer the fraction of neutral hydrogen at the EoR \cite[e.g.,][]{ouchi17}. Nevertheless, if the clustering dependence on Ly$\alpha$ luminosity continues to $z\approx6$, this comparison should be performed with caution. Because the observed high redshift bins ($z\geq6$) mainly contain high-luminosity LAEs, a strong clustering signal at $z\approx6$ may be wrongly interpreted as incomplete reionization when, in fact, it may only reflect the natural relation between Ly$\alpha$ luminosity and clustering strength.

We speculate that our results also play a role in the amount of escaping Ly$\alpha$ photons (Ly$\alpha$ $f_{\rm{esc}}$). \cite{durkalec18} observed a dependence between halo mass and absolute UV magnitude ($M_{\rm{UV}}$). The interpretation of their relation goes as follows: $M_{\rm{UV}}$ traces star formation rate (SFR; e.g., \citealt{walter}), which, in turn, tracks stellar mass ($M_*$; e.g., \citealt{salmon}), which correlates with halo mass (e.g., \citealt{moster}). Because we observe a similar relation of $M_h$ with $L_{\rm{Ly}\alpha}$, $L_{\rm{Ly}\alpha}$ is presumably also a tracer of star formation. If this is correct, the object-to-object variations in Ly$\alpha$ escape fraction cannot be so large that they obscure the trend of SFR --$\;M_*\;$--$\;M_h$. Given the typical Ly$\alpha$ luminosities of our sample, this is in agreement with the model suggestions of
\cite{schaerer11a,garel15}, where the Ly$\alpha$ $f_{\rm{esc}}$ is of the order of unity for sources with $\rm{SFR} \approx 1\; M_{\odot}\; \rm{yr}^{-1}$. The Ly$\alpha$ luminosity would then be a good tracer of the SFR for less luminous LAEs. 

\section{Conclusions}
\label{sec:conclusions}


We report a strong clustering dependence on Ly$\alpha$ luminosity from the clustering measurements of three MUSE Ly$\alpha$ emitting galaxy (LAE) samples at $3 < z < 6$. Following the pencil-beam design of MUSE surveys from spatially large and shallow observation to spatially small and deep observation, we use 1030 LAEs from the full MUSE-Wide survey (1 h exposure time), 679 LAEs from MUSE-Deep (10 h), and 367 LAEs from MXDF (140 h). We thus  connect the clustering properties of $L^{\star}$ LAEs with those of much fainter ones in the MXDF. We applied an optimized version of the K-estimator as the clustering statistic, coupled to state-of-the-art halo occupation distribution (HOD) modeling. 

From our full HOD analysis, we derive constraints on the HOD of high-luminosity ($\log (L_{\rm{Ly}\alpha}/\rm{erg\;s}^{-1})\approx42.34$), intermediate ($\log (L_{\rm{Ly}\alpha}/\rm{erg\;s}^{-1})\approx41.64$) and low-luminosity ($\log (L_{\rm{Ly}\alpha}/\rm{erg\;s}^{-1})\approx41.22$) LAEs. We modeled the LAE HOD with three parameters: the threshold dark matter halo (DMH) mass for hosting a central LAE ($M_{\rm{min}}$), for hosting (on average) one satellite LAE ($M_1$), and the power-law slope of the number of satellites per halo ($\alpha$) as a function of halo mass. For the high-luminosity sample we derived a typical DMH mass of $\log (M_h/[h^{-1}M_\odot])=11.09^{+0.10}_{-0.09}$, corresponding to a bias factor of $b=2.65^{+0.13}_{-0.11}$. These findings, although more accurate, are in agreement with the results based on the two-halo term only HOD modeling performed in \cite{yohana} for a subset of our MUSE-Wide sample. For the lower luminosity samples we found  lower DMH masses. While for the $\log (L_{\rm{Ly}\alpha}/\rm{erg\;s}^{-1})\approx41.64$ dataset we inferred $\log (M_h/[h^{-1}M_\odot])=10.89^{+0.09}_{-0.09}$ ($b=2.42^{+0.10}_{-0.09}$), for the low-luminosity LAE sample we computed $\log (M_h/[h^{-1}M_\odot])=10.77^{+0.13}_{-0.15}$ ($b=2.43^{+0.15}_{-0.15}$). 

We also derived threshold DMH masses for centrals and satellites for each sample. We found that the minimum DMH mass to host a central LAE is $\log (M_{\rm{min}}/[h^{-1}M_\odot])=10.3^{+0.2}_{-0.3}, \;10.5^{+0.2}_{-0.1},\;10.7^{+0.2}_{-0.3}$  for low-, intermediate-, and high-luminosity LAEs, respectively. The threshold halo mass for satellites and the power-law slope of the number of satellite LAEs also increase with Ly$\alpha$ luminosity, from $\log (M_1/[h^{-1}M_\odot])=11.7^{+0.3}_{-0.2}$ and $\alpha=1.5\pm0.5$ to $\log (M_1/[h^{-1}M_\odot])=12.4^{+0.3}_{-0.2}$ and $\alpha=3.0^{+0.4}_{-0.5}$ and to $\log (M_1/[h^{-1}M_\odot])=12.4^{+0.4}_{-0.6}$ and $\alpha=2.8^{+0.9}_{-0.7}$. 
These HOD constraints imply a decreasing number of detected satellite LAEs with luminosity. Indeed we infer satellite fractions of $f_{\rm{sat}}\lesssim10,20$\% (at 3$\sigma$ confidence level) for high- and low-luminosity LAEs, respectively. This suggests that the most common scenario for current MUSE surveys is that in which DMHs mainly host a single detected LAE.

Motivated by these results, we aimed to further explore the clustering dependence on Ly$\alpha$ luminosity. Exploiting the large dynamic range of $L_{\rm{Ly}\alpha}$ from MXDF, we split the main LAE sample at its median $L_{\rm{Ly}\alpha}$. We found a $3.9\sigma$ difference between the clustering of the low-luminosity ($\log (L_{\rm{Ly}\alpha}/\rm{erg\;s}^{-1})\approx40.97$, $b_{\rm{low}}=1.79^{+0.08}_{-0.06}$) and the high-luminosity subset ($\log (L_{\rm{Ly}\alpha}/\rm{erg\;s}^{-1})\approx41.54$, $b_{\rm{high}}=3.10^{+0.24}_{-0.22}$). We then selected the highest luminosity LAE subset from the MUSE-Wide survey ($\log (L_{\rm{Ly}\alpha}/\rm{erg\;s}^{-1})\approx42.53$) and the lowest luminosity LAE subsample from MXDF ($\log (L_{\rm{Ly}\alpha}/\rm{erg\;s}^{-1})\approx40.97$), resulting in a clear clustering dependence where the high-luminosity LAEs from MUSE-Wide cluster more strongly ($b_{\rm{high}}=3.13^{+0.08}_{-0.15}$ or $\log(M_{\rm{h}} / [h^{-1}\rm{M}_{\odot}])=11.43^{+0.04}_{-0.10}$) than the low-luminosity ones from MXDF ($b_{\rm{low}}=1.79^{+0.08}_{-0.06}$  or $\log(M_{\rm{h}} / [h^{-1}\rm{M}_{\odot}])=10.00^{+0.12}_{-0.09}$) at  $8\sigma$ significance, excluding cosmic variance effects. The ongoing Hobby-Eberly Telescope Dark Energy Experiment (HETDEX; \citealt{hetdex2}) survey will complement these results at the high-luminosity end and at somewhat lower redshifts ($1.9 < z < 3.5$).

The implications of this framework are however not only relevant for LAE clustering studies, but also for reported measurements of evolving Ly$\alpha$ luminosity functions, detections of incomplete reionization at $z\approx6$, and the relation between Ly$\alpha$ escape fraction and halo mass. Our results are also crucial for the much debated relevance of unresolved satellite LAEs (fainter than those in MXDF) for the measured Ly$\alpha$ surface brightness profiles.

\begin{acknowledgements}
      The authors give thanks to the staff at ESO for extensive support during the visitor-mode campaigns at Paranal Observatory. We thank the eScience group at AIP for help with the functionality of the MUSE-Wide data release webpage. T.M. and H.A. thank
      for financial support by CONACyT Grant Cient\'ifica B\'asica \#252531 and by UNAM-DGAPA (PASPA, PAPIIT IN111319 and IN114423). L.W. and T.U. by the Deutsche Forschungsgemeinschaft through grant Wi 1369/32-1. M.K. acknowledges support by DLR grant 50OR1904 and DFG grant KR 3338/4-1.
      The data were obtained with the European Southern Observatory Very Large Telescope, Paranal, Chile, under Large Program 185.A-0791. This research made use of Astropy, a community-developed core Python package for Astronomy \citep{astropy}. 
\end{acknowledgements}


\begin{appendices} 

\section{Effect of different fields on the clustering measurements}
\label{appendix:k-fields}

In this work, we have analyzed the clustering of LAEs in the full MUSE-Wide sample, including the CANDELS/COSMOS fields and the HUDF parallel fields. Here, we explore the possible effects on the MUSE-Wide clustering results when including or excluding various sets of fields. In appendix A of \cite{yohana}, we showed that the HUDF parallel fields did not alter the clustering results, their exclusion or inclusion mainly affected the clustering uncertainties. We therefore explore the effect of including the CANDELS/COSMOS region by comparing the clustering of the full MUSE-Wide survey with that present in a subsample without the CANDELS/COSMOS fields. The number of LAEs in the CANDELS/COSMOS region is 250.

It is clear from Fig.~\ref{fig:with-withoutfields} that the clustering in both samples is in good agreement. The large-scales bias factors derived from the two curves are indistinguishable (within 1$\sigma$). The uncertainties corresponding to the smaller sample are (on average) 20\% larger than in the full MUSE-Wide sample. We conclude that the inclusion of these fields has no notable effect on our clustering results but helps in reducing cosmic sample variance uncertainties.

 \renewcommand\thefigure{A.\arabic{figure}}
\setcounter{figure}{0}
\begin{figure}[tb]
\centering
\includegraphics[width=\columnwidth,height=6cm]{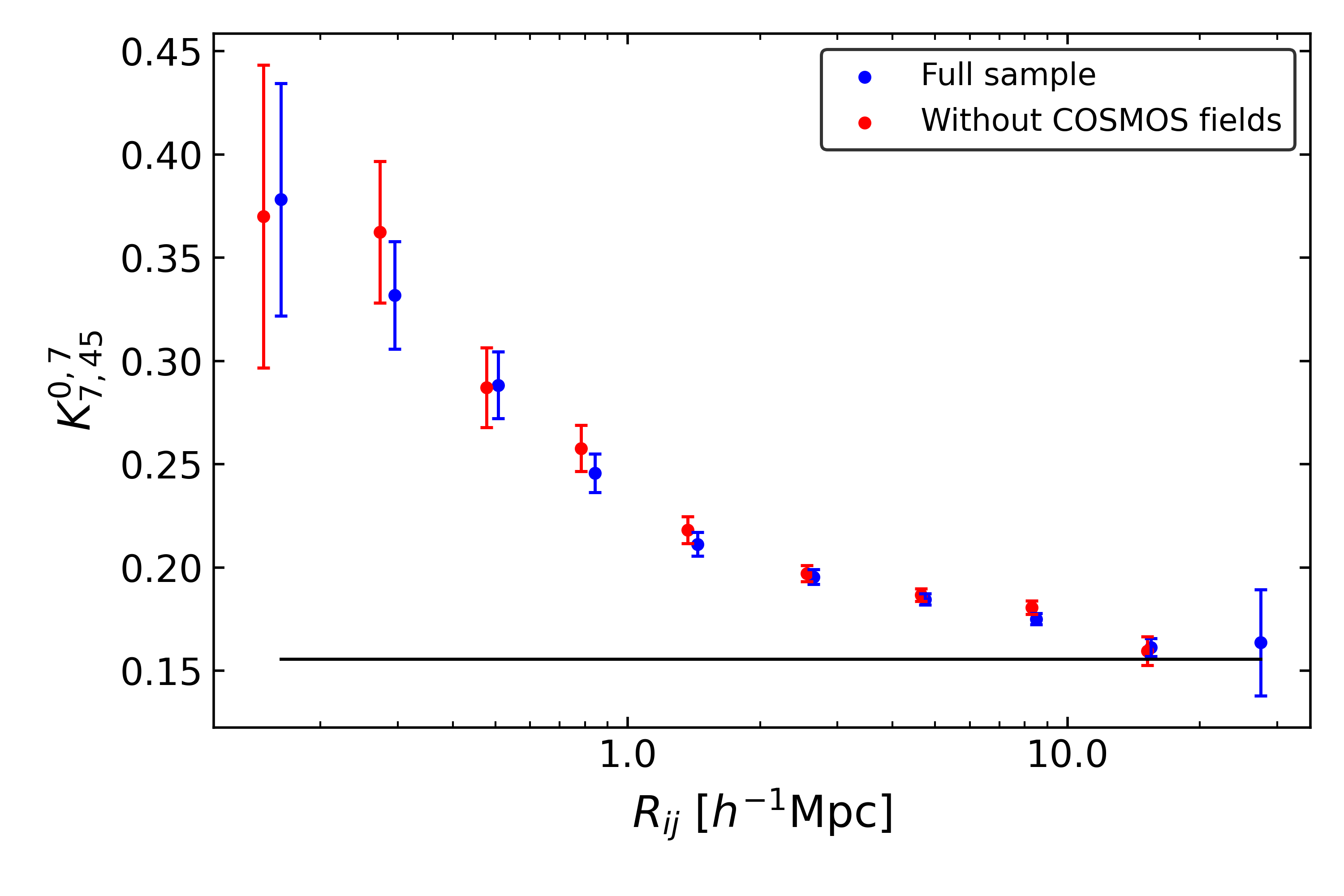}
\caption{Clustering of the LAEs in the full MUSE-Wide sample (blue, see Fig.~\ref{fig:RA-DEC}) and without the CANDELS/COSMOS fields
 (red, see right panel of Fig.~\ref{fig:RA-DEC}). The black baseline represents the expected clustering of an unclustered sample. The error bars are Poissonian. The red measurements have been shifted along the x-axis for visual purposes. }
\label{fig:with-withoutfields}
\end{figure}

\section{Covariance matrix}
\label{appendix:covariance}

A common approach to quantify the correlation of the clustering data points is to resample the set of galaxies with the jackknife technique, followed by the calculation of the covariance matrix. To apply the jackknife method, we find a compromise between the number and the size of the jackknife zones. Thus, we split the sky area into ten independent regions (see Fig.~\ref{fig:jackknife}) with a spatial extent of $\approx 4 \: h^{-1}$Mpc in both RA and Dec directions. We then construct ten different subsamples, each of them excluding one jackknife zone, and compute the K-estimator in each subset. These measurements are then used to build up the covariance matrix using Eq.~\ref{eq:covariance} (see Sect.~\ref{sec:errorsK}). 

Considering that the probability of one galaxy pair to contribute to various adjacent bins is higher than that to contribute to several distant bins, one would naively expect a higher correlation in the former
 case. This is indeed what the (normalized) covariance matrix reflects in the left panel of Fig.~\ref{fig:covariance}. In fact, the noise in the matrix elements corresponding to notably separate bins is substantial. In the right panel of Fig.~\ref{fig:covariance}, we plot the normalized matrix elements as a function of bin $i$ for each bin $j$ to better illustrate the high level of noise in the matrix, especially for bins $i>6$, where most curves become negative. This is likely due to the limited spatial size of the survey, which does not allow neither for a higher number of jackknife zones nor for spatially larger zones.

As a result of the considerable noise in the matrix on account of barely correlated bins significantly apart from each other, the minimization of the $\chi^2$ values (Eq.~\ref{eq:chi2}) including the full covariance fails 
(i.e., various $\chi^2$ values become negative). We therefore limit the use of the covariance matrix to its main diagonal and two adjacent diagonals (see red section in the left panel of Fig.~\ref{fig:covariance}; our so-called reduced covariance matrix). This means we set the negative part of the curves in the right panel of Fig.~\ref{fig:covariance} to zero (i.e., no correlation between those bins), in an attempt to smooth out the noise. 
\renewcommand\thefigure{B.\arabic{figure}}
\setcounter{figure}{0}
\begin{figure*}
\centering
\begin{tabular}{c c}
  \centering
  \includegraphics[width=.46\linewidth]{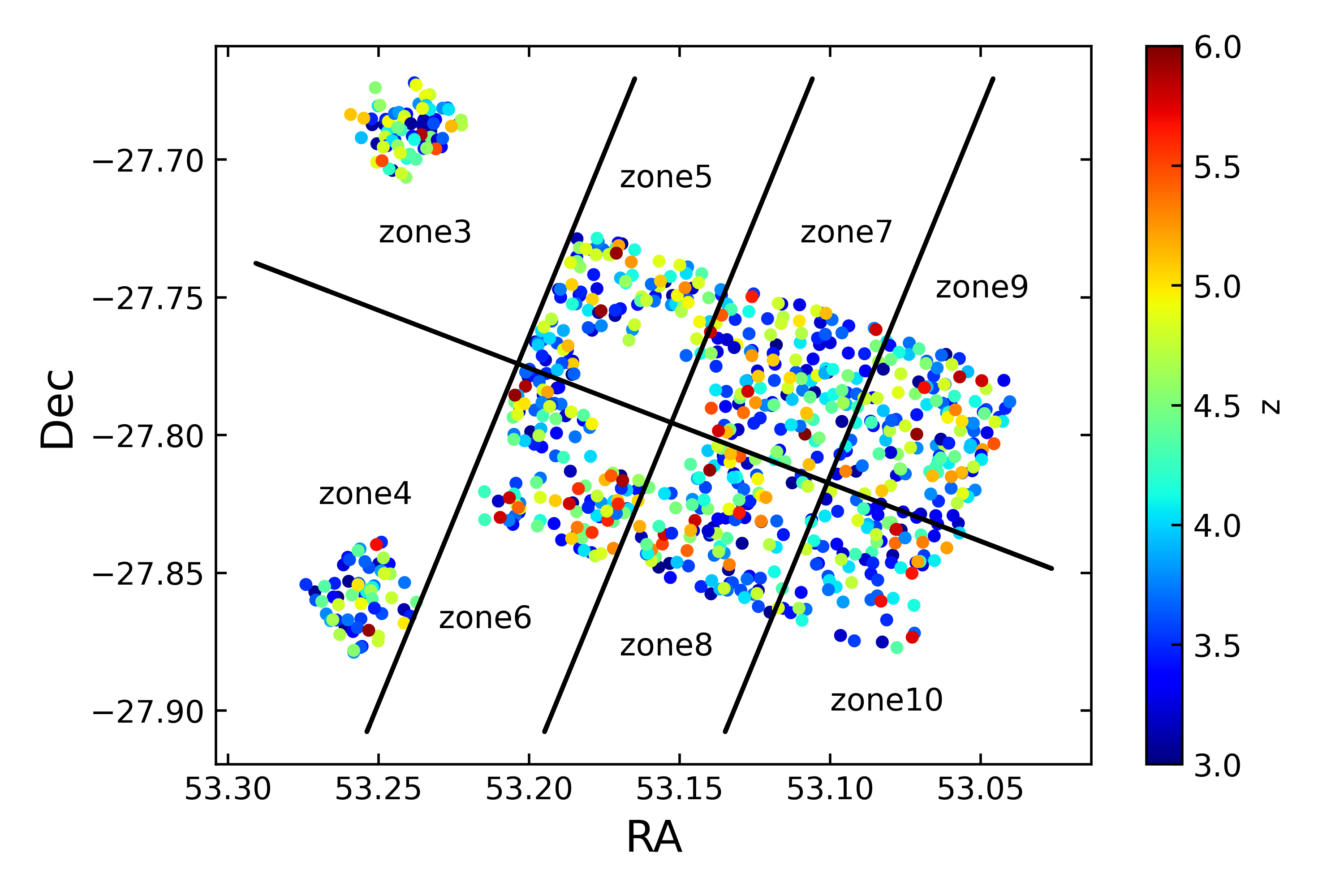}
\end{tabular}
\begin{tabular}{c c}
  \centering
  \includegraphics[width=.48\linewidth]{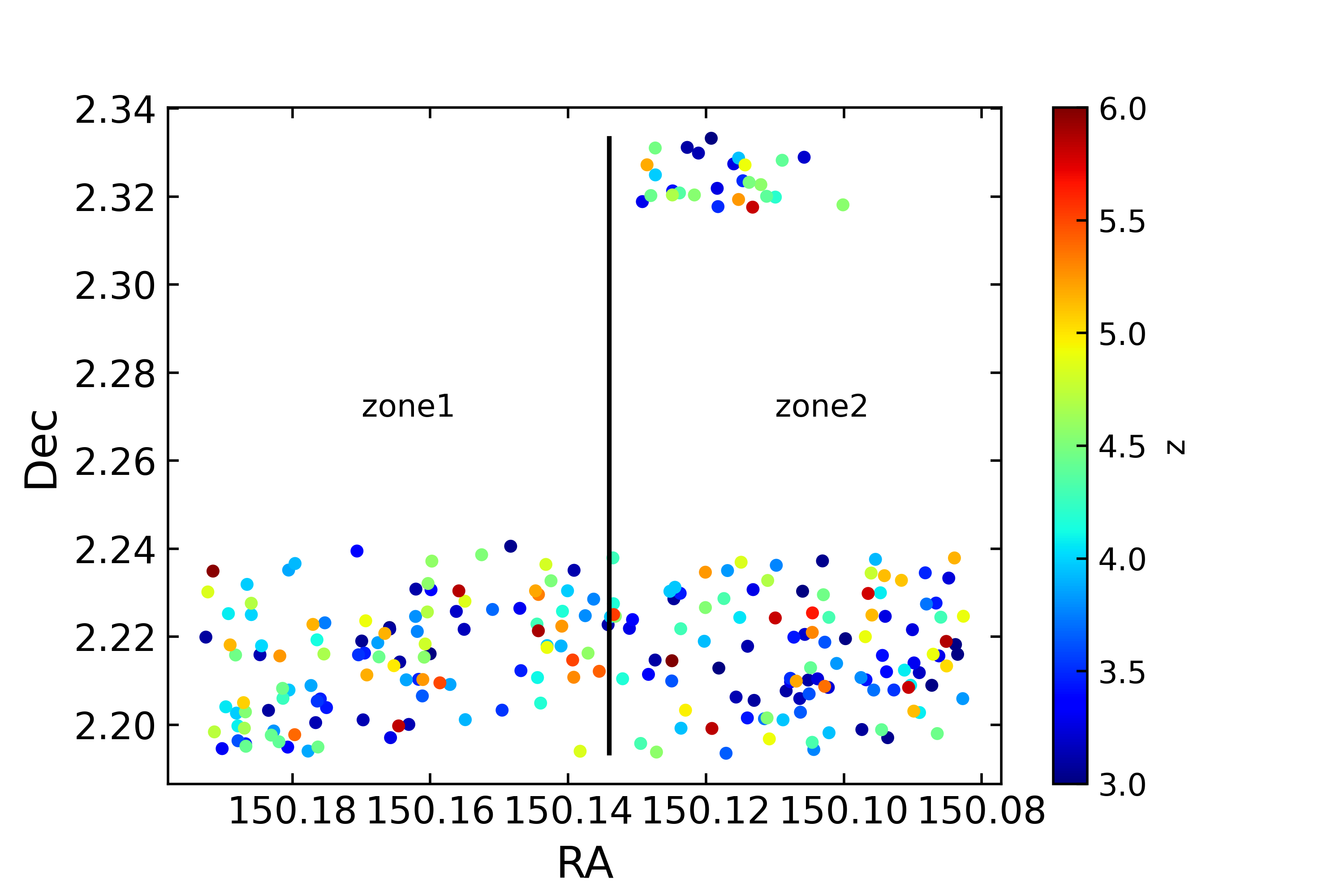}
\end{tabular}
\caption{Ten Jackknife zones in the spatial coverage of the full MUSE-Wide survey (83.52 arcmin$^2$). Each Jackknife zone has a spatial extent of $\approx4\; h^{-1}$Mpc in both RA and Dec directions. }
\label{fig:jackknife}
\end{figure*}
\begin{figure*}
\centering
\begin{tabular}{c c }
  \centering
  \includegraphics[width=.45\linewidth]{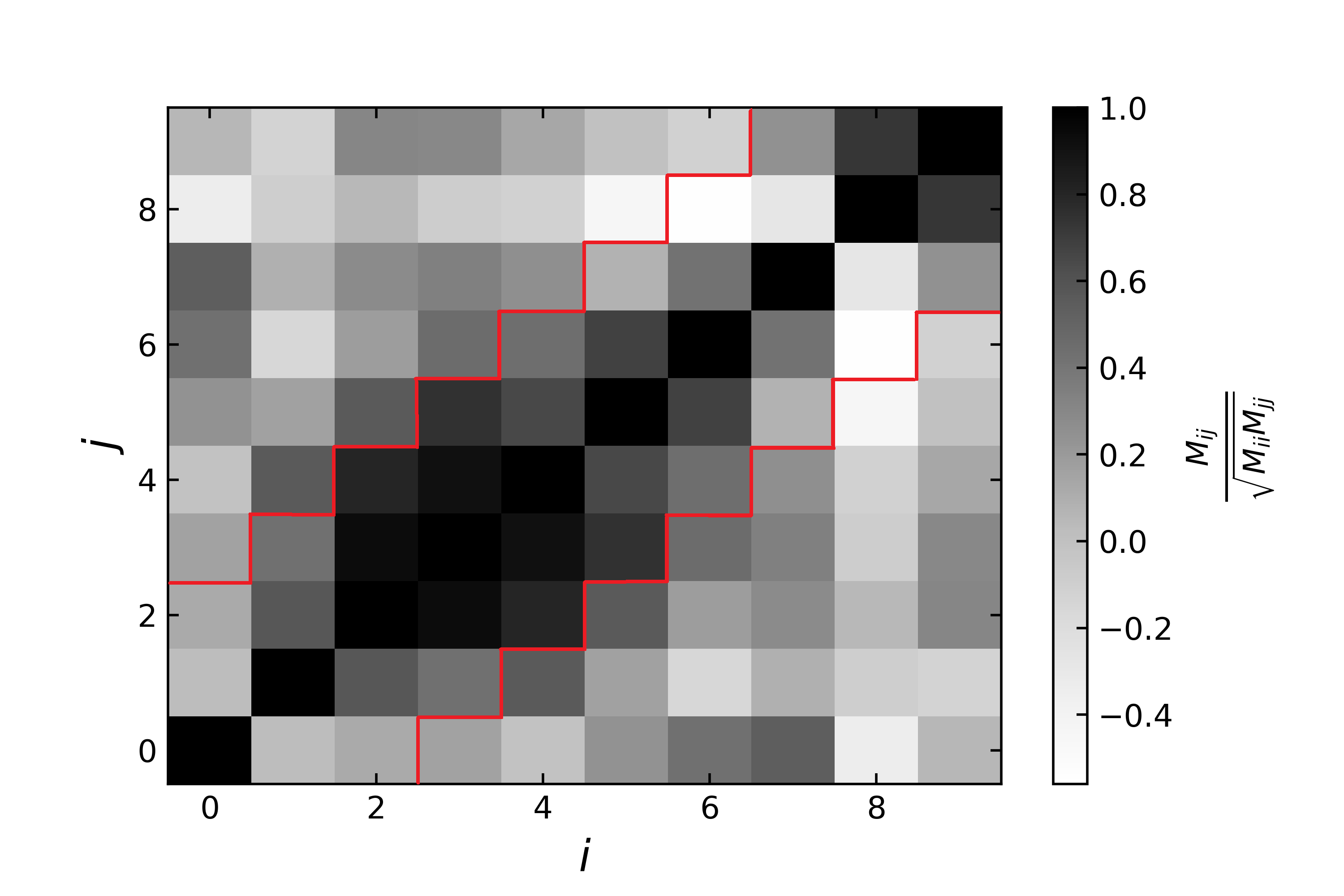}
\end{tabular}%
\begin{tabular}{c c }
  \centering
  \includegraphics[width=.45\linewidth]{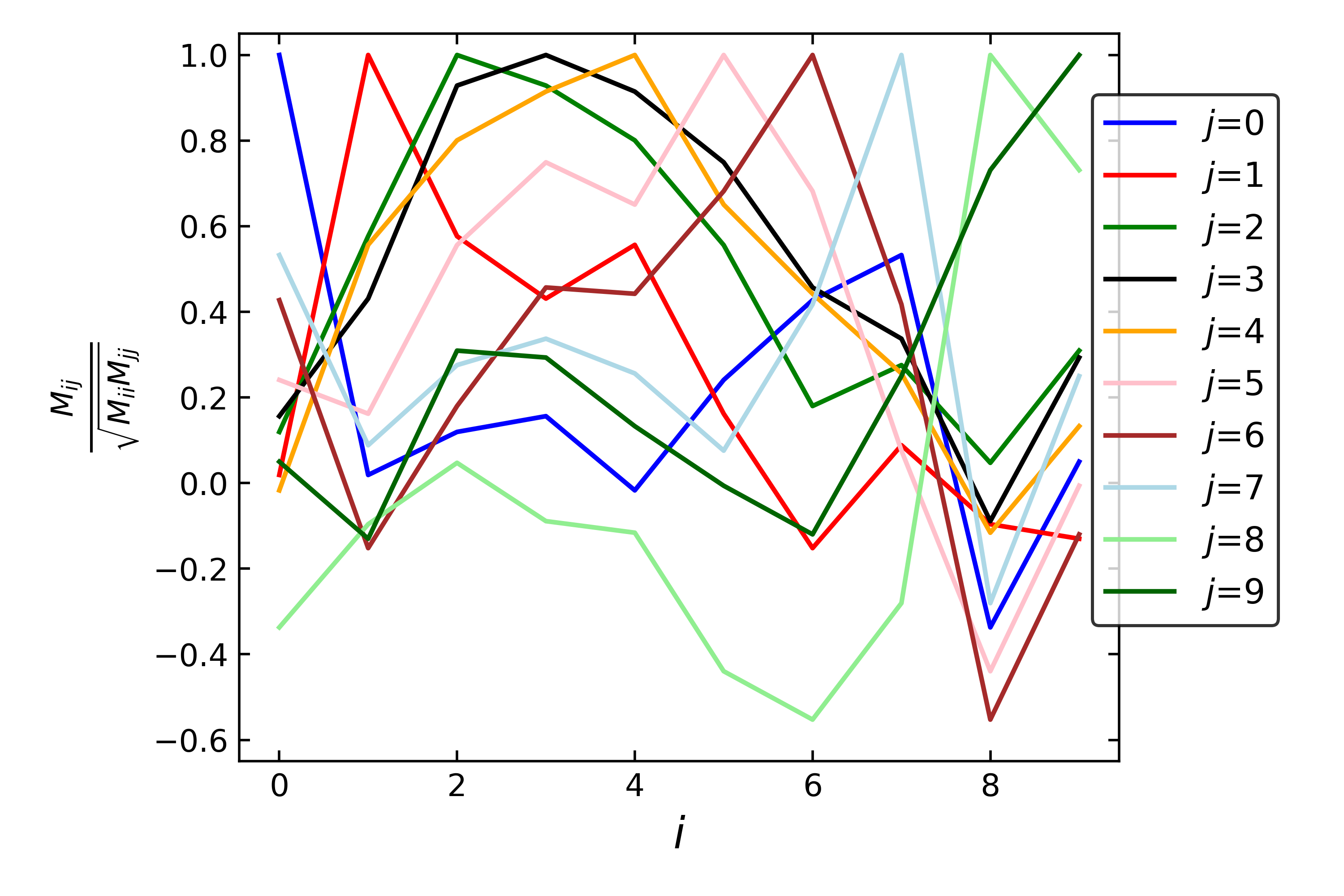}
\end{tabular}
\caption{Covariance matrix computed from ten independent K-estimator measurements from the jackknife resampling technique. Left: Normalized covariance matrix for bins $i$ and $j$. The red region defines the main diagonal and the two adjacent diagonals used for our reduced covariance matrix. Right: Normalized covariance matrix elements as a function of bin $i$ for each bin $j$ (colored). }
\label{fig:covariance}
\end{figure*}
While incorporating more diagonals results mathematically problematic for the $\chi^2$ minimization, we have verified that the number of adjacent diagonals (one or two) slightly modifies
the $\chi^2$ values but the probability contours represented in Fig.~\ref{fig:contours} remain unaltered. Thus, so do the best-fit HOD parameters.

\renewcommand\thefigure{C.\arabic{figure}}
\setcounter{figure}{0}
\noindent
\begin{minipage}{\linewidth}
  \strut\newline
  \centering
 \includegraphics[width=0.9\columnwidth,height=6cm]{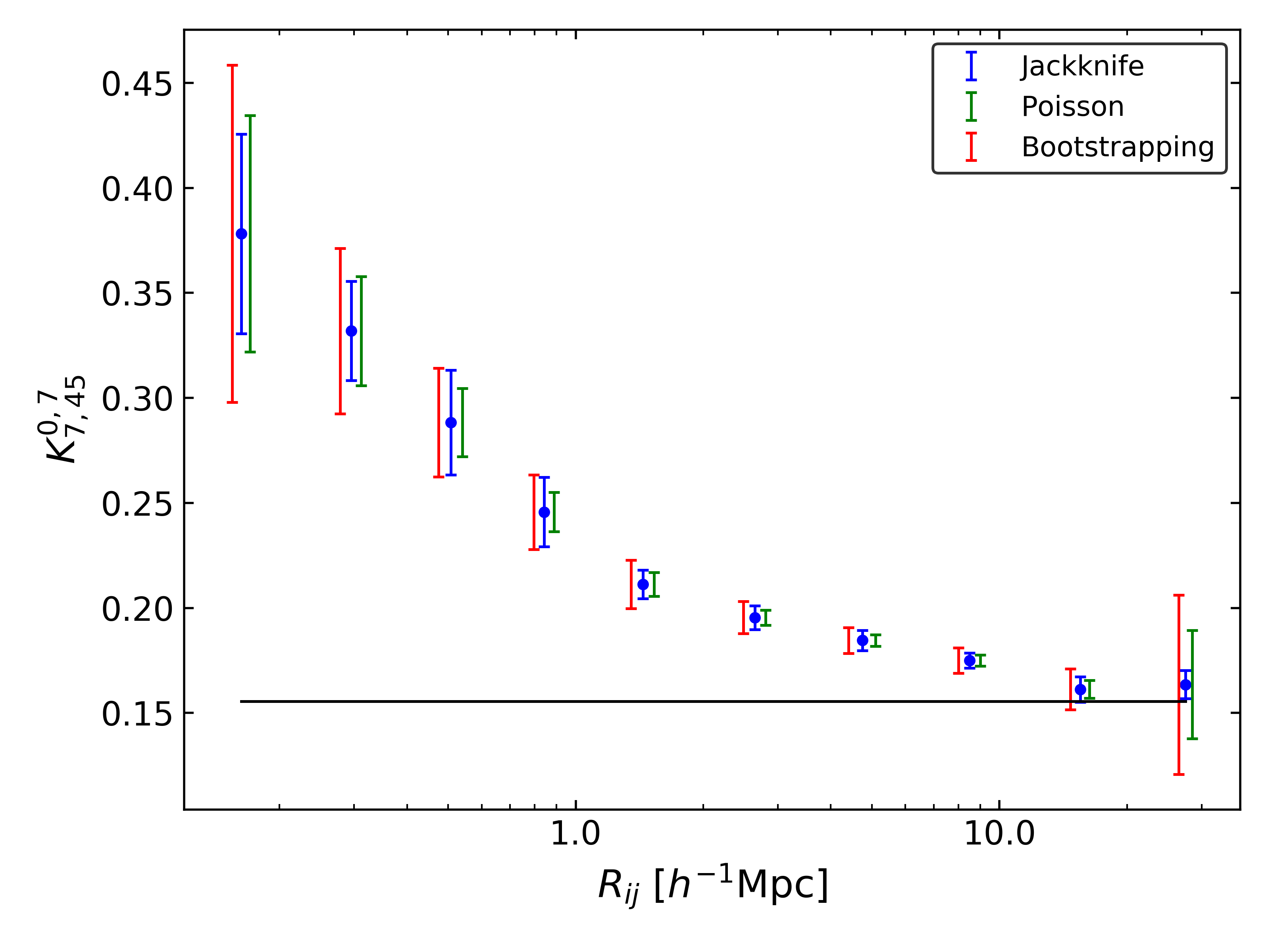}
  \captionof{figure}{Error estimation method comparison for the sample of LAEs in the MUSE-Wide survey. Uncertainties from the covariance matrix and the jackknife resampling technique described in Sect.~\ref{sec:errorsK} are colored in blue, those from the bootstrapping approach used in \cite{yohana} in red, and Poisson uncertainties in green.}\label{fig:errors}
\end{minipage}

Despite current limitations, jackknife is still the most robust method to compute the K-estimator uncertainties. While galaxy bootstrapping or Poisson error bars do not account for bin to bin correlations, our reduced covariance matrix only neglects the correlation between bins remarkably separated (expected to be minimal), but accounts for the correlation between nearby bins.

\section{Error estimation comparison}
\label{appendix:err_comparison}

In order to quantify the correlation between the K-estimator bins, the covariance matrix must be computed. By splitting the sky area into independent regions, following the jackknife resampling technique, we create as many subsamples from the MUSE-Wide sample as jackknife zones (see Sect.~\ref{sec:errorsK}). The K-estimator is then computed  in each subset and the measurements are used to quantify the covariance matrix, whose diagonal provides the variance of each clustering data point. The square root of the diagonal represents the 1$\sigma$ uncertainties and are represented in blue in Fig.~\ref{fig:errors} (same along the main paper).

\renewcommand\thefigure{D.\arabic{figure}}
\setcounter{figure}{0}
\begin{figure*}[h]
\centering
\begin{tabular}{c c c}
  \centering
  \includegraphics[width=.3\linewidth]{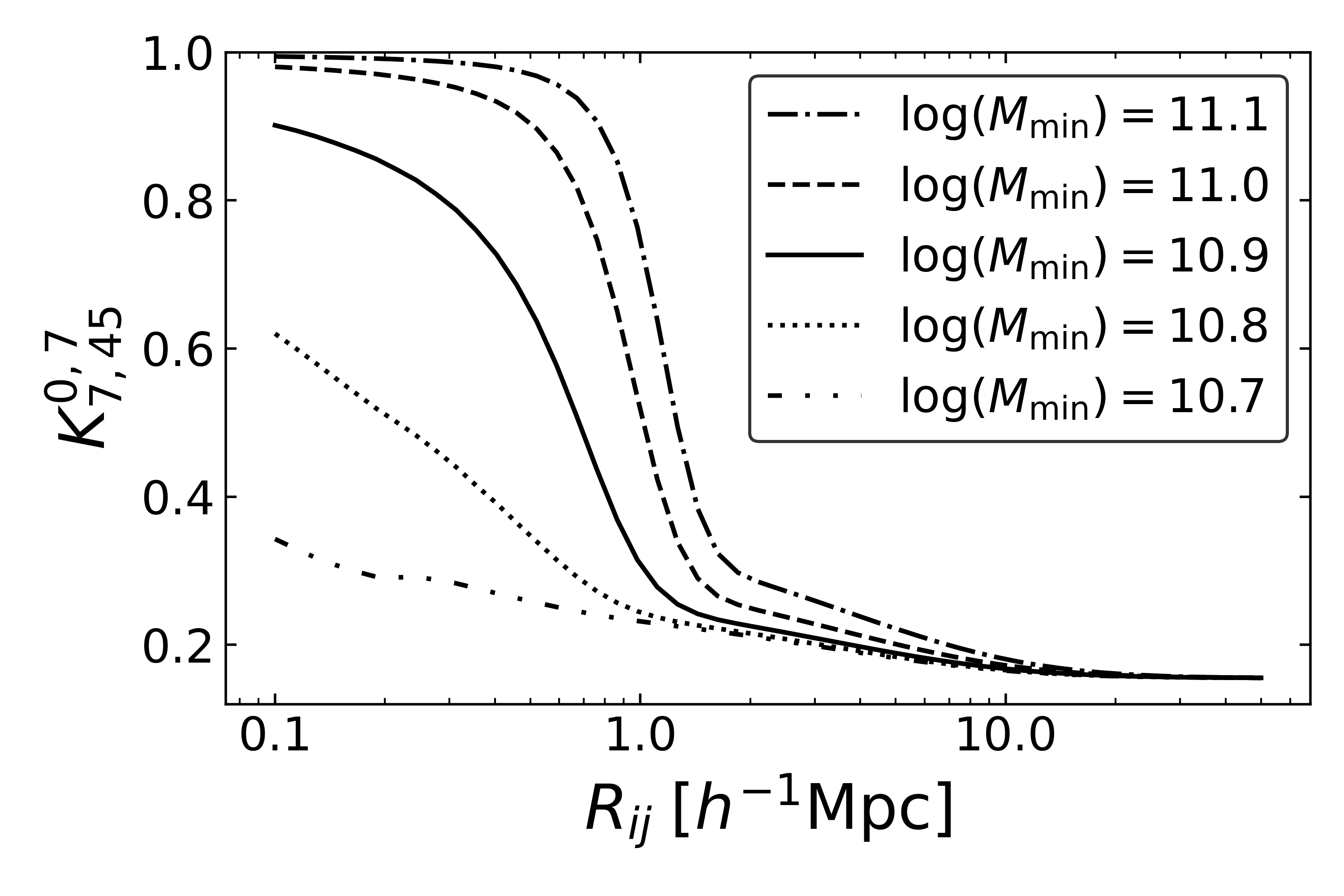}
\end{tabular}%
\begin{tabular}{c c c}
  \centering
  \includegraphics[width=.3\linewidth]{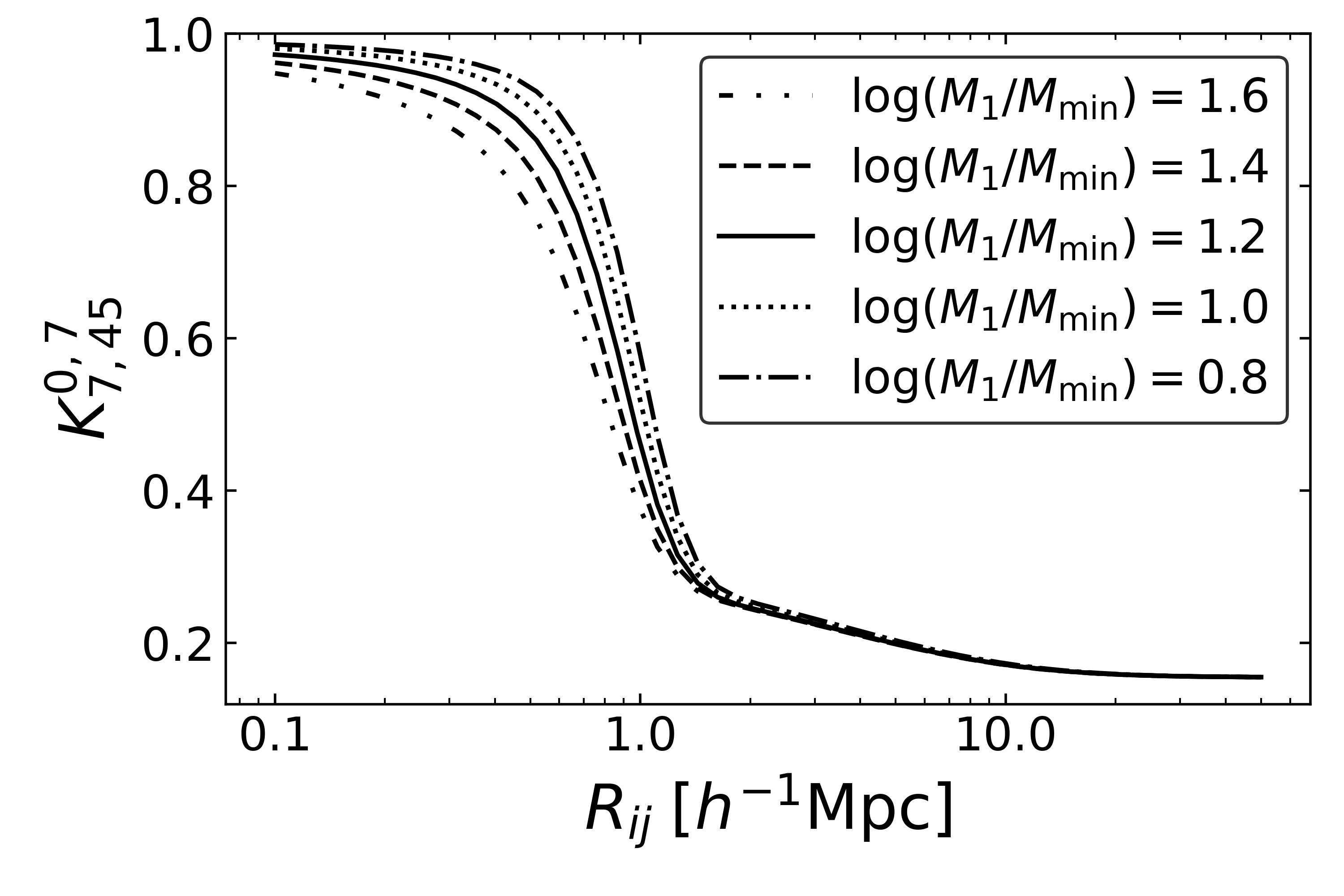}
\end{tabular}
\begin{tabular}{c c c}
  \centering
  \includegraphics[width=.3\linewidth]{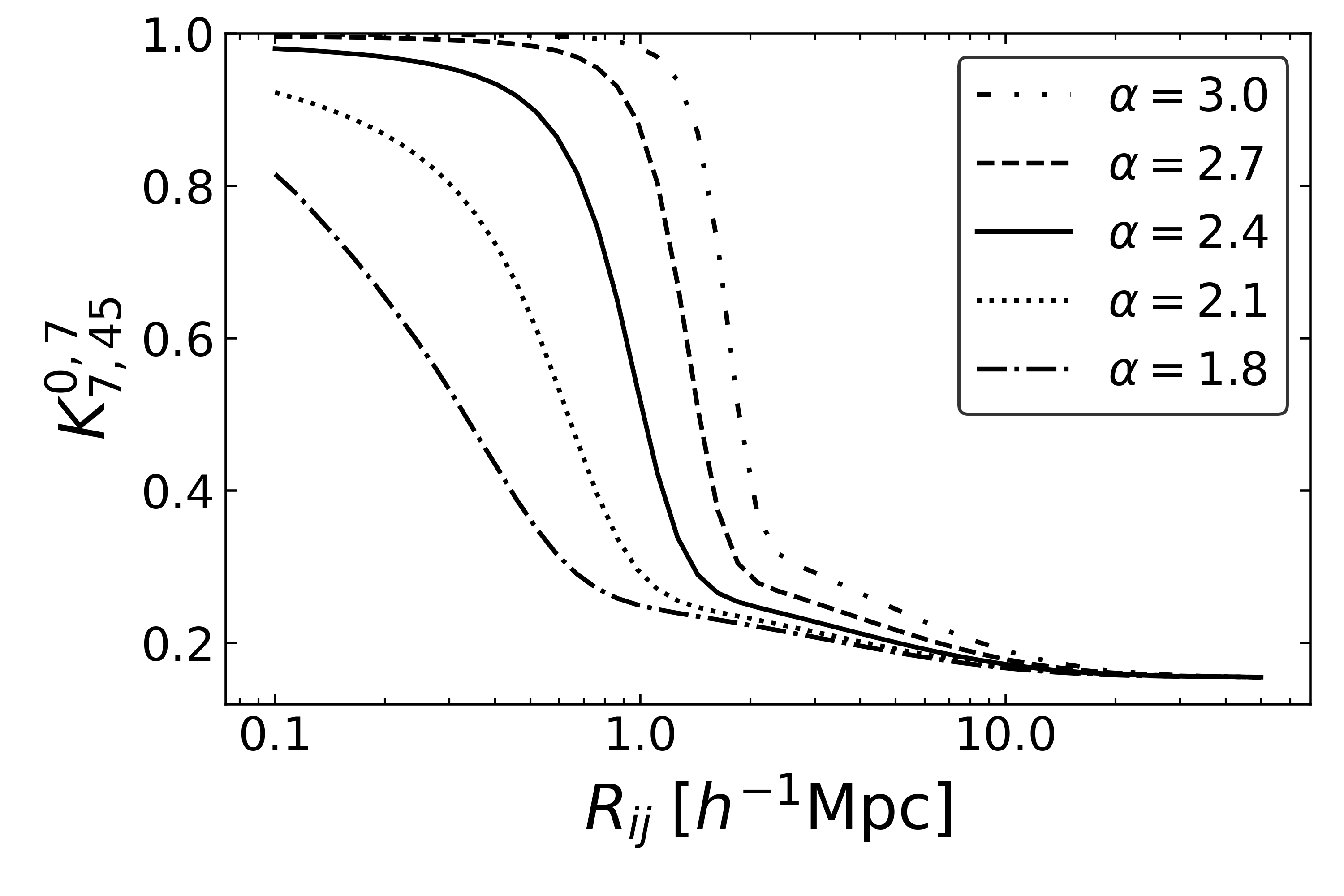}
\end{tabular}
\caption{Effect of HOD parameters on the shape of the K-estimator. Left: Dependence on $\log (M_{\rm{min}})$ for fixed $\log (M_1/M_{\rm{min}})=1.2$ and $\alpha=2.4$. Middle: Dependence on $\log( M_1/M_{\rm{min}})$ for fixed $\log (M_{\rm{min}}/[h^{-1}M_\odot])=10.9$ and $\alpha=2.4$. Right: Dependence on $\alpha$ for fixed $\log( M_{\rm{min}}/[h^{-1}M_\odot])=10.9$ and $\log( M_1/M_{\rm{min}})=1.2$. }
\label{fig:k-parameters}
\end{figure*}

The jackknife resampling method requires a division of the sky area into several independent regions, each of which should ideally be  large enough to cover the full range of scales under 
consideration. Out of the three samples examined in this study, this can only be partially achieved in the MUSE-Wide dataset. MUSE-Deep and MXDF do not allow for a spatial split  into independent zones. We are thus left with two options for the deeper samples: the bootstrapping technique applied in \cite{yohana}, shown in red in Fig.~\ref{fig:errors}, and Poisson uncertainties, shown in green.

We find that Poisson (bootstrapping) errors are, on average, 7\% (46\%) larger than those computed with the jackknife technique. These findings corroborate the results from \cite{norberg09}, who found that the bootstrapping approach overestimates the uncertainties.

Similarly as for the MUSE-Wide survey, we find that bootstrapping uncertainties are $\approx40$\% (on average) larger than Poisson in both MUSE-Deep and MXDF. We thus decide to use Poisson errors for the deeper samples in an attempt to least overvalue the uncertainties.

We verified that the error estimation method does not significantly affect our clustering results. The best-fit parameters from MUSE-Deep and MXDF using bootstrapping error bars and the $\chi^2$ minimization described in Sect.~3.1.3 of \cite{yohana} are consistent with those delivered from Poisson statistics. Although in agreement, bootstrapping delivers $\approx45$\% larger uncertainties than Poisson for the best-fit HOD parameters.

We last perform the same experiment in MUSE-Deep and MXDF but using scaled Poisson error bars. We decreased the Poisson in 7\% (excess found in MUSE-Wide) and find that the best-fit parameters are $\approx10$\% less uncertain than if  Poisson errors are directly applied. 


\section{Dependence of HOD parameters on the shape of the K-estimator}
\label{appendix:HODparameters}

Here we visualize and  qualitatively describe the effect of the HOD parameters on the K-estimator.  Figure~\ref{fig:k-parameters} shows the K-estimator for numerous HOD models. Each panel represents the result of varying one HOD parameter with the other two parameters fixed. Before detailing the major effects, it should be pointed out that the exact change in the shape of the K-estimator does not only depend on the varied parameter but also on the specific choice of the other two. Hence, these panels should merely be seen as illustrative examples.

The left panel of Fig.~\ref{fig:k-parameters} shows the dependence of the K-estimator on $ M_{\rm{min}}$. Higher values of $\log M_{\rm{min}}$ (i.e., more massive halos) raise the expected K-estimator at all $R_{ij}$ scales (one- and two- halo terms). At large scales, this occurs because more massive halos present larger bias factors, whereas at small scales, this is due to the decline in the contribution from less massive DMHs.

The middle panel of Fig.~\ref{fig:k-parameters} shows the dependence of the K-estimator on $ M_1/M_{\rm{min}}$. Larger $\log (M_1/M_{\rm{min}})$ values (i.e., more massive halos) reduce the one-halo term clustering amplitude because of the decrease in the contribution from less massive DMHs. The clustering in the two-halo term does not depend on $M_1$.

The right panel of Fig.~\ref{fig:k-parameters} shows the dependence of the K-estimator on $ \alpha$. Higher values of $\alpha$ increase the fraction of galaxies in massive DMHs with respect to smaller mass DMHs. Given that more massive halos are more strongly biased, the amplitude of the two-halo term increases. The change observed in the one-halo term is explained because galaxies hosted by massive DMHs can contribute to the one-halo term on its largest scales, while galaxies residing in less massive halos can only contribute to the one-halo term at smaller $R_{ij}$ scales. Since $\alpha$ modifies the fraction of galaxies in massive DMHs to less mass DMHs, the corresponding fraction of the clustering contribution also varies. This alters the slope of the one-halo term.

\end{appendices}


\begin{thebibliography}{}

    \bibitem[Adelberger et al.(2005)]{adelberger} Adelberger, K. L., Steidel, C. C., Pettini, M., Shapley, A. E., Reddy, N. A., \& Erb, D. K. 2005, ApJ, 619, 697-713

    
    \bibitem[Astropy Collaboration et al.(2013)]{astropy} Adelberger, Astropy Collaboration, Robitaille, T. P., Tollerud, E. J., et al. 2013, A\&A, 558, A33
    
    \bibitem[Bacon et al.(2017)]{bacon17} 
    Bacon, R., Conseil, D., Mary, D., et al. 2017, A\&A, 608, A1
    

    \bibitem[Bacon et al.(2022)]{bacon22} 
    Bacon, R., Brinchmann, J., Conseil, S., et al. 2022, arXiv e-prints, arXiv:2211.08493
    
    \bibitem[Bielby et al.(2016)]{bielby} 
    Bielby, R. M., Tummuangpak, P., Shanks, T., et al. 2016, MNRAS, 456, 4061
    
    \bibitem[Davis \& Peebles(1983)]{davispeebles} Davis, M. \& Peebles, P. J. E. 1983, ApJ, 267, 465
    
    \bibitem[Diener et al.(2017)]{catrina} 
    Diener, C., Wisotzki, L., Schmidt, K. B., et al. 2017, MNRAS, 471, 3186-3192
    
    \bibitem[Durkalec et al.(2014)]{durkalec14} 
    Durkalec, A., Le F\`evre, O., Pollo, A., et al. 2014, A\&A, 583, A128
    
    \bibitem[Durkalec et al.(2018)]{durkalec18} 
    Durkalec, A., Le F\`evre, O., Pollo, A., et al. 2018, A\&A, 612, A42
    
    \bibitem[Furlanetto et al.(2006)]{furlanetto} 
     Furlanetto, S. R., Zaldarriaga, M.,  Hernquist, L., et al. 2006, MNRAS, 365, 1012
     
     \bibitem[Garel et al.(2015)]{garel15} 
    Garel, T., Blaizot, J., Guiderdoni, B., et al. 2015, MNRAS, 450, 1279
   
    \bibitem[Gawiser et al.(2007)]{gawiser07} 
    Gawiser, E., Francke, H., Lai, K., et al. 2007, ApJ, 671, 278

    \bibitem[Gebhardt et al.(2021)]{hetdex2} 
    Gebhardt, K., Cooper, E. M., Ciardullo, R., et al. 2021, ApJ, 923, 217

     
     \bibitem[Hatfield et al.(2018)]{hatfield}
     Hatfield, P. W., Bowler, R. A. A., Jarvis, M. J., et al. 2018, MNRAS, 477, 3760
    
    \bibitem[Herenz et al.(2017)]{herenz17} 
     Herenz, E. C., Urrutia, T., Wisotzki, L., et al. 2017, A\&A, 606, A12

     \bibitem[Herenz et al.(2019)]{herenz19} 
     Herenz, E. C., Wisotzki, L., Saust, R., et al. 2019, A\&A, 621, A107
     
     \bibitem[Harikane et al.(2018)]{harikane18} 
     Harikane, Y., Ouchi, M., Ono, Y., et al. 2018, Publications of the Astronomical Society of Japan, 70, S11
     
     \bibitem[Herrero Alonso et al.(2021)]{yohana}
     Herrero Alonso, Y., Krumpe, M., Wisotzki, L., et al. 2021, A\&A, 653, A136
    
    \bibitem[Hinshaw et al.(2013)]{constants} 
     Hinshaw, G., Larson, D., Komatsu, E., et al. 2013, AJSS, 208, 19
     
     \bibitem[Hinton et al.(2016)]{hinton} 
     Hinton, S. R., Davis, T. M., Lidman, C., et al. 2016, Astronomy and Computing, 15, 61

    \bibitem[Hu et al.(1998)]{hu98} 
     Hu, E. M., Cowie, L. L., McMahon, \& R. G. 1998, ApJ, 502, L99
     
     \bibitem[Hutter et al.(2015)]{hutter15} 
     Hutter, A., Dayal, P., \& M{\"u}ller, V. 2015, mnras, 450, 4025

     \bibitem[Inami et al.(2017)]{inami} 
     Inami, H., Bacon, R., Brinchmann, J., et al. 2017, A\&A, 608, A2
     
     \bibitem[Jenkins et al.(1998)]{jenkins}
     Jenkins, A., Frenk, C. S., Pearce, F. R., et al. 1998, ApJ, 499, 20
    
    \bibitem[Kaiser(1987)]{kaiser} 
     Kaiser, N. 1987, MNRAS, 227, 1-21
     
     \bibitem[Khostovan et al.(2018)]{khostovan18} Khostovan, A. A., Sobral, D., Mobasher, B., et al. 2018, MNRAS, 478, 2999--3015
     
     \bibitem[Khostovan et al.(2019)]{khostovan19}
     Khostovan, A. A., Sobral, D., Mobasher, B., et al. 2019, MNRAS, 489, 555-573
     
     \bibitem[Konno et al.(2014)]{konno14}
     Konno, A., Ouchi, M., Ono, Y., et al. 2014, ApJ, 797, 16
     
     \bibitem[Krumpe et al.(2010)]{mirko10} 
     Krumpe, M., Miyaji, T. \& Coil, A. L. 2010, ApJ, 713, 558
     
     \bibitem[Krumpe et al.(2012)]{mirko12} 
     Krumpe, M., Miyaji, T., Coil, A. L. \& Aceves H. 2012, ApJ, 746, 1
     
     \bibitem[Krumpe et al.(2015)]{mirko15} 
     Krumpe, M., Miyaji, T. Husemann, B., et al. 2015, ApJ, 815, 21
     
     \bibitem[Krumpe et al.(2018)]{mirko18} 
     Krumpe, M., Miyaji, T. Coil, A. L. \& Aceves, H. 2018, MNRAS, 474, 1773
     
     \bibitem[Kusakabe et al.(2018)]{haruka} 
     Kusakabe, H., Shimasaku, K., Ouchi, M., et al. 2018, Publications of the Astronomical Society of Japan, 70, 4

    \bibitem[Lee et al.(2006)]{lee06} 
     Lee, K.-S., Giavalisco, M., Gnedin, O., et al. 2006, ApJ, 642, 63
    
     \bibitem[Limber(1953)]{limber}
     Limber, D. N. 1953, ApJ, 117, 134

     \bibitem[Luo et al.(2017)]{luo17} 
     Luo, B., Brandt, W. N., Xue, Y. Q., et al. 2017, ApJSS, 228, 2


     \bibitem[Malkan et al.(2017)]{malkan} 
     Madau, M. A., Cohen, D. P, Maruyama, M., et al. 2017, ApJ, 850, 5
     
     \bibitem[Mary et al.(2020)]{mary} 
     Mary, D., Bacon, R., Conseil, S., et al. 2020, A\&A, 635,A194
     
     \bibitem[Matthee et al.(2015)]{matthee15} 
     Matthee, J., Sobral, D., Santos, S., et al. 2015, MNRAS, 451, 400
     
     \bibitem[McQuinn et al.(2007)]{mcquinn} 
     McQuinn, M., Hernquist, L., Zaldarriaga, M., et al. 2007, MNRAS, 381, 75
     
     \bibitem[Miyaji et al.(2011)]{miyaji11} 
     Miyaji, T., Krumpe, M., Coil, A. \& Aceves, H. 2011, ApJ, 726, 83

    \bibitem[Moster et al.(2010)]{moster} 
     Moster, B. P., Somerville, R. S., Maulbetsch, C., et al. 2010, ApJ, 710, 903
     
     \bibitem[Navarro, Frenk \& White(1997)]{nfw97} 
     Navarro, J. F., Frenk, C. S. \& White, S.D.M 1997, ApJ, 490, 493
     
     \bibitem[Norberg et al.(2009)]{norberg09} 
     Norberg, P., Baugh, C. M., Gaztañaga, E., \& Croton, D. J. 2009, MNRAS, 396, 19
     
     \bibitem[Ouchi et al.(2003)]{ouchi03} 
     Ouchi, M., Shimasaku, K., Furusawa, H., et al. 2003, ApJ, 582, 60 
   
     \bibitem[Ouchi et al.(2010)]{ouchi10} 
     Ouchi, M., Shimasaku, K., Furusawa, H., et al. 2010, ApJ, 723, 869
 
     \bibitem[Ouchi et al.(2018)]{ouchi17} 
     Ouchi, M., Harikane, Y., Shibuya, T., et al. 2018, PASJ, 70, S13

    \bibitem[Salmon et al.(2015)]{salmon} 
     Salmon, B., Papovich, C., Finkelstein, S. L, et al. 2015, ApJ, 799, 183
     
     \bibitem[Santos et al.(2016)]{santos16}
     Santos, S., Sobral, D. \& Matthee, J. 2016, MNRAS, 463, 1678
     
     \bibitem[Schaerer et al.(2011a)]{schaerer11a}
     Schaerer, D., Hayes, M., Verhamme, A., et al. 2011a, A\&A, 531, A12
     
     \bibitem[Sheth et al.(2001)]{sheth} 
     Sheth, R., Mo, H. J. \& Tormen, G. 2001, 323, 1-12
    

    \bibitem[Spinoso et al.(2020)]{spinoso} 
    Spinoso, D., Orsi, A., L{\'o}pez-Sanjuan, C. 2020, A\&A, 643, A149

    \bibitem[Steidel et al.(1996)]{steidel96} 
    Steidel, C. C., Giavalisco, M., Pettini, M. 1996, ApJ, 462, L17

    \bibitem[Tilvi et al.(2020)]{tilvi20} 
    Tilvi, V., Malhotra, S., Rhoads, J. E., et al. 2020, ApJL, 891, L10
      
    \bibitem[Tinker et al.(2005)]{tinker} 
    Tinker, J. L., Weinberg, D. H. \& and Zheng, Z. 2005, MNRAS, 368, 85
    
    \bibitem[Tinker(2007)]{tinker07} Tinker, J.~L.\ 2007, MNRAS, 374, 477 
    

    \bibitem[Urrutia et al.(2019)]{urrutia19} 
    Urrutia, T., Wisotzki, L., Kerutt, J., et al. 2019, A\&A, 624, 24
    
    \bibitem[van den Bosch et al.(2013)]{vandenbosch13} 
    Van Den Bosch, F. C., More, S., Cacciato, M., et al. 2013, MNRAS, 430, 725

    \bibitem[Walter et al.(2012)]{walter} 
    Walter, F., Decarli, R., Carilli, C., et al. 2012, ApJ, 752, 93
     
     \bibitem[Wechsler \& Tinker(2018)]{wechsler} Wechsler, R. H. \& Tinker, J. L. 2018, Annual Review of Astronomy and Astrophysics, 56, 435
     
     \bibitem[Yoshioka et al.(2022)]{yoshioka}
     Yoshioka, T., Kashikawa, N., Inoue, A., et al. 2022, ApJ, 927, 32
     
     \bibitem[Yajima et al.(2018)]{yajima}
     Yajima, H., Sugimura, K., \& Hasegawa, K. 2018, MNRAS, 477, 5406

    
    \bibitem[Zheng et al.(2007)]{zheng07} 
    Zheng, Z., Coil, A. \& Zehavi, I. 2007, ApJ, 667, 760-779
    
\end{thebibliography}
\end{document}